\documentclass{JHEP3}
\usepackage{epsfig,multicol,bbm}
\usepackage{makeidx}
\usepackage{amsmath}
\usepackage{amsfonts}
\usepackage{amssymb}
\usepackage{graphicx}
\usepackage{accents}
\input{tcilatex}

\newcommand\fverb{\setbox\fverbbox=\hbox\bgroup\verb}
\newcommand\fverbdo{\egroup\medskip\noindent%
            \fbox{\unhbox\fverbbox}\ }
\newcommand\fverbit{\egroup\item[\fbox{\unhbox\fverbbox}]}
\newbox\fverbbox

\title{Entropy of Pairs of
Dual Attractors \\in Six and Seven Dimensions}

\author{El Hassan Saidi\thanks{College SPC, Academie Hassan II des Sciences et Techniques, Rabat, Morocco}\\
   \small Lab/UFR- Physique des Hautes Energies, Facult\'{e} des Sciences,
Rabat, Morocco,\\  E-mail: \email{h-saidi@fsr.ac.ma}}

\author{Antonio Segui \\
   \small Departamento de Fisica Teorica, Universidad de Zaragoza,
50009-Zaragoza, Spain,\\
    E-mail: \email{segui@unizar.es}}
 \preprint{Lab/UFR-HEP/0804-rev, GNPHE/0804-rev}

\abstract{We study the attractor mechanism of dual pairs of black brane bounds in $%
\mathcal{N}=2$ supergravity in six and seven dimensions. First, we
consider the effective potentials of the \emph{6D} and \emph{7D}
black branes as well as their entropies. The contribution coming
from the $SO\left( 1,1\right) $ factor of the moduli spaces $M_{6D}$
and $M_{7D}$ of these theories is carefully analyzed and it is used
to motivate the study of the dual black branes bounds; which in turn
allow to fix the critical value of the dilaton at horizon. The
attractor eqs of the black branes and the bound pairs are derived by
combining the criticality conditions of the corresponding effective
potentials and the Lagrange multiplier method capturing constraints
eqs on the fields moduli.}


\begin{document}


\section{Introduction}
Supersymmetric and non supersymmetric black attractors have received
an increasing interest in the framework of supergravity theories \textrm{\cite%
{a1}-\cite{a4}}; especially in the case of those supergravity models
embedded in \emph{10D} superstrings and \emph{11D} M-theory
compactifications \textrm{\cite{a5}-\cite{a15}}. New solutions to
the attractor equations describing BPS and non-BPS states have been
obtained and
many results concerning supergravity theories in \emph{four} and \emph{higher%
} dimensional space times have been derived \textrm{\cite{a16}}-\textrm{\cite%
{a17}. }For reviews, see for instance
\textrm{\cite{B1}-\cite{B3}}.\newline In this paper, we contribute
to this matter by studying the attractor mechanism and the entropy
$\mathcal{S}$ of the two following \emph{6D/7D} black brane systems:

\ \newline
\textbf{1}) the first system we consider concerns generic \emph{%
Electrically\ }charged\emph{\ Black Branes} (\emph{EBB }for short) in $%
\mathcal{N}=2$ supergravity theory in six and seven dimensional
space times. If most of the basic properties of these \emph{EBB}s
with electric charges
\begin{equation}
\begin{tabular}{llllll}
$q_{\Lambda }$ & $\neq $ & $0$ & , & $\Lambda =1,...$ & ,%
\end{tabular}%
\end{equation}%
but no magnetic charges
\begin{equation}
\begin{tabular}{llllll}
$g_{\Lambda }$ & $=$ & $0$ & , & $\Lambda =1,...$ & ,%
\end{tabular}%
\end{equation}%
are quite known; there are nevertheless some specific properties
that need more studies. Here, we would like to shed more light on
the \emph{EBB} entropy and the electric/magnetic duality which, as
we will show, turn out to be strongly related:\newline \textbf{a})
Concerning the entropy \emph{S}$_{EBB}=$\emph{S}$_{EBB}\left(
q\right) $ of the \emph{EBB} black attractors in \emph{6D} and
\emph{7D}, it
turns out that it takes a very remarkable value\textrm{\footnote{%
Black holes could be either small or large depending on whether the
corresponding classical horizon area is zero or non zero \textrm{\cite{sen,a,b,c}}%
. If we naively apply the Bekenstein-Hawking entropy-area formula to
the small black holes, their entropy vanishes and the expected
quantum degrees of freedom seem to totally \emph{disappear}. This
discrepancy comes from the fact that the general relativity is only
a classical effective theory of quantum gravity opening then a way
to deal with small black holes in connection with $R^{2}$
corrections and supersymmetry enhancement in near horizon geometry
\textrm{\cite{sen1,a,b,c,sen2}}. Small black holes exist also in
higher dimensions. In $5D$, an explicit study of small black holes in $%
\mathcal{N}=2$ and $\mathcal{N}=4$ supergravity can be found in \textrm{\cite%
{m} }and refs therein.}} namely,
\begin{equation}
\begin{tabular}{llll}
$\emph{S}_{EBB}$ & $=$ & $0$ & .%
\end{tabular}
\label{se}
\end{equation}%
This degenerate value will be analyzed in details throughout this
study by using the criticality method; but to fix the ideas think
about it as given by the $g_{\Lambda }\rightarrow 0$ limit of the
following relation to be explicitly derived in this work,
\begin{equation}
\emph{S}_{EBB}=\frac{1}{2}\lim_{g_{\Lambda }\rightarrow 0}\left( \sqrt{%
\left\vert q^{2}g^{2}\right\vert }\right) =0,  \label{qp}
\end{equation}%
with $q^{2}=\sum_{\Lambda }\left( q_{\Lambda }q^{\Lambda }\right) $ and $%
g^{2}=\sum_{\Lambda }\left( g_{\Lambda }g^{\Lambda }\right) $.
\newline In an attempt to analyze what kind of information we can
extract from the \emph{classical} relation $\emph{S}_{EBB}\left(
q\right) =0$, we ended with the conclusion that this degenerate
value could be interpreted as \emph{a singular limit} of a bound
state of the following pair of dual black
attractor%
\begin{equation}
\begin{tabular}{ll}
$\text{\emph{EBB}}\emph{-}\text{\emph{MBB}}$ & ,%
\end{tabular}
\label{eb}
\end{equation}%
where \emph{MBB} stands for the magnetic dual of \emph{EBB}.
\newline The black attractor bound state \emph{EBB-MBB} will be
introduced and commented succinctly below; see the point
(\textbf{2}) of this motivating presentation. But explicit details
and extensive comments will be given in the \emph{section 6} of this
work. \newline \textbf{b}) Concerning the electric/magnetic duality,
it is used to deal with the \emph{Magnetically} charged \emph{Black
Branes} (\emph{MBB}). Roughly, this duality exchanges the charges of
the $\emph{EBB}$ and the
corresponding \emph{MBB} dual along the standard correspondence,%
\begin{equation}
\begin{tabular}{llll}
\emph{EBB} & $\underleftrightarrow{\text{{\small electric/magnetic
duality}}}
$ & \emph{MBB} & .%
\end{tabular}%
\end{equation}%
In this study, we will show that electric/magnetic duality is in
fact a \textit{universal} symmetry of \emph{EBB} and \emph{MBB
}attractors. It exchanges not only the electric $\left\{ q_{\Lambda
}\right\} $ and magnetic $\left\{ g_{\Lambda }\right\} $ charges
($q_{\Lambda }\leftrightarrow g_{\Lambda }$); but also the effective
scalar potentials $\mathcal{V}_{EBB}$
and $\mathcal{V}_{MBB}$ as well as the corresponding entropies\emph{\ }$%
\mathcal{S}_{EBB}$ and $\mathcal{S}_{MBB}$ as illustrated below,%
\begin{equation*}
\begin{tabular}{llllll}
\underline{\emph{EBB}} &  & $\underleftrightarrow{\text{{\small %
electric/magnetic duality}}}$ &  & \underline{\emph{MBB}} &  \\
$q_{\Lambda }$ &  & $\ \ \ \ \ \ \ \ \ \ \ \ \ \longleftrightarrow $ &  & $%
g_{\Lambda }$ &  \\
$\mathcal{V}_{EBB}$ &  & $\ \ \ \ \ \ \ \ \ \ \ \ \
\longleftrightarrow $ &
& $\mathcal{V}_{MBB}$ &  \\
$\mathcal{S}_{EBB}$ &  & $\ \ \ \ \ \ \ \ \ \ \ \ \
\longleftrightarrow $ &
& $\mathcal{S}_{MBB}$ & .%
\end{tabular}%
\end{equation*}%
From this correspondence, we immediately conclude that the entropy $\mathcal{%
S}_{MBB}=\mathcal{S}_{MBB}\left( g\right) $ of the magnetically
charged
black brane \emph{MBB} should be identically zero; in agreement with eq(\ref%
{qp}). The relation $\mathcal{S}_{MBB}=0$ will be rigourously
derived in \emph{sub-section 5.2}. \newline
We also learn that the scalar potentials $\mathcal{V}_{EBB}$ and $\mathcal{V}%
_{MBB}$ are intimately related as it will be explicitly shown in \emph{%
section 5}.\

\ \newline \textbf{2}) the second system that we want to study in
this paper concerns precisely generic \emph{Dual Black Brane Pairs}
( dual pairs \emph{DP }for short). \newline A \emph{DP} attractor
can defined as a bound state consisting of an electrically charged
brane \emph{EBB} and its magnetic dual \emph{MBB}.
Formally, we can represent a generic \emph{DP} bound state either as in eq(%
\ref{eb}) or roughly, by using group theory representation language,
like a
doublet\emph{\ }%
\begin{equation}
\begin{tabular}{llll}
$DP$ & $\sim $ & $\left(
\begin{array}{c}
\text{\emph{EBB}} \\
\text{\emph{MBB}}%
\end{array}%
\right) $ & .%
\end{tabular}
\label{dp}
\end{equation}%
In this set up, the \emph{EBB} attractor considered in point
(\textbf{1}), with the degenerate entropy \emph{S}$_{EBB}=0$, can be
thought of as
corresponding to the \emph{singular} limit%
\begin{equation}
\begin{tabular}{llll}
$DP$ & $\underrightarrow{g_{\Lambda }\rightarrow 0}$ & $\left(
\begin{array}{c}
EBB \\
0%
\end{array}%
\right) $ & ,%
\end{tabular}
\label{dm}
\end{equation}%
describing a singular geometry where the horizon area
\emph{A}$_{MBB}$ of
the \emph{MBB} attractor shrinks to a singular point (\emph{A}$%
_{MBB}\rightarrow 0$). \newline The same picture is valid for the
dual \emph{MBB} attractor which
corresponds to the degenerate limit,%
\begin{equation}
\begin{tabular}{llll}
$DP$ & $\underrightarrow{q_{\Lambda }\rightarrow 0}$ & $\left(
\begin{array}{c}
0 \\
MBB%
\end{array}%
\right) $ & ,%
\end{tabular}%
\end{equation}%
describing the electric/magnetic dual of eq(\ref{dm}).\newline
Before proceeding ahead, we would like to notice that the results we
will derive below for the \emph{DP} attractors apply as well
to:\newline \textbf{i)} the dyonic \emph{Black String }(\emph{BS
}for short) of the \emph{6D} $\mathcal{N}=2$ supergravity,\newline
\textbf{ii)} the \emph{DP} brane bounds in all supergravity theories
with
scalar manifolds of the form $SO\left( 1,1\right) \times \left( G/H\right) $%
. \newline Regarding the \emph{6D} black string \emph{BS}, it can be
viewed as a
particular representation of the electric magnetic duality group. The \emph{%
BS} is a pure singlet while \emph{DP }is based on the\emph{\ pair }(\ref{dp}%
). \newline Moreover, it is interesting to have in mind that,
despite their geometric differences, the $\emph{BS}$ and \emph{DP}
entropy formulas are also
comparable. This feature can be explicitly checked by comparing the $%
\mathcal{S}_{\emph{DP}}$ formula (\ref{qp}) and the \emph{BS}
entropy
relation $\mathcal{S}_{\emph{BS}}$ to be derived in \emph{section 4} eq(\ref%
{bfe}), and which we recall below,%
\begin{equation}
\mathcal{S}_{\emph{BS}}=\frac{\left\vert q_{0}g_{0}\right\vert }{2}=\frac{1}{%
2}\sqrt{q_{0}^{2}g_{0}^{2}}\text{ },  \label{bse}
\end{equation}%
with $q_{0}$ and $g_{0}$ being respectively the electric and
magnetic
charges of the \emph{6D} black string. As we see, the above $\mathcal{S}_{%
\emph{BS}}$ expression and the $\mathcal{S}_{\emph{DP}}$ relation
(\ref{qp}) have more a less the same charge dependence structure.
\newline
Concerning the second feature; it has been pointed\textrm{\ }out in \textrm{%
\cite{B3,BDSS,S2}}, that supergravity theories with scalar manifolds
having
an $SO\left( 1,1\right) $ factor would have zero entropies. The result $%
\mathcal{S}_{\text{EBB}}=0$ of eq(\ref{se}) and, up on using
electric/magnetic duality
\begin{equation}
\mathcal{S}_{\text{MBB}}=0\text{ ,}  \label{sm}
\end{equation}%
should be then thought of as special relations that are valid as
well for supergravity theories beyond those embedded in \emph{10D}
type IIA superstring and \emph{11D} M-theory on K3 we are
considering in this work.

On the other hand, we will also take the opportunity of the use of
the criticality condition of the black branes effective potentials
to develop a tricky approach to get the BPS and non BPS states
solutions by using an adapted Lagrange multiplier method. Details on
this issue will be given in \emph{section 5} of this study. BPS and
non BPS black holes as well as black membranes are distinguished by
the values (\ref{sol1}-\ref{sol3},\ref{cr}) of the Lagrange
multipliers at the minimum of the effective potential. The Lagrange
multipliers $\left\{ \lambda ^{\Lambda \Sigma }\right\} $ given by
eq(\ref{lan}) capture the constraint eqs(\ref{ort},\ref{art}) on the
fields coordinates $\left\{ L_{\Lambda \Sigma }\right\} $
(\ref{lac}) that are used to parameterize the moduli space of the
theory.\newline

The organization of this paper is as follows: In section 2, we
describe briefly some useful tools; in particular the derivation of
the singular value (\ref{se}). In section 3, we study the \emph{EBB}
and \emph{MBB} attractors as well as the \emph{DP} black attractor
bounds in \emph{6D} and \emph{7D} $\mathcal{N}=2$ supergravity
theories. In section 4, we consider with details the \emph{6D} black
string. We show, amongst others, that its entropy is invariant under
electric/magnetic duality and conclude with a general result on
dyonic duals pairs of black branes. In section 5, we study the BPS
and non BPS black attractors in \emph{6D} by using a new method.
This approach is based on combining the criticality of the effective
potential and the Lagrange multiplier method capturing constraint
eqs on the
field moduli. Using this approach we derive the attractors eqs of the \emph{%
6D}\ black hole \emph{BH} and \emph{6D} black membrane \emph{BM}. We
also give the explicit solutions as well as their entropies. In
section 6, we derive the effective potential
$\mathcal{V}_{\emph{DP}}$ of the dyonic dual pair bound \emph{DP
}$\equiv $ \emph{BH-BM}.\emph{\ }Then we study the attractor
mechanism for the dyonic \emph{DP} and derive the general formula
for its entropy $\mathcal{S}_{\emph{DP}}$. This result obtains for
the 6D apply as well to the dyonic black hole-black 3- brane
(\emph{BH-B3B}) and the (\emph{BS-BM}) bound state of the 7D theory.
In section 7, we give the conclusion and discussion and in section
8, we give an appendix.

\section{General tools}
To exhibit explicitly the particular features
\begin{equation}
\begin{tabular}{llllllll}
$\mathcal{S}_{\text{EBB}}^{6D}$ & $=$ & $0$ & \qquad ,\qquad \qquad & $%
\mathcal{S}_{\text{EBB}}^{7D}$ & $=$ & $0$ & , \\
$\mathcal{S}_{\text{MBB}}^{6D}$ & $=$ & $0$ & \qquad ,\qquad \qquad & $%
\mathcal{S}_{\text{MBB}}^{7D}$ & $=$ & $0$ & ,%
\end{tabular}%
\end{equation}%
of the entropies of the electrically charged \emph{EBB} and the
magnetically charged \emph{MBB }in six and seven space time
dimensions, it is interesting to start by describing briefly some
useful results.\newline We begin by recalling that the moduli space
$\boldsymbol{M}_{6D}^{N=2}$ of the \emph{6D} $\mathcal{N}=2$
supergravity theory embedded in \emph{10D}
type IIA superstring on K3 is given by the following Lie group coset%
\begin{equation}
\begin{tabular}{llll}
$\boldsymbol{M}_{6D}^{N=2}$ & $=$ & $SO\left( 1,1\right) \times
G_{6}$ & ,
\\
$G_{6}$ & $=$ & $\frac{SO\left( 4,20\right) }{SO\left( 4\right)
\times
SO\left( 20\right) }$ & .%
\end{tabular}
\label{0}
\end{equation}%
A quite similar factorization holds for the scalar manifold $\boldsymbol{M}%
_{7D}^{N=2}$ of the $\mathcal{N}=2$ supergravity theory in \emph{7D}
space
time embedded in \emph{11D} M-theory on K3. It reads as follows%
\begin{equation}
\begin{tabular}{llll}
$\boldsymbol{M}_{7}$ & $=$ & $SO\left( 1,1\right) \times G_{7}$ & , \\
$G_{7}$ & $=$ & $\frac{SO\left( 3,19\right) }{SO\left( 3\right)
\times
SO\left( 19\right) }$ & .%
\end{tabular}
\label{01}
\end{equation}%
As we see, the two scalar manifolds $\boldsymbol{M}_{6D}^{N=2}$ and $%
\boldsymbol{M}_{7D}^{N=2}$ are given by the product of two factors namely $%
SO\left( 1,1\right) $ and $G_{n}$ with $n=6,7$. \newline The real
one dimensional factor $SO\left( 1,1\right) $ is parameterized by a
real field variable $\sigma $, to be interpreted as the \emph{6D}
(resp. \emph{7D}) dilaton $\sigma =\sigma \left( x\right) $.
\newline The factor $G_{6}$ is real \emph{80} dimensional manifold
parameterized by
the real field coordinates,%
\begin{equation}
\begin{tabular}{llll}
$\phi ^{aI}\left( x\right) $ & $\simeq $ & $\left( \underline{\mathbf{4}},%
\underline{\mathbf{20}}\right) $ & ,%
\end{tabular}%
\end{equation}%
transforming in the bi-fundamental of the $SO\left( 4\right) \times
SO\left( 20\right) $ isotropy symmetry with $a=1,...,4$ and
$I=1,...,20$. \newline The factor $G_{7}$ is real \emph{57}
dimensional manifold parameterized by
the field coordinates%
\begin{equation}
\begin{tabular}{llll}
$\xi ^{\alpha i}\left( x\right) $ & $\simeq $ & $\left( \underline{\mathbf{3}%
},\underline{\mathbf{19}}\right) $ & ,%
\end{tabular}%
\end{equation}%
transforming in the bi-fundamental of the $SO\left( 3\right) \times
SO\left( 19\right) $ isotropy symmetry with $\alpha =1,2,3$ and
$i=1,...,19$.\newline As the technical analysis of eqs(\ref{0}) and
(\ref{01}) is quite similar, we will focus our attention mainly on
the \emph{6D} theory and just give the results for the \emph{7D}
case.

\subsection{Effective potential}

\textrm{The effective potential }$\mathcal{V}$\textrm{\ of black
attractors
in generic space time D- dimensional extended supergravity, including \emph{%
6D} }$\mathcal{N}=2$, \textrm{\ have been studied in \cite{B3}; see
also \cite{BDSS} as well as the appendix of this paper}. \textrm{The
general form
of this potential reads formally, in terms of the geometric $\mathcal{Z}$}$%
_{geo}$\textrm{\ and the matter }$\mathcal{Z}_{matter}$\textrm{\
central
charges, as follows}%
\begin{equation*}
\mathcal{V}\left( \phi \right) \sim \left\vert \mathrm{\mathcal{Z}}%
_{geo}\left( \phi \right) \right\vert ^{2}+\left\vert \mathcal{Z}%
_{matter}\left( \phi \right) \right\vert ^{2}\text{.}
\end{equation*}%
Notice that $\mathrm{\mathcal{Z}}_{geo}$ has contributions coming
from the physical charges of the various gauge fields of the gravity
supermultiplet while $\mathcal{Z}_{matter}$ has contributions
\textrm{coming from the gauge fields in the matter sector. }\newline
\textrm{In the case of 6D }$\mathcal{N}=2$\textrm{\ non chiral
supergravity,
we have the following gauge field strengths, }%
\begin{equation}
\begin{tabular}{llllll}
gravity multiplet & : & $\mathcal{H}_{3}=d\mathcal{B}_{2}$ & , & $\mathcal{F}%
_{2}^{a}=d\mathcal{A}_{1}^{a}$ & , \\
matter multiplets & : & $\mathcal{F}_{2}^{I}=d\mathcal{A}_{1}^{I}$ &  &  & ,%
\end{tabular}
\label{bb}
\end{equation}%
together with their magnetic duals $\mathcal{G}_{3}$,
$\mathcal{G}_{4}^{a}$
and $\mathcal{G}_{4}^{I}$. So \textrm{\ }$\mathcal{Z}_{geo}^{6D,N=2}$\textrm{%
\ and }$\mathcal{Z}_{matter}^{6D,N=2}$ have contributions from the
charges
of $\left( \mathcal{H}_{3},\mathcal{G}_{3}\right) $ , $\left( \mathcal{F}%
_{2}^{a},\mathcal{F}_{2}^{I}\right) $ and $\left( \mathcal{G}_{4}^{a},%
\mathcal{G}_{4}^{I}\right) $; and then the full effective potential $%
\mathcal{V}^{6D,N=2}$ involves three blocks namely $\mathcal{V}_{{\small %
black\ string}}$, $\mathcal{V}_{{\small black\ hole}}$ and $\mathcal{V}_{%
{\small black\ membrane}}$. Notice also that in eq(\ref{bb}), $\mathcal{B}%
_{2}=\frac{1}{2}dx^{\mu }dx^{\nu }$\textrm{\emph{B}}$_{\left[ \mu
\nu \right]
}$ is \textrm{the usual NS-NS \emph{B}}$_{\mu \nu }$\textrm{\emph{- field }}%
in 6D\textrm{, the gauge fields }$\mathcal{A}_{\mu }^{a}$\textrm{\
stand for the four graviphotons and }$\mathcal{A}_{\mu
}^{I}$\textrm{\ for the twenty Maxwell fields of the non chiral 6D
supergravity embedded in type IIA superstring on K3, see
eqs(\ref{fff}-\ref{ggg}) to fix the ideas. }\newline
\textrm{Following \cite{B3} and \cite{BDSS}, we can compute
explicitly the
various contributions }$\mathcal{V}_{{\small black\ string}}$\textrm{, }$%
\mathcal{V}_{{\small black\ hole}}$\textrm{\ and }$\mathcal{V}_{{\small %
black\ membrane}}$ by using the following generic relations,\textrm{\ }%
\begin{equation*}
\begin{tabular}{llllll}
$\mathcal{V}_{{\small black\ string}}$ & $\sim $ & $\left\vert \mathrm{%
\mathcal{Z}}_{geo}^{BS}\right\vert ^{2}$ & $+$ & $\left\vert \mathcal{Z}%
_{matter}^{BS}\right\vert ^{2}$ & , \\
$\mathcal{V}_{{\small black\ hole}}$ & $\sim $ & $\left\vert \mathrm{%
\mathcal{Z}}_{geo}^{BH}\right\vert ^{2}$ & $+$ & $\left\vert \mathcal{Z}%
_{matter}^{BH}\right\vert ^{2}$ & , \\
$\mathcal{V}_{{\small black\ membrane}}$ & $\sim $ & $\left\vert \mathrm{%
\mathcal{Z}}_{geo}^{BM}\right\vert ^{2}$ & $+$ & $\left\vert \mathcal{Z}%
_{matter}^{BM}\right\vert ^{2}$ & .%
\end{tabular}%
\end{equation*}%
\textrm{\ These contributions, which are respectively associated with }$%
\left( \mathcal{H}_{3},\mathcal{G}_{3}\right) $ , $\left( \mathcal{F}%
_{2}^{a},\mathcal{F}_{2}^{I}\right) $ and $\left( \mathcal{G}_{4}^{a},%
\mathcal{G}_{4}^{I}\right) $, \textrm{will be studied later on; they
are given by eqs(\ref{vp}), (\ref{blh}), (\ref{b2b}). With these
relations in
mind, we turn now to study some specific properties of these potentials.}%
\newline
One of the consequences of the factorization (\ref{0}) of the manifold $%
\boldsymbol{M}_{6D}^{N=2}$ is that the \emph{EBB} (resp. \emph{MBB})
effective scalar potential
\begin{equation}
\begin{tabular}{llll}
$\mathcal{V}_{6D}^{\text{SBB}}$ & $=$ &
$\mathcal{V}_{6D}^{\text{SBB}}\left(
\sigma ,\phi \right) $ & ,%
\end{tabular}
\label{sb}
\end{equation}%
where the upper index \emph{SBB} stands either for \emph{EBB} or
\emph{MBB}, factorizes as well like
\begin{equation}
\begin{tabular}{llll}
$\mathcal{V}_{6D}^{\text{SBB}}$ & $=$ & $\mathcal{V}_{SO\left(
1,1\right)
}\times \mathcal{V}_{G_{6}}$ & .%
\end{tabular}
\label{10}
\end{equation}%
The term in the right hand of the above relation,%
\begin{equation}
\begin{tabular}{llll}
$\mathcal{V}_{{\small SO}\left( 1,1\right) }$ & $=$ & $\mathcal{V}%
_{dil}\left( \sigma \right) $ & ,%
\end{tabular}%
\end{equation}%
is the dilaton contribution to eq(\ref{10}); it has no dependence in
the local field coordinates $\phi ^{aI}$; that is no dependence in
the matter fields of the Maxwell sector of the theory,
\begin{equation}
\frac{\partial \mathcal{V}_{{\small SO}\left( 1,1\right) }}{\partial
\phi ^{aI}}=0.
\end{equation}%
We will see later on that this contribution is given by the typical
remarkable relation%
\begin{equation}
\begin{tabular}{llll}
$\mathcal{V}_{{\small SO}\left( 1,1\right) }\left( \sigma \right) $ & $%
\simeq $ & $\exp \left( \mathrm{n}\sigma \right) $ & ,%
\end{tabular}
\label{11}
\end{equation}%
where the number $\mathrm{n}$ depends on the type of the black brane
we are
dealing with. More precisely, we have the following values \textrm{\cite%
{BDSS,S2}},%
\begin{equation}
\begin{tabular}{llllll}
{\small 6D\ black string} & $:$ & $\mathrm{n}$ & $=$ & $\pm 4$ & , \\
{\small 6D black hole} & $:$ & $\mathrm{n}$ & $=$ & $-2$ & , \\
{\small 6D black membrane} & $:$ & $\mathrm{n}$ & $=$ & $+2$ & .%
\end{tabular}%
\end{equation}%
The factor $\mathcal{V}_{G_{6}}$ of (\ref{10}) has no dependence in
the dilaton
\begin{equation}
\begin{tabular}{llll}
$\mathcal{V}_{G_{6}}$ & $=$ & $\mathcal{V}_{G_{6}}\left( \phi
\right) $ & ,
\\
$\frac{\partial \mathcal{V}_{G_{6}}}{\partial \sigma }$ & $=$ & $0$ & ,%
\end{tabular}%
\end{equation}%
it describes the contribution of the matter fields $\left\{ \phi
^{aI}\right\} $ in the Maxwell sector of the 6D $\mathcal{N}=2$
supergravity theory. The explicit field expression of
$\mathcal{V}_{G_{6}}$ in terms of the $\phi ^{aI}$ will be given
later on.

\subsection{Criticality condition}

First we study the electrically (\emph{resp}. magnetically) charged
black brane \emph{EBB} (\emph{resp}. \emph{MBB}). Then we consider
the case of the dyonic black string \emph{BS}.\newline The critical
values $\left( \sigma ,\phi \right) =\left( \sigma _{c},\phi
_{c}\right) $ of the effective scalar potential $\mathcal{V}_{6D}^{\text{SBB}%
}$ (\ref{sb}-\ref{10}) are obtained by solving the constraint equations%
\begin{equation}
\begin{tabular}{llll}
$\frac{\partial \mathcal{V}_{6D}^{\text{SBB}}}{\partial \sigma }$ &
$=$ & $0$
& , \\
$\frac{\partial \mathcal{V}_{6D}^{\text{SBB}}}{\partial \phi ^{aI}}$
& $=$ &
$0$ & ,%
\end{tabular}%
\end{equation}%
which, due to the factorization property (\ref{10}), simplify like,
\begin{equation}
\begin{tabular}{llll}
$\frac{\partial \mathcal{V}_{{\small SO}\left( 1,1\right)
}}{\partial \sigma
}$ & $=$ & $0$ & , \\
$\frac{\partial \mathcal{V}_{G_{6}}}{\partial \phi ^{aI}}$ & $=$ & $0$ & .%
\end{tabular}%
\end{equation}%
The critical value $\sigma _{c}$ of the dilaton that extremize the
potential
$\mathcal{V}_{SO\left( 1,1\right) }$, and then $\mathcal{V}_{6D}^{\text{SBB}%
} $, is obtained by computing
\begin{equation}
\frac{\partial \mathcal{V}_{SO\left( 1,1\right) }}{\partial \sigma
}\simeq
\frac{\partial \left[ e^{n\sigma }\right] }{\partial \sigma }=ne^{n\sigma }=0%
\text{ },
\end{equation}%
from which we learn that the critical point corresponds to the
following infinite value,
\begin{equation}
n\sigma _{c}\longrightarrow -\infty .
\end{equation}%
For $n>0$, $\sigma _{c}\longrightarrow -\infty $ and for $n<0$,
$\sigma _{c}\longrightarrow +\infty $.\newline
Putting this value back into $\mathcal{V}_{SO\left( 1,1\right) }$ eq(\ref{11}%
), we see that the value of the potential $\mathcal{V}_{SO\left(
1,1\right)
} $ at the critical point vanishes identically; i.e,%
\begin{equation}
\begin{tabular}{llll}
$\left[ \mathcal{V}_{SO\left( 1,1\right) }\right] _{\sigma =\sigma _{c}}$ & $%
=$ & $0$ & .%
\end{tabular}%
\end{equation}%
Because of the factorization (\ref{10}), we also have%
\begin{equation}
\begin{tabular}{llll}
$\left[ \mathcal{V}_{6D}^{\text{SBB}}\right] _{\sigma =\sigma _{c}}$
& $=$ &
$0$ & ,%
\end{tabular}%
\end{equation}%
leading as well to the zero entropy relation,
\begin{equation}
\begin{tabular}{llll}
$\mathcal{S}_{6D}^{\text{SBB}}$ & $=$ & $0$ & $,$%
\end{tabular}
\label{ebd}
\end{equation}%
in agreement with eq(\ref{se}).\newline For the \emph{dyonic} 6D
black string, the situation is different. The form of the
corresponding effective potential $\mathcal{V}_{BS}$ has the
following field moduli factorization,%
\begin{equation}
\begin{tabular}{llll}
$\mathcal{V}_{BS}$ & $=$ & $\mathcal{V}_{{\small SO}\left(
1,1\right)
}\left( \sigma \right) \times \mathcal{V}_{6}\left( \phi \right) +\mathcal{V}%
_{{\small SO}\left( 1,1\right) }\left( -\sigma \right) \times \mathcal{%
\tilde{V}}_{6}\left( \phi \right) $ & $,$%
\end{tabular}%
\end{equation}%
where $\mathcal{V}_{6}\left( \phi \right) $ stands for the
contribution coming from the electric charge and
$\mathcal{\tilde{V}}_{6}\left( \phi \right) $ the contribution
coming from the magnetic charge.\newline As we will see in details
later, it turns out that the solving of the criticality condition of
$\mathcal{V}_{BS}$ leads to a finite critical value of the dilaton
\begin{equation}
\left\vert \sigma _{c}\right\vert <\infty \text{ .}
\end{equation}%
Substituting this value back into $\mathcal{V}_{BS}$, we obtain a
positive definite value of the effective potential at the minimum,
\begin{equation}
\left[ \mathcal{V}_{BS}\left( \sigma \right) \right] _{\sigma =\sigma _{c}}>0%
\text{ },
\end{equation}%
leading in turn to a on a zero value of the entropy
$\mathcal{S}_{BS}$ for
the dyonic 6D black string. The value of $\mathcal{S}_{BS}$ is given by eq(%
\ref{bse}); it will be computed explicitly later on, see
eq(\ref{bfe}).

\section{Duality symmetry and entropy}

First, we describe some useful aspects on:\newline
(\textbf{1}) the \emph{6D} and\emph{7D} attractors and the dual pairs,%
\newline
(\textbf{2}) the gauge invariant n-forms in generic d\emph{-}
dimensions; in particular the electric/magnetic duality
\textrm{\cite{EM1,EM2,EM3}} and the fluxes used to define the
various electric and magnetic charges.\newline Then, we study the
\emph{"dyonic"} attractors in \emph{6D} and \emph{7D}. We will
distinguish the two following cases:\newline (\textbf{a}) the
\emph{6D} Black String \emph{BS}; behaving as a singlet
under electric/magnetic duality%
\begin{equation}
\left( BS\right) \text{ }.
\end{equation}%
No analogous object exists in \emph{7D}.\newline (\textbf{b}) Bound
states of dual \emph{EBB}\ and \emph{MBB} behaving as
pairs under electric magnetic duality%
\begin{equation}
\left(
\begin{array}{c}
EBB \\
MBB%
\end{array}%
\right) .
\end{equation}%
The possible candidates for these bound pairs are:\newline
(\textbf{i}) the \emph{6D} Black Hole - Black Membrane
(\emph{BH-BM}),
\newline
(\textbf{ii}) the \emph{7D} Black Hole - Black 3- Brane (\emph{BH-B3B}),%
\newline
(\textbf{iii}) the \emph{7D} Black String - Black Membrane
(\emph{BS-BM}).\
\newline
Below, we shall focus our attention in a first step on the special
\emph{6D} dyonic string \emph{BS} and its entropy
$\mathcal{S}_{BS}$. \newline Then, we study the basic properties of
the \emph{BH} and \emph{BM} black
attractors separately. This study can be viewed as a prelude to \emph{BH-BM }%
bound.\newline More details on the dual pair \emph{BH-BM} in six
dimensions and its analogues in \emph{7D} will be considered in
section 6 and the discussion section.

\subsection{6D and 7D black attractors}

\qquad Electric/magnetic duality permutes electrically charged
objects and
their magnetic charged duals. In \emph{10D}\ type II superstrings and \emph{%
11D} M-theory compactifications down to \emph{d- dimensions}, this
discrete symmetry relates those pairs of $p_{1}$- and $p_{2}$-
dimensional black
objects with the condition%
\begin{equation}
p_{1}+p_{2}=d-4\text{\qquad },\qquad d\geq 4\text{ }.  \label{pd}
\end{equation}%
From this relation, one recognizes:\ \newline
(\textbf{1}) the \emph{4D} dyonic black hole corresponding to $p_{1}+p_{2}=0$%
. \newline
(\textbf{2}) the \emph{6D} dyonic black string corresponding to $%
p_{1}+p_{2}=2$.\newline
(\textbf{3}) the \emph{8D} dyonic black membrane corresponding to $%
p_{1}+p_{2}=4$.\newline
In six and seven dimensions we are interested in we have the following:%
\newline

\emph{6D case}\newline In the non chiral 6D $\mathcal{N}=2$
supergravity theory embedded in 10D
type IIA superstring on K3, the relation (\ref{pd}) reads as,%
\begin{equation}
p_{1}+p_{2}=2\text{ },
\end{equation}%
and can be solved in three ways like:\newline (\textbf{a}) the case
$\left( p_{1},p_{2}\right) =\left( 1,1\right) $ describing a dyonic
black string (\emph{BS}). \newline The 6D \emph{BS} attractor
carries both an electric charge \textrm{q}$_{0}$ and a magnetic
charge \textrm{g}$_{0}$ associated with the gauge invariant
3-form field strength%
\begin{equation}
\mathcal{H}_{3}=d\mathcal{B}_{2}\text{ ,}
\end{equation}%
of the $\mathcal{N}=2$ supergravity multiplet.\newline (\textbf{b})
the case $\left( p_{1},p_{2}\right) =\left( 0,2\right) $ describing
a magnetic black hole (\emph{BH}).\newline In \emph{6D}, the
\emph{BH} attractor carries \emph{24} magnetic charges
\textrm{g}$_{\Lambda }^{BH}$ (\textrm{g}$_{\Lambda }$ for short)
associated with the gauge invariant fields strengths
\begin{equation}
\mathcal{F}_{2}^{\Lambda }=d\mathcal{A}_{1}^{\Lambda }\text{ ,}
\end{equation}%
of the $\mathcal{N}=2$ supergravity theory. The \emph{6D} \emph{BH}
hole has no electric charge,
\begin{equation}
\mathrm{q}_{\Lambda }^{BH}=0.
\end{equation}%
(\textbf{c}) the case $\left( p_{1},p_{2}\right) =\left( 2,0\right)
$
describing an electric \emph{6D} black membrane (\emph{BM}) carrying \emph{24%
} electric charges \textrm{q}$_{\Lambda }^{BM}$
(\textrm{q}$_{\Lambda }$ for short) related to \textrm{g}$_{\Lambda
}^{BH}$ under electric magnetic duality. \newline The \emph{6D}
black membrane has no magnetic charge
\begin{equation}
\mathrm{g}_{\Lambda }^{BM}=0.
\end{equation}%
The above \emph{BH} and the \emph{BM} attractors are related by
electric/magnetic duality in six dimensions. As such, the bound
state made
of the \emph{6D }black hole \emph{BH} and the \emph{6D} black membrane \emph{%
BM}
\begin{equation}
\begin{tabular}{llllll}
$6D$ & $:$ & $\emph{BH-BM}$ & $\equiv $ & $\left(
\begin{array}{c}
\emph{BH} \\
\emph{BM}%
\end{array}%
\right) $ & ,%
\end{tabular}
\label{hm}
\end{equation}%
form a \emph{dyonic pair} of\ black attractors with 24 electric and
24
magnetic charges%
\begin{equation}
\left\{ \mathrm{q}_{\Lambda },\mathrm{g}_{\Lambda }\right\} \qquad
,\qquad \Lambda =1,...,24.
\end{equation}%
Viewed as a single entity, the composite state $\emph{BH-BM}$
should, \`{a} priori, share the basic features of the dyonic black
string \emph{BS}; in particular sharing aspects of the effective
potentials and their entropies. We will study these features details
later on.\newline

\emph{7D case}\newline In the case of \emph{7D} $\mathcal{N}=2$
supergravity theory embedded in
\emph{11D} M-theory on K3, the relation (\ref{pd}) becomes%
\begin{equation}
p_{1}+p_{2}=3
\end{equation}%
and it is solved in four manners as follows: \newline (\textbf{a})
the case $\left( p_{1},p_{2}\right) =\left( 0,3\right) $ describing
a magnetic \emph{7D} black hole (\emph{BH}),\newline \textbf{(b)}
the case $\left( p_{1},p_{2}\right) =\left( 3,0\right) $ describing
an electric \emph{7D} black 3-brane (\emph{B3B}), dual to the black
hole.\newline \textbf{(c)} the case $\left( p_{1},p_{2}\right)
=\left( 1,2\right) $ describing a magnetic \emph{7D} black string
(\emph{BS}).\newline \textbf{(d)} the case $\left(
p_{1},p_{2}\right) =\left( 2,1\right) $ describing an electric
\emph{7D} black 2-brane (\emph{BM}), dual to the black
string.\newline The 7D $\mathcal{N}=2$ supergravity theory embedded
in \emph{11D} M-theory
on K3 has the following abelian gauge symmetry,%
\begin{equation}
U_{NS}\left( 1\right) \times U^{3}\left( 1\right) \times
U^{19}\left( 1\right) .
\end{equation}%
The 7D\ black hole \emph{BH} and black 3-brane \emph{B3B} are
charged under the $U^{22}\left( 1\right) $ gauge symmetry of the
supergravity theory while the 7D black string \emph{BS} and black
membrane \emph{BM} are charged under the gauge invariant 3-form and
its dual 4-form.

\ \newline Notice that in \emph{7D} $\mathcal{N}=2$ supergravity
theory, we have no dyonic singlet; but rather two kinds of dyonic
pairs: \newline \textbf{(i)} the pair
\begin{equation}
\begin{tabular}{llll}
\emph{BH-B3B} & $\equiv $ & $\left(
\begin{array}{c}
\emph{BH} \\
\emph{B3B}%
\end{array}%
\right) $ & ,%
\end{tabular}
\label{za}
\end{equation}%
carrying \emph{22} electric charges $\left\{
q_{1},...,q_{22}\right\} $ and \emph{22} magnetic ones $\left\{
g_{1},...,g_{22}\right\} $.\newline
\textbf{(ii)}\emph{\ }the pair\emph{\ }%
\begin{equation}
\begin{tabular}{llll}
\emph{BS-BM} & $\equiv $ & $\left(
\begin{array}{c}
\emph{BS} \\
\emph{BM}%
\end{array}%
\right) $ & ,%
\end{tabular}
\label{az}
\end{equation}%
\emph{\ }carrying an electric charge q$_{0}$ and a magnetic charge
g$_{0}$.
\newline
Notice also that there is a correspondence between the attractors in
6D and 7D space time dimensions. We have,
\begin{equation}
\begin{tabular}{llll}
$6D:\emph{BS}$ & $\qquad \longleftrightarrow \qquad $ &
$7D:\emph{BS-BM}$ & ,
\\
$6D:\emph{BH-BM}$ & $\qquad \longleftrightarrow \qquad $ &
$7D:\emph{BH-B3B}$
& .%
\end{tabular}%
\end{equation}%
This property is a consequence following from the relation between
\emph{11D} M- theory and \emph{10D} type IIA superstring; which
after compactification on K3, descends to the \emph{6D} and the
\emph{7D} space times.

\subsection{Dyonic attractors in 6D supergravity}

\qquad To start recall that in the d- dimensional space time, a
gauge invariant $\left( p+2\right) $- form field strength ($p\leq
d-2$),
\begin{equation}
\begin{tabular}{llll}
$\mathcal{F}_{p+2}$ & $=$ & $\frac{1}{\left( p+2\right) !}dx^{\mu
_{p+2}}\ldots dx^{\mu _{1}}\mathcal{F}_{\mu _{1}...\mu _{p+2}}$ & $,$%
\end{tabular}%
\end{equation}%
\ with a $\left( p+1\right) $- form gauge connection
\begin{equation}
\begin{tabular}{llll}
$\mathcal{A}_{p+1}$ & $=$ & $\frac{1}{\left( p+1\right) !}dx^{\mu
_{p+1}}\ldots dx^{\mu _{1}}\mathcal{A}_{\mu _{1}...\mu _{p+1}}$ & $,$%
\end{tabular}%
\end{equation}%
has a Poincar\'{e} (magnetic) dual given by
\begin{equation}
\begin{tabular}{llll}
$\mathcal{G}_{d-p-2}$ & $=$ & $^{\star }\mathcal{F}_{p+2}$ & .%
\end{tabular}%
\end{equation}%
with the usual property
\begin{equation}
^{\ast }\mathcal{G}_{d-p-2}=-\left( -\right) ^{\left( p+2\right)
\left( d-p-2\right) }\mathcal{F}_{p+2}.
\end{equation}%
Expanding $\mathcal{G}_{d-p-2}$,
\begin{equation}
\begin{tabular}{llll}
$\mathcal{G}_{d-p-2}$ & $=$\  & $\frac{1}{\left( p+2\right) !\left(
d-p-2\right) !}dx^{\mu _{D}}\cdots dx^{\mu _{p+3}}\mathcal{G}_{\mu
_{p+3}...\mu _{d}}$ & $,$%
\end{tabular}%
\end{equation}%
we also have%
\begin{equation}
\begin{tabular}{llll}
$\mathcal{G}^{\mu _{p+3}...\mu _{d}}$ & $=$ & $\mathcal{F}_{\mu
_{1}...\mu
_{p+2}}\varepsilon ^{\mu _{1}...\mu _{p+2}\mu _{p+3}...\mu _{d}}$ & $,$%
\end{tabular}%
\end{equation}%
with $\varepsilon ^{\mu _{1}...\mu _{p+2}\mu _{p+3}...\mu _{d}}$ being the $%
d $- dimensional completely antisymmetric tensor. \newline The
magnetic charge $\mathrm{g}$ (resp electric charge $\mathrm{q}$)
associated with the field strength $\mathcal{F}_{p+2}$ (resp. $\mathcal{G}%
_{d-p-2}$) can be defined as
\begin{equation}
\begin{tabular}{ll}
$\mathrm{g}=\int_{S^{p+2}}\mathcal{F}_{p+2}$ & $,$ \\
$\mathrm{q}=\int_{S^{d-p-2}}\mathcal{G}_{d-p-2}$ & $.$%
\end{tabular}
\label{q}
\end{equation}%
Using the normalized n- volume form $\Omega _{n}$ of the real n- sphere $%
\mathbb{S}^{n}$,
\begin{equation}
\int_{\mathbb{S}^{n}}\Omega _{n}=1\text{ },\qquad n=p+2\text{ \ \ or \ \ }%
d-p-2\text{ },
\end{equation}%
we can also express the gauge invariant field strengths as follows,%
\begin{equation}
\begin{tabular}{llll}
$\mathcal{F}_{p+2}$ & $=$ & $\mathrm{g}\text{ }\Omega _{p+2}$ & $,$ \\
$\mathcal{G}_{d-p-2}$ & $=$ & $\mathrm{q}\text{ }\Omega _{d-p-2}$ & ,%
\end{tabular}%
\end{equation}%
with
\begin{equation}
\begin{tabular}{llll}
$\Omega _{p+2}\wedge \Omega _{d-p-2}$ & $\simeq $ & $V_{d}$ & $.$%
\end{tabular}%
\end{equation}%
where $V_{d}$ is the volume real d- form of the space time. We also
have
\begin{equation}
\begin{tabular}{llll}
$\mathcal{F}_{p+2}\wedge \mathcal{G}_{d-p-2}$ & $\simeq $ & $\mathrm{gq}%
V_{d} $ & ,%
\end{tabular}%
\end{equation}%
with the following quantization condition relating electric and
magnetic
sectors,%
\begin{equation}
\begin{tabular}{llllll}
$\mathrm{gq}$ & $\mathrm{=}$ & $\text{ }2\pi k$ & , & $k\text{ \
integer}$ &
.%
\end{tabular}%
\end{equation}%
Seen that the analysis for \emph{6D} and the analysis for \emph{7D}
are
quite similar, we shall fix our attention in what follows on the \emph{6D} $%
\mathcal{N}=2$ non chiral supergravity theory; too particularly on
the case of \emph{6D} supergravity models embedded in \emph{10D}
type IIA superstring on K3. There, the field theory spectrum
following from the compactification of \emph{10D} type IIA
superstring on K3, involves the two supersymmetric
multiplets namely the gravity supermultiplet and the Maxwell supermultiplets:%
\newline

(\textbf{1}) \emph{the gravity supermultiplet} \newline This
supermultiplet contains, in addition to fermions, the following
bosonic fields:

\ \ \newline
\begin{tabular}{llllll}
(\textbf{i}) & the 6D gravity field & : & \ $g_{\mu \nu }=g_{\mu \nu
}\left(
x\right) $ & , $\ \mu ,\nu =0,...,5$ & $,$ \\
(\textbf{ii}) & the NS-NS B- field & : & $B_{\mu \nu }=B_{\mu \nu
}\left(
x\right) $ & , &  \\
(\textbf{iii}) & four $U\left( 1\right) $ gauge fields & : & $\mathcal{A}%
_{\mu }^{a}=\mathcal{A}_{\mu }^{a}\left( x\right) $ & $,$ $\
a=1,...,4$ & $,$
\\
(\textbf{iv}) & the 6D dilaton & : & $\sigma =\sigma \left( x\right)
$ & . &
\end{tabular}

\ \

(\textbf{2}) \emph{the} \emph{Maxwell gauge sector}\newline This
sector involves twenty Maxwell supermultiplets with the following
bosons:

\ \ \newline
\begin{tabular}{lllll}
(\textbf{i}) & twenty abelian gauge fields & : & $\mathcal{A}_{\mu }^{I}=%
\mathcal{A}_{\mu }^{I}\left( x\right) $ & $,$ $\ I=1,...,20,$ \\
(\textbf{ii}) & twenty quartets of scalars & : & $\phi _{aI}=\phi
_{aI}\left( x\right) $ & .%
\end{tabular}

\ \ \ \newline The abelian gauge symmetry group of the \emph{6D}
$\mathcal{N}=2$ supergravity theory that we are considering here can
be cast as follows
\begin{equation}
U_{NS}\left( 1\right) \times U^{4}\left( 1\right) \times
U^{20}\left( 1\right) .
\end{equation}%
The $U_{NS}\left( 1\right) $ factor is the abelian gauge symmetry
associated
with the NS-NS gauge field $B_{\mu \nu }$ and field strength $\mathcal{F}%
_{3} $ that we have denoted earlier as $\mathcal{H}_{3}$. \newline
The abelian factor $U^{4}\left( 1\right) $ is the gauge symmetry
with the
four gauge fields $\mathcal{A}_{\mu }^{a}$ and the field strengths $\mathcal{%
F}_{2}^{a}$ of the supergravity multiplet. \newline The factor
$U^{20}\left( 1\right) $ is associated with $\mathcal{A}_{\mu }^{I}$
and the field strength $\mathcal{F}_{2}^{I}$ of the Maxwell-matter
sector. \ \newline Along with the gauge invariant fields strengths
\begin{equation}
\begin{tabular}{llllll}
$\mathcal{F}_{3}$ & $,$ & $\mathcal{F}_{2}^{a}$ & $,$ &
$\mathcal{F}_{2}^{I}$
& .%
\end{tabular}
\label{fff}
\end{equation}%
We also have their Poincare duals namely
\begin{equation}
\begin{tabular}{llllll}
$\mathcal{G}_{3}$ & , & $\mathcal{G}_{4}^{a}$ & , & $\mathcal{G}_{4}^{I}$ & .%
\end{tabular}
\label{ggg}
\end{equation}%
The $\left( 1+24\right) $ electric charges and the $\left(
1+24\right) $ magnetic charges associated with these gauge invariant
field strengths are as follows:\ \newline

(\textbf{a}) \emph{dyonic black string BS}: \newline The 6D dyonic
\emph{BS} has an electric charge $\mathrm{q}_{0}$ and a magnetic
charge $\mathrm{g}_{0}$ with the following quantization condition
\begin{equation}
\begin{tabular}{llll}
$\mathrm{q}_{0}\mathrm{g}_{0}$ & $=$ & $2\pi k_{0}$ & $,$%
\end{tabular}
\label{k0}
\end{equation}%
where $k_{0}$ is an integer ($k_{0}\in Z$). \ \newline

(\textbf{b}) \emph{6D black hole BH}: \newline
The six dimensional \emph{BH} has magnetic charges\textrm{\footnote{%
For a geometric derivation of the explicit relation between the bare charge $%
g_{\Lambda }$ and the physical charges $\left( m_{a},m_{I}\right) $,
see
\textrm{\cite{S2}}}} under the gauge symmetry $U^{24}\left( 1\right) $.%
\begin{equation}
\begin{tabular}{llll}
$\mathrm{g}_{\Lambda }$ & $,$ & $\Lambda =1,...,24$ & .%
\end{tabular}
\label{gl}
\end{equation}

(\textbf{c}) \emph{6D black membrane BM}:\newline The six
dimensional \emph{BM} is the dual of the black hole and is
electrically charged under the $U^{24}\left( 1\right) $ gauge
symmetry:
\begin{equation}
\begin{tabular}{llll}
$\mathrm{q}_{\Lambda }$ & $,$ & $\Lambda =1,...,24$ & .%
\end{tabular}
\label{qa}
\end{equation}%
The electric and magnetic charges of the 6D black hole and the 6D
black
membrane are related by the quantization condition,%
\begin{equation}
\begin{tabular}{llllll}
$\mathrm{q}_{\Lambda }\mathrm{g}_{\Lambda }$ & $=$ & $2\pi k_{\Lambda }$ & $%
\qquad ,\qquad $ & $k_{\Lambda }\in Z$ & $.$%
\end{tabular}
\label{ka}
\end{equation}%
In the brane language of \emph{10D} type IIA superstring on
Calabi-Yau manifolds, the electric and/or the magnetic charges are
associated with branes wrapping cycles of Calabi-Yau manifold (CY).
In the \emph{6D}\ case we are considering, the CY manifold in
question is given by K3 with a homology containing, in addition to
the 0- cycle $C_{0}$ (K3 points) and the real 4- cycle $C_{4}$, real
\emph{twenty- two} 2- cycles $C_{2}^{I}$. We also have the following
features

\begin{equation*}
\begin{tabular}{|llllll|}
\hline
6D black attractors & : & electric/magnetic & near horizon geometry & $%
\qquad $Entropy &  \\
&  &  &  &  &  \\
$\text{dyonic black string}$ & : & $\qquad \left( \mathrm{q}^{0}\mathrm{,g}%
_{0}\right) $ & $\qquad AdS_{3}\times S^{3}$ & $\qquad R_{H_{1}}^{3}G_{N}^{-%
\frac{3}{4}}$ &  \\
&  &  &  &  &  \\
$\text{black hole}$ & : & $\qquad \left( 0,\mathrm{g}_{\Lambda }\right) $ & $%
\qquad AdS_{2}\times S^{4}$ & $\qquad R_{H_{2}}^{4}G_{N}^{-1}$ &  \\
&  &  &  &  &  \\
$\text{black membrane}$ & : & $\qquad \left( \mathrm{q}^{\Lambda
},0\right) $ & $\qquad AdS_{4}\times S^{2}$ & $\qquad
R_{H_{3}}^{2}G_{N}^{-\frac{1}{2}}$
&  \\
&  &  &  &  &  \\ \hline
\end{tabular}%
\end{equation*}%
\ \newline \textrm{where the last column stands for the 6D
generalized Bekenstein-Hawking entropy formula expressed in terms of
the 6D Newton constant }$G_{N}$\ and the radius of the horizon
geometry.\textrm{\ In the
case of black string for instance, we have }%
\begin{equation*}
S_{BFS}^{\text{entropy}}\sim G_{N}^{-\frac{3}{4}}\times \mathcal{A}_{\text{ha%
}},
\end{equation*}%
\textrm{where }$\mathcal{A}_{\text{ha}}$\textrm{\ is the 3d- horizon
"hyper-area" in agreement with dimensional arguments and black
object thermodynamics laws}. \newline Notice in passing that the
\emph{6D} black hole \emph{BH} is made of:\newline - D0 branes,
\newline - D2 brane wrapping the \emph{twenty- two} 2-cycles of K3,
and \newline - D4 wrapping K3. \newline The dual black membrane
\emph{BM} is made of:\newline - D2 branes, \newline - D4 brane
wrapping\ the 2- cycles of K3, and \newline - D6 brane wrapping
K3.\newline These \emph{6D} black objects have different
$AdS_{p+2}\times S^{4-p}$ near horizon geometries; they are
schematically represented on the figure 1.\
\begin{figure}[tbph]
\begin{center}
\epsfxsize=10cm\epsfbox{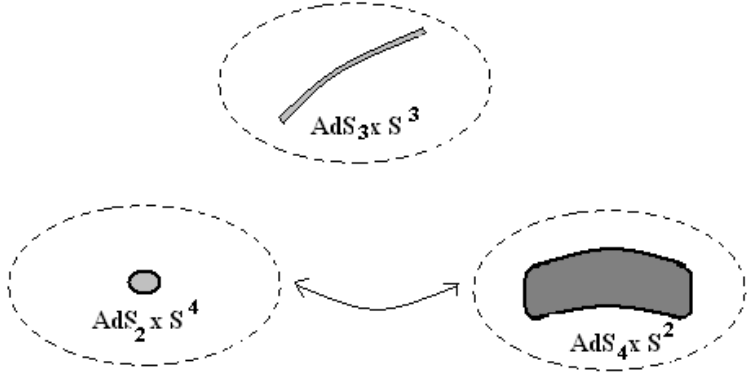}
\end{center}
\caption{{\protect\small This figure represents the black attractors
in 6D N=2 supergravity. Dashed loops refer to the near horizon
geometries: (i) On top: we represent a black string with near
horizon geometry }$AdS_{3}\times
S^{3}${\protect\small . (ii) Bottom-left: a BH with its near horizon }$%
AdS_{2}\times S^{4}${\protect\small . (iii) Bottom-right: a black\
membrane with }$AdS_{4}\times S^{2}$ {\protect\small geometry .}}
\end{figure}

\subsection{Entropy of 6D black attractors}

Using dimensional arguments and the near horizon geometry, the
entropy formula $\mathcal{S}_{p}$ of the black\ p-brane attractor in
six space time dimensions, that describes the analogue of the
Hawking Bekenstein entropy of
the \emph{4D} black hole, can be written by as follows%
\begin{equation}
\begin{tabular}{llllll}
$\mathcal{S}_{p}$ & $=$ & $R_{H_{1}}^{4-p}G_{N}^{-\frac{4-p}{4}}$ &
$\qquad
,\qquad $ & $p=0,1,2$ & $.$%
\end{tabular}%
\end{equation}%
Here $G_{N}\sim l_{\text{{\small Planck}}}^{4}$ is the \emph{6D}
Newton
constant scaling as $\left( lenght\right) ^{4}$ and where $R_{H_{1}}^{2},$\ $%
R_{H_{2}}^{3}$\ and $R_{H_{3}}^{4}$ stand respectively for the
horizon \textrm{"hyper-areas"} of the black hole \emph{BH} ($p=0$),
the dyonic black string \emph{BS} ($p=1$), and the black membrane
\emph{BM} ($p=2$). \newline The entropy $\mathcal{S}_{p}$ is
completely specified by the electric $q$
and magnetic $g$ charges of the black attractor,%
\begin{equation}
\begin{tabular}{llll}
$\mathcal{S}_{p}$ & $=$ & $\mathcal{S}_{p}\left( q,g\right) $ & $.$%
\end{tabular}%
\end{equation}%
In the next sections, we will first show that the entropies
\begin{equation}
\begin{tabular}{llll}
$\mathcal{S}_{BH}$ & $=$ & $\mathcal{S}_{0}\left( g_{\Lambda
}\right) $ & ,
\\
$\mathcal{S}_{BS}$ & $=$ & $\mathcal{S}_{1}\left( q_{0},g_{0}\right)
$ & ,
\\
$\mathcal{S}_{BM}$ & $=$ & $\mathcal{S}_{2}\left( q_{\Lambda }\right) $ & ,%
\end{tabular}%
\end{equation}%
are indeed specified by the appropriate electric $q_{0}$ and
$q_{\Lambda }$ as well as the corresponding magnetic $g_{0}$ and
$g_{\Lambda }$ ones.
\newline
Then we use this result to check explicitly that, at supergravity
level, the entropy $\mathcal{S}_{1}\left( q_{0},g_{0}\right) $ of
the 6D black string is invariant under electric/magnetic duality.
\newline
This property is also used to conjecture that the invariance of $\mathcal{S}%
_{BS}$ under the electric/magnetic duality is a \emph{general
feature} of dyonic objects including black brane bound states.
\newline As such, invariance under electric/magnetic duality should
holds also for the entropy
\begin{equation}
\begin{tabular}{llll}
$\mathcal{S}_{EBB-MBB}$ & $=$ & $\mathcal{S}_{DP}\left( q_{\Lambda
},g_{\Lambda }\right) $ & ,%
\end{tabular}%
\end{equation}%
of the dual pair bounds \emph{DP }$\equiv $ $\left(
\emph{EBB-MBB}\right) $ given by
eqs(\ref{hm},\ref{za},\ref{az}).\newline
By using this natural conjecture, it follows that under the dual change%
\begin{equation}
\begin{tabular}{llll}
$q_{A}$ & $\rightarrow $ & $q_{A}^{\prime }=g_{A}$ & , \\
$g_{A}$ & $\rightarrow $ & $g_{A}^{\prime }=q_{A}$ & ,%
\end{tabular}
\label{qqg}
\end{equation}%
we should also have%
\begin{equation}
\begin{tabular}{llll}
$\mathcal{S}_{EBB}\left( q_{A}\right) $ & $\leftrightarrow $ & $\mathcal{S}%
_{MBB}\left( g^{A}\right) $ & ,%
\end{tabular}
\label{ss}
\end{equation}%
where $\mathcal{S}_{EBB}$ and $\mathcal{S}_{MBB}$ stand respectively
for the entropies of an electrically charged black brane \emph{EBB}
and its magnetic dual \emph{MBB}. \newline In this view, invariance
of the entropy $\mathcal{S}_{1}$ of the dyonic black string
\emph{BS} under the change (\ref{qqg}) as well as of the dual pair
bounds \emph{DP},
\begin{equation}
\begin{tabular}{llll}
$\mathcal{S}_{1}\left( q_{0}^{\prime },g_{0}^{\prime }\right) $ & $=$ & $%
\mathcal{S}_{1}\left( q_{0},g_{0}\right) $ & , \\
$\mathcal{S}_{DP}\left( q_{\Lambda }^{\prime },g_{\Lambda }^{\prime
}\right)
$ & $=$ & $\mathcal{S}_{DP}\left( q_{\Lambda },g_{\Lambda }\right) $ & ,%
\end{tabular}%
\end{equation}%
follows straightforwardly. The result for the case
$\mathcal{S}_{DP}\left( q_{\Lambda },g_{\Lambda }\right) $ will be
explicitly derived in section 6. The case of the \emph{6D} black
string \emph{BS} will be studied below.

\section{Black string}

\qquad The six dimensional black string is a dyonic black attractor
solution
of the $\mathcal{N}=2$ non chiral supergravity with near horizon geometry $%
AdS_{3}\times S^{3}$. The magnetic charge $m=\mathrm{g}^{0}$ and the
electric charge $e=q_{0}$ carried by the BS are those of the gauge
invariant
3- form field strengths $\mathcal{F}_{3}$ and $\mathcal{G}_{3}=$ $^{\star }%
\mathcal{F}_{3}$. The $\star $ conjugation stands for the usual six
dimensional Hodge duality interchanging $n$- forms with $\left(
6-n\right) $ ones. \newline The field strengths $\mathcal{F}_{3}$
and $\mathcal{G}_{3}$ are associated with the NS-NS 2- form
$\mathcal{B}_{\mu \nu }$ fields in six dimensions.
Using eqs(\ref{q}), we have%
\begin{equation}
\begin{tabular}{llll}
$\mathrm{g}_{0}=\int_{S^{3}}\mathcal{F}_{3}$ & $\qquad ,\qquad $ & $\mathrm{q%
}^{0}=\int_{S^{3}}\mathcal{G}_{3}$ & $.$%
\end{tabular}%
\end{equation}%
The electric $\mathrm{q}^{0}$ and magnetic $\mathrm{g}_{0}$ charges
obey the quantization condition (\ref{k0}).

\subsection{Entropy of the black string}

\qquad The effective potential
$\mathcal{V}_{BS}=\mathcal{V}_{BS}\left( \sigma
,\mathrm{g}_{0},\mathrm{q}_{0}\right) $ of the \emph{BS} is given by
the the \emph{6D} extension of the \emph{4D}\ \emph{Weinhold}
relation \textrm{\cite{B3,b31,b32}}. It reads in terms of the
dilaton field $\sigma =\sigma \left( x\right) $ of the 6D
$\mathcal{N}=2$ supergravity multiplet and the electric/magnetic
charges like \textrm{\cite{BDSS}},
\begin{equation}
\begin{tabular}{llllll}
$\mathcal{V}_{BS}$ & $=$ & $\frac{\mathrm{g}_{0}^{2}}{2}\exp \left(
-4\sigma \right) $ & $+$ & $\frac{\mathrm{q}_{0}^{2}}{2}\exp \left(
4\sigma \right) $
& $.$%
\end{tabular}
\label{vp}
\end{equation}%
In addition to the exponential behavior, the potential
$\mathcal{V}_{BS}$
has the remarkable invariance under the following change,%
\begin{equation}
\begin{tabular}{llll}
$\sigma \rightarrow -\sigma $ & $\qquad \text{and }$\qquad & $\mathrm{g}%
_{0}\leftrightarrow \mathrm{q}_{0}$ & .%
\end{tabular}
\label{si}
\end{equation}%
At the horizon $r=R_{\text{horizon}}^{BS}$ of the dyonic \emph{BS},
the above potential $\mathcal{V}_{BS}$ is at its minimum. The value
of the
dilaton at horizon%
\begin{equation}
\sigma _{1}=\sigma \left( r=R_{\text{horizon}}^{BS}\right) ,
\end{equation}%
is obtained by solving the following constraint equation%
\begin{equation}
\frac{d\mathcal{V}_{BS}\left( \sigma \right) }{d\sigma }=0,
\end{equation}%
which in turns leads to,%
\begin{equation}
\begin{tabular}{lllll}
$2\mathrm{q}_{0}^{2}\exp \left( 4\sigma \right) $ & $-$ & $2\mathrm{g}%
_{0}^{2}\exp \left( -4\sigma \right) $ & $=0$ & $.$%
\end{tabular}
\label{gq}
\end{equation}%
The critical value $\sigma _{1}$ of the dilaton at the horizon $R_{\text{%
horizon}}^{BS}$ is given by%
\begin{equation}
\begin{tabular}{llll}
$\exp \left( 4\sigma _{1}\right) $ & $=$ & $\pm \frac{\mathrm{g}_{0}}{%
\mathrm{q}_{0}}>0$ & ,%
\end{tabular}%
\end{equation}%
or equivalently%
\begin{equation}
\begin{tabular}{llll}
$\sigma _{1}$ & $=$ & $\frac{1}{4}\ln \left( \left\vert \frac{\mathrm{g}_{0}%
}{\mathrm{q}_{0}}\right\vert \right) $ & .%
\end{tabular}
\label{sg}
\end{equation}%
From this solution, we learn two interesting information: \newline
The first information, noted previously in \textrm{\cite{S2}},
concerns the
electric/magnetic duality. The latter requires interchanging the electric $%
\mathrm{q}_{0}$ and magnetic $\mathrm{g}_{0}$ charges; but also
performing the change
\begin{equation}
\sigma \rightarrow -\sigma
\end{equation}%
in the moduli space. This property is manifestly exhibited by eqs(\ref{si}-%
\ref{gq}).\newline The second information we learn concerns the
critical value $\sigma _{1}$ of the dilaton at the horizon of the
black string eq(\ref{sg}). Finite critical
values $\sigma _{1}$ of the dilaton requires that both the electric $\mathrm{%
q}_{0}$ and the magnetic $\mathrm{g}_{0}$ charges have to be non
zero, i.e
\begin{equation}
\mathrm{q}_{0}\mathrm{g}_{0}\neq 0\text{ },  \label{nz}
\end{equation}%
or equivalently by using eq(\ref{k0})%
\begin{equation}
\mathrm{k}_{0}\neq 0\text{ }.
\end{equation}%
We will see later that this is a general property valid also for of
the \emph{6D} dyonic pair \emph{BH-BM}.\newline
Moreover, the value $\mathcal{V}_{BS}^{\min }$ of the BS\ potential $%
\mathcal{V}_{BS}\left( \sigma \right) $ at the minimum of the black
string potential is
\begin{equation}
\mathcal{V}_{BFS}^{\min }\left( \sigma _{1}\right) =\left\vert \mathrm{q}_{0}%
\mathrm{g}_{0}\right\vert >0\text{ },
\end{equation}%
and so the BS entropy reads as%
\begin{equation}
\mathcal{S}_{1}=\frac{\left\vert \mathrm{q}_{0}\mathrm{g}_{0}\right\vert }{4}%
\text{ }.  \label{bfe}
\end{equation}%
Up on using eq(\ref{k0}), $\mathcal{S}_{1}$ can be also expressed as
\begin{equation}
\mathcal{S}_{1}=\pi \frac{\left\vert k_{0}\right\vert }{2}\text{ }.
\label{bff}
\end{equation}%
where $k_{0}$ is an integer. Notice that because of the constraint eq(\ref%
{nz}), the entropy $\mathcal{S}_{1}$ given by the above eqs is
necessarily positive definite. \newline Below, we want to discuss
what happens to the entropy $\mathcal{S}_{1}$ if we try to go beyond
the constraint eq(\ref{nz}).

\subsection{Regular and singular representations}

\qquad For later use, we make two comments concerning the effective
potential of the dyonic black string. The first comment concerns the
generic
case where $k_{0}\neq 0$ and the second deals with the singular case $%
k_{0}=0 $.

(\textbf{1}) \emph{case} $k_{0}\neq 0$:\newline This case
corresponds to the dyonic black string of the 6D $\mathcal{N}=2$ non
chiral supergravity we have been studding. In fact it is interesting
to distinguish two situations:

\textbf{(a)} \emph{Regular representation: }$\mathrm{g}_{0}\neq 0,$ $\mathrm{%
q}_{0}\neq 0$\newline Here, the extremum of the black string
potential at
\begin{equation}
\sigma _{1}=\frac{1}{4}\ln \left( \left\vert \frac{\mathrm{g}_{0}}{\mathrm{q}%
_{0}}\right\vert \right) \text{ ,}
\end{equation}%
is well defined and is precisely a minimum. Since $\mathrm{g}_{0}$ and $%
\mathrm{q}_{0}$ are related as in eq(\ref{k0}), we can be expressed
$\sigma _{1}$ either as
\begin{equation}
\sigma _{1}=\frac{1}{4}\ln \left\vert \frac{2\pi k_{0}}{\mathrm{q}_{0}^{2}}%
\right\vert \text{ },
\end{equation}%
by using the electric charge $\mathrm{q}_{0}$ and the integer
$k_{0}$, or equivalently by using the magnetic charge
$\mathrm{g}_{0}$ like
\begin{equation}
\sigma _{1}=\frac{1}{4}\ln \left\vert \frac{\mathrm{g}_{0}^{2}}{2\pi k_{0}}%
\right\vert \text{ }.
\end{equation}%
The value of the potential at the minimum depends remarkably on the integer $%
k_{0}$ as shown below,%
\begin{equation}
\mathcal{V}_{BFS}^{\min }\left( \sigma _{1}\right) =2\pi \left\vert
k_{0}\right\vert >0\text{ }.  \label{min}
\end{equation}%
This is an interesting property that let understand that a non zero
entropy
value seems to need dyonic charged black branes since if taking for example $%
\mathrm{g}_{0}\neq 0$ and finite but $\mathrm{q}_{0}=0$,
eq(\ref{min}) vanishes identically.

\textbf{(b)} \emph{Singular representation}\newline
Notice that, strictly speaking, the condition $k_{0}\neq 0$ corresponds to $%
\mathrm{g}_{0}\mathrm{q}_{0}\neq 0$. But this condition could be
solved in general in two ways:\newline
First by using regular finite charges $\mathrm{g}_{0}\neq 0$ and $\mathrm{q}%
_{0}\neq 0$ as just discussed above. \newline Second by considering
the singular situation where we have an infinite
number of electric charges and no magnetic charge; i.e%
\begin{equation}
\mathrm{q}_{0}\rightarrow \infty \qquad \text{and}\qquad \mathrm{g}%
_{0}\rightarrow 0,
\end{equation}%
together with the constraint
$\mathrm{g}_{0}=\frac{k_{0}}{\mathrm{q}_{0}}$.
\newline
We can also have the symmetric case where we do have an infinite
number of magnetic charges and no electric charge:
\begin{equation}
\mathrm{q}_{0}\rightarrow 0\qquad \text{and}\qquad
\mathrm{g}_{0}\rightarrow \infty .
\end{equation}%
These particular and singular configurations are in some sense
formal; but very suggestive. They will be used later on to approach
the dyonic pair \emph{BH-BM}.

(\textbf{2}) \emph{Case} $k_{0}=0$:\newline
Using eq(\ref{k0}\textbf{), }this\textbf{\ }case is solved as%
\begin{equation}
\mathrm{g}_{0}\neq 0\qquad ,\qquad \mathrm{q}_{0}=0.
\end{equation}%
or like
\begin{equation}
\mathrm{g}_{0}=0\qquad ,\qquad \mathrm{q}_{0}\neq 0.
\end{equation}%
They can be also associated with the (self dual and anti-self dual
part) black string of the 6D $\mathcal{N}=\left( 2,0\right) $ chiral
supergravity. There, the \emph{NS-NS }B- field field
$\mathcal{B}_{\left[ \mu \nu \right] } $ splits into a self dual
part
\begin{equation}
\mathcal{B}_{\left[ \mu \nu \right] }^{+}\text{ ,}
\end{equation}
and anti-self dual part
\begin{equation}
\mathcal{B}_{\left[ \mu \nu \right] }^{-}\text{ .}
\end{equation}
The strength $\mathcal{H}_{\left[ \lambda \mu \nu \right] }^{+}$\
associated with the self dual part $B_{\left[ \mu \nu \right]
}^{+}$,
\begin{equation}
\mathcal{H}_{3}^{+}=dB_{2}^{+},
\end{equation}%
is in the gravity supermultiplet while the field strength $\mathcal{H}%
_{\lambda \mu \nu }^{-}$\ of the anti-self dual part
$\mathcal{B}_{\left[ \mu \nu \right] }^{-}$,
\begin{equation}
\mathcal{H}_{3}^{-}=dB_{2}^{-},
\end{equation}%
together with the field $\sigma $, are in the tensor multiplet.\ \
\newline The minimum $\mathcal{V}_{BS}^{\min }\left( \sigma
_{1}\right) $ of the black string potential is at infinity; that is
either at
\begin{equation}
\sigma _{1}\rightarrow +\infty \text{ ,}
\end{equation}%
or at
\begin{equation*}
\sigma _{1}\rightarrow -\infty \text{ }.
\end{equation*}%
In both cases, $\mathcal{V}_{BS}^{\min }\left( \sigma _{1}\right) $
takes a zero value in agreement with eq(\ref{min}). For
illustration; we give in \emph{figure 2} the general behavior of the
black string potential in terms of the dilaton $\sigma $.\newline
\begin{figure}[tbph]
\begin{center}
\epsfxsize=13cm\epsfbox{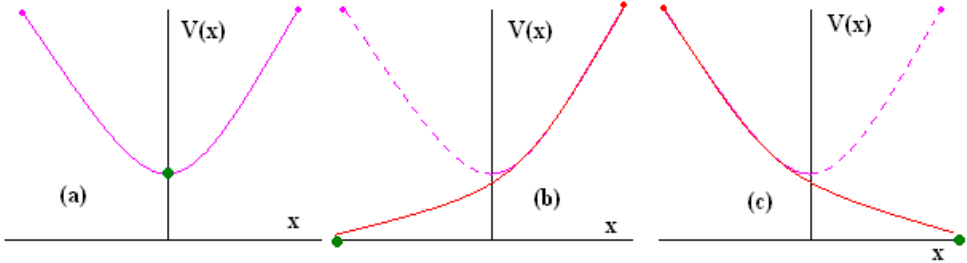}
\end{center}
\caption{{\protect\small Variation of the effective potential with
to the
dilaton field. (a) Case of black string in \emph{6D} }$\mathcal{N}=2$%
{\protect\small \ non chiral supergravity. (b/c) black string in \emph{6D} }$%
\mathcal{N}=\left( {\protect\small 2,0}\right) ${\protect\small \
chiral supergravity.}}
\end{figure}

\subsection{Electric/magnetic duality}

\qquad The key property of the above dyonic black string entropy relation (%
\ref{bfe}) is that $\mathcal{S}_{BS}^{\text{ }}$ is invariant under
the following electric-magnetic change
\begin{equation}
\begin{tabular}{llll}
$\mathrm{q}_{0}$ & $\qquad \rightarrow \qquad $ & $\mathrm{q}_{0}^{\prime }=%
\mathrm{g}_{0}$ & , \\
$\mathrm{g}_{0}$ & $\qquad \rightarrow \qquad $ & $\mathrm{g}_{0}^{\prime }=%
\mathrm{q}_{0}$ & ,%
\end{tabular}
\label{em}
\end{equation}%
with%
\begin{equation}
\begin{tabular}{llll}
$\mathcal{S}_{BS}^{\text{ }}\left( \mathrm{q}_{0},\mathrm{g}_{0}\right) $ & $%
\rightarrow $ & $\mathcal{S}_{BS}^{\text{ }}\left( \mathrm{q}_{0}^{\prime },%
\mathrm{g}_{0}^{\prime }\right) =\mathcal{S}_{BS}^{\text{ }}\left( \mathrm{q}%
_{0},\mathrm{g}_{0}\right) $ & .%
\end{tabular}
\label{dy}
\end{equation}%
This property, which can be explicitly checked on previous eqs, was
expected since we are dealing with a dyonic object. \newline
Moreover, using the discussion of sub-section 4.2 (\emph{singular
representation}), the relation (\ref{dy}) can be extended to the 6D
dyonic pair \emph{BH-BM}. \newline

\emph{Black hole/black membrane}\newline Denoting by
$\mathcal{S}_{BH}$ the entropy of magnetically charged black hole
\emph{BH}; i.e
\begin{equation}
\mathcal{S}_{BH}=\mathcal{S}_{BH}\left( \mathrm{g}_{\Lambda }\right)
\text{ \ },
\end{equation}%
and by $\mathcal{S}_{BM}$ the entropy of the electrically charged
black membrane \emph{BM}; i.e
\begin{equation}
\mathcal{S}_{BM}=\mathcal{S}_{BM}\left( \mathrm{q}_{\Lambda }\right)
\text{ \ }.
\end{equation}%
Then performing the electric/magnetic duality (\ref{em}), the entropies $%
\mathcal{S}_{BH}\left( \mathrm{g}_{\Lambda }\right) $ and $\mathcal{S}%
_{BM}\left( \mathrm{q}_{\Lambda }\right) $ are interchanged as follows%
\begin{equation}
\begin{tabular}{llll}
$\mathcal{S}_{BH}\left( \mathrm{g}_{\Lambda }\right) $ & $\rightarrow $ & $%
\mathcal{S}_{BH}\left( \mathrm{g}_{\Lambda }^{\prime }\right) =\mathcal{S}%
_{BM}\left( \mathrm{q}_{\Lambda }\right) $ & , \\
$\mathcal{S}_{BM}\left( \mathrm{q}_{\Lambda }\right) $ & $\rightarrow $ & $%
\mathcal{S}_{BM}\left( \mathrm{q}_{\Lambda }^{\prime }\right) =\mathcal{S}%
_{BH}\left( \mathrm{g}_{\Lambda }\right) $ & .%
\end{tabular}%
\end{equation}%
From these relations we learn that, like for the dyonic \emph{BS}
singlet, the entropy
\begin{equation}
\mathcal{S}_{DP}=\mathcal{S}_{BH-BM}\left( \mathrm{g}_{\Lambda }\mathrm{,q}%
_{\Lambda }\right) ,
\end{equation}%
of the dyonic pair \emph{DP }$\equiv $ \emph{BH-BM} obeys the same
identity as eq(\ref{dy}). It is invariant under electric/magnetic
duality transformation.
\begin{equation}
\begin{tabular}{llll}
$\mathcal{S}_{DP}\left( \mathrm{g}_{\Lambda }\mathrm{,q}_{\Lambda
}\right) $
& $=$ & $\mathcal{S}_{DP}\left( \mathrm{g}_{\Lambda }^{\prime }\mathrm{,q}%
_{\Lambda }^{\prime }\right) $ & $.$%
\end{tabular}%
\end{equation}%
In what follows, we want to prove this statement by computing
explicitly the
expressions of $\mathcal{S}_{DP}\left( \mathrm{g}_{\Lambda }\mathrm{,q}%
_{\Lambda }\right) $ and $\mathcal{S}_{DP}\left( \mathrm{g}_{\Lambda
}^{\prime }\mathrm{,q}_{\Lambda }^{\prime }\right) $. But before
that we need to study the effective potentials and the attractor eqs
of the following:\newline
(\textbf{a}) the \emph{BPS} and non \emph{BPS} black holes in \emph{6D} $%
\mathcal{N}=2$ non chiral supergravity \newline
(\textbf{b}) the \emph{BPS} and non \emph{BPS} black membranes in \emph{6D} $%
\mathcal{N}=2$ non chiral supergravity \newline (\textbf{c}) the
\emph{BPS} and non \emph{BPS} dyonic pairs \emph{DP}$\equiv $
$\left( \emph{BH-BM}\right) $.\newline The effective potential and
the attractor mechanism of the \emph{6D} black string \emph{BS} has
been explicitly studied in \textrm{\cite{BDSS}}. We will then just
give the results. \newline We also take this opportunity to develop
a \emph{new method} to deal with the computation of the critical
values of the effective potentials of the \emph{BH} and the \emph{BM
}in six dimensions. \newline
This new method relies on enlarging the moduli space of the \emph{6D} $%
\mathcal{N}=2$ supergravity by including Lagrange multipliers,%
\begin{equation}
\lambda ^{\Lambda \Sigma }=\lambda ^{\Sigma \Lambda },  \label{lan}
\end{equation}%
capturing the constraint eqs on the field matrix
\begin{equation}
L_{\Lambda \Sigma }=L_{\Lambda \Sigma }\left( x\right) ,
\label{lac}
\end{equation}%
used to parameterize the $SO\left( 4,20\right) $ orthogonal group
manifold involved in the moduli space $\frac{SO\left( 4,20\right)
}{SO\left( 4\right) \times SO\left( 20\right) }$.

\section{Attractor eqs and Lagrange multiplier method}

\qquad We first describe the effective potential $\mathcal{V}_{BH}$
of the \emph{6D} black hole. Then, we use the electric/magnetic
duality and results
from \textrm{\cite{S2}} to determine the effective potential $\mathcal{V}%
_{BM}$ of the black membrane. After that, we study the attractor eqs
and their solutions by combining the approach of the criticality of
the potential and the Lagrange multiplier method.

\subsection{Effective potential $\mathcal{V}_{BH}$}

\qquad In the \emph{6D} $\mathcal{N}=2$ non chiral supergravity
embedded in \emph{10D} type IIA superstring on K3, the black hole
\emph{BH} is magnetically charged under the $U^{4}\left( 1\right)
\times U^{20}\left( 1\right) $ gauge group symmetry. \newline
The bare magnetic charges $\mathrm{g}^{\Lambda }$ are given by,%
\begin{equation}
\begin{tabular}{llllll}
$\mathrm{g}^{\Lambda }$ & $=$ &
$\int_{S^{2}}\mathcal{F}_{2}^{\Lambda }$ & ,
& $\Lambda =1,...,24$ & .%
\end{tabular}%
\end{equation}%
The\ magnetic charges $\mathrm{g}^{\Lambda }$ form a charge vector
the group
$SO\left( 4,20\right) $ with signature \emph{4}$\left( +\right) $ and \emph{%
20}$\left( -\right) $ captured by the diagonal flat metric $\eta
_{\Lambda \Sigma }$ of the tangent space $\mathbb{R}^{\left(
4,20\right) }$.

\subsubsection{Potential of the BH}

The effective scalar potential $\mathcal{V}_{BH}$ of the black hole
is given by the Weinhold relation, expressed in the flat coordinate
frame,
\begin{equation}
\mathcal{V}_{BH}=\left( \sum_{a=1}^{4}\delta
_{ab}Z^{a}Z^{b}+\sum_{I=1}^{20}\delta _{IJ}Z^{I}Z^{J}\right) .
\label{blh}
\end{equation}%
In this relation, the central charges $Z_{a}$ and $Z_{I}$ are
respectively the \emph{dressed} charges describing respectively the
physical charges of
the \emph{four} Maxwell fields in the gravity supermultiplet and the \emph{%
twenty} Maxwell fields of the matter sector. \newline
The dressing of the charges is given by the following linear combination,%
\begin{equation}
\begin{tabular}{llll}
$Z_{a}$ & $=$ & $\sum_{\Lambda =1}^{24}U_{a\Lambda
}\mathrm{g}^{\Lambda }$ &
$,$ \\
$Z_{I}$ & $=$ & $\sum_{\Lambda =1}^{24}U_{I\Lambda
}\mathrm{g}^{\Lambda }$ &
,%
\end{tabular}
\label{u}
\end{equation}%
where $U_{\Sigma \Lambda }$ parameterize the moduli space $\boldsymbol{M}%
_{6D}^{N=2}$ of \emph{10D} type IIA superstring on K3,%
\begin{equation}
\begin{tabular}{llll}
$\boldsymbol{M}_{6D}^{N=2}$ & $=$ & $SO\left( 1,1\right) \times
G_{6}$ & $,$
\\
$G_{6}$ & $=$ & $\frac{SO\left( 4,20\right) }{SO\left( 4\right)
\times
SO\left( 20\right) }$ & .%
\end{tabular}
\label{mds}
\end{equation}%
Notice that the real matrix $U_{\Lambda \Sigma }$ obeys a set of
constraint relations that can be used to put $U_{\Lambda \Sigma }$
in a more convenient form. We have the following properties:\newline
(\textbf{i}) the factorization property which allows to factorize $%
U_{\Lambda \Sigma }$ as follows:
\begin{equation}
\begin{tabular}{llll}
$U_{\Lambda \Sigma }$ & $=$ & $e^{-\sigma }L_{\Lambda \Sigma }$ & , \\
$U_{\Lambda \Sigma }^{-1}$ & $=$ & $e^{\sigma }L_{\Lambda \Sigma
}^{-1}$ & ,
\\
$L^{-1}$ & $=$ & $\eta L^{t}\eta $ & .%
\end{tabular}%
\
\end{equation}%
Here $e^{-\sigma }$ parameterizes the factor $SO\left( 1,1\right) $ of $%
\boldsymbol{M}_{6D}^{N=2}$ and $L_{\Lambda \Sigma }$ defines
$G_{6}$. Multiplying this equation by the magnetic charge vector
$\mathrm{g}^{\Sigma } $, we obtain the dressed magnetic charge
vector $Z_{\Lambda }=\left( Z_{a},Z_{I}\right) $,
\begin{equation}
\begin{tabular}{llllll}
$Z_{a}$ & $=$ & $\dsum\limits_{\Sigma =1}^{24}U_{a\Sigma
}\mathrm{g}^{\Sigma
}$ & $=$ & $e^{-\sigma }\dsum\limits_{\Sigma =1}^{24}L_{a\Sigma }\mathrm{g}%
^{\Sigma }$ & , \\
$Z_{I}$ & $=$ & $\dsum\limits_{\Sigma =1}^{24}U_{I\Sigma
}\mathrm{g}^{\Sigma
}$ & $=$ & $e^{-\sigma }\dsum\limits_{\Sigma =1}^{24}L_{I\Sigma }\mathrm{g}%
^{\Sigma }$ & .%
\end{tabular}%
\end{equation}%
(\textbf{ii}) the \emph{orthogonality} property of the elements of the $%
SO\left( 4,20\right) $ group which requires that the real $24\times
24$
matrices $L_{\Lambda \Sigma }$ should obey the orthogonality condition:%
\begin{equation}
\begin{tabular}{llllll}
$\dsum\limits_{c,d=1}^{4}\delta ^{cd}L_{c\Lambda }L_{d\Sigma }$ & $-$ & $%
\dsum\limits_{K,L=1}^{20}\delta ^{KL}L_{K\Lambda }L_{L\Sigma }$ & $=$ & $%
\eta _{\Lambda \Sigma }$ & ,%
\end{tabular}%
\end{equation}%
with $\Lambda $ and $\Sigma =1,\cdots ,24$. This relation can be
also
rewritten as%
\begin{equation}
\begin{tabular}{llllll}
$\left( L^{t}\eta L\right) _{\Lambda \Sigma }$ & $=$ & $\dsum\limits_{%
\Upsilon ,\Gamma =1}^{24}L_{\Lambda }^{\Upsilon }\eta _{\Upsilon
\Gamma
}L_{\Sigma }^{\Gamma }$ & $=$ & $\eta _{\Lambda \Sigma }$ & .%
\end{tabular}
\label{ort}
\end{equation}%
Multiplying both sides of this equation by $\mathrm{g}^{\Lambda }\mathrm{g}%
^{\Sigma }$, we obtain the following constraint eq
\begin{equation}
\begin{tabular}{llll}
$\dsum\limits_{a,b=1}^{4}\delta
_{ab}Z^{a}Z^{b}-\dsum\limits_{I,J=1}^{20}\delta _{IJ}Z^{I}Z^{J}$ & $=$ & $%
e^{-2\sigma }\mathrm{g}^{2}$ & ,%
\end{tabular}
\label{or}
\end{equation}%
with%
\begin{equation}
\begin{tabular}{llll}
$\dsum\limits_{\Lambda ,\Sigma =1}^{24}\mathrm{g}^{\Lambda }\eta
_{\Lambda
\Sigma }\mathrm{g}^{\Sigma }$ & $=$ & $\mathrm{g}^{2}$ & .%
\end{tabular}%
\end{equation}%
This relation expresses the orthogonality condition in terms of the
magnetic
charges. We will refer to it as the "magnetic orthogonality" relation.%
\newline
(\textbf{iii}) the isotropy invariance under $SO\left( 4\right)
\times SO\left( 20\right) $ which acts on the matrix $L_{\Lambda
\Sigma }\in
SO\left( 4,20\right) $ as a gauge group symmetry,%
\begin{equation}
L=hLh^{-1},
\end{equation}%
where $h\in SO\left( 4\right) \times SO\left( 20\right) $, the
maximal compact subgroup of $SO\left( 4,20\right) $.

\subsubsection{Implementing the Lagrange multiplier}

\qquad Notice that the matrix variable $L_{\Lambda \Sigma }$ has
$24\times 24 $ real parameters which is much larger that the
\emph{80} moduli required by (\ref{mds}). The properties
(\textbf{ii}) and (\textbf{iii}) are then constraint eqs on
$L_{\Lambda \Sigma }$ which is convenient to cast as
follows:%
\begin{equation}
L_{\Lambda \Sigma }=\left(
\begin{array}{cc}
L_{ab} & L_{aJ} \\
L_{Ib} & L_{IJ}%
\end{array}%
\right) .  \label{lab}
\end{equation}%
To deal with the undesired degrees of freedom in $L_{\Lambda \Sigma
}$, we proceed as follows:\newline (\textbf{1}) we fix the $SO\left(
4\right) \times SO\left( 20\right) $ gauge symmetry by working in
the gauge where the sub-matrices $L_{ab}$ and $L_{IJ}$ are taken
symmetric:
\begin{equation}
L_{ab}=L_{ba}\qquad ,\qquad L_{IJ}=L_{JI}.
\end{equation}%
(\textbf{2}) the orthogonality property eq(\ref{ort})
\begin{equation}
\left( L^{t}\eta L\right) _{\Lambda \Sigma }=\eta _{\Lambda \Sigma }
\label{art}
\end{equation}%
will be imposed by using the Lagrange multiplier method. This method
should be understood as an alternative way to the usual
Maurer-Cartan equation generally used to deal with this matter
\textrm{\cite{B3}}. \newline Eq(\ref{ort}) suggests that the
Lagrange multipliers should be a symmetric matrix field $\lambda
^{\Lambda \Sigma }$ like in eq(\ref{lan}); but the equivalent
reduced form eq(\ref{or}) of the constraints suggests that it is
more convenient to take the Lagrange parameters $\lambda ^{\Lambda
\Sigma }$ as follows,
\begin{equation}
\lambda ^{\Lambda \Sigma }=\mathrm{\lambda }g^{\Lambda }g^{\Sigma },
\label{la}
\end{equation}%
where now we have only one Lagrange multiplier $\mathrm{\lambda
}$.\newline
Therefore the previous expression of the effective scalar potential $%
\mathcal{V}_{BH}$ of the black hole potential can be put into the
following
form%
\begin{equation}
\widetilde{\mathcal{V}}_{BH}\left( \sigma ,Z,\lambda \right) =\left(
Z_{a}Z^{a}+Z_{I}Z^{I}\right) +\mathrm{\lambda }\left(
Z_{a}Z^{a}-Z_{I}Z^{I}-e^{-2\sigma }\mathrm{g}^{2}\right) \text{ },
\label{lm}
\end{equation}%
where we have set%
\begin{equation}
\begin{tabular}{llll}
$Z_{a}Z^{a}$ & $=$ & $\dsum\limits_{a,b=1}^{4}\delta _{ab}Z^{a}Z^{b}$ & , \\
$Z_{I}Z^{I}$ & $=$ & $\dsum\limits_{I,J=1}^{20}\delta _{IJ}Z^{I}Z^{J}$ & .%
\end{tabular}%
\end{equation}%
In this relation, we have an extra dependence on the Lagrange multiplier $%
\mathrm{\lambda }$. Moreover, setting
\begin{equation}
\begin{tabular}{llll}
$Z_{a}$ & $=$ & $e^{-\sigma }R_{a}$ & , \\
$Z_{I}$ & $=$ & $e^{-\sigma }R_{I}$ & ,%
\end{tabular}
\label{fz}
\end{equation}%
with%
\begin{equation}
\begin{tabular}{llll}
$R_{a}$ & $=$ & $L_{a\Lambda }\mathrm{g}^{\Lambda }$ & , \\
$R_{I}$ & $=$ & $L_{I\Lambda }\mathrm{g}^{\Lambda }$ & ,%
\end{tabular}%
\   \label{ry}
\end{equation}%
we can factorize out the dilaton field dependence in the effective
potential. We have:%
\begin{equation}
\widetilde{\mathcal{V}}_{BH}\left( \sigma ,R,\lambda \right) =e^{-2\sigma }%
\mathcal{V}_{0}\left( R,\mathrm{\lambda ,g}\right)  \label{bh}
\end{equation}%
with $\mathcal{V}_{0}\left( R,\mathrm{\lambda ,g}\right) $,
\begin{equation}
\mathcal{V}_{0}\left( R,\mathrm{\lambda ,g}\right) =\left(
R_{a}R^{a}+R_{I}R^{I}\right) +\mathrm{\lambda }\left( R_{a}R^{a}-R_{I}R^{I}-%
\mathrm{g}^{2}\right) ,  \label{bo}
\end{equation}%
being the potential of the black hole at $\sigma =0$. The factor $\mathcal{V}%
_{0}$ has no dependence in the dilaton field.

\subsubsection{Attractor eqs and their solutions}

Because of the structure of the effective black hole potential%
\begin{equation}
\begin{tabular}{llll}
$\widetilde{\mathcal{V}}_{BH}\left( Z,\lambda \right) $ & $=$ & $\widetilde{%
\mathcal{V}}_{BH}\left( \sigma ,R,\lambda \right) $ & ,%
\end{tabular}%
\end{equation}%
with%
\begin{equation}
\begin{tabular}{llll}
$Z$ & $=$ & $Z\left( \sigma ,R\right) $ & , \\
$R$ & $=$ & $R\left( L_{\Lambda \Sigma }\right) $ & ,%
\end{tabular}%
\end{equation}%
the attractor eqs for the six dimensional magnetic black hole can be
written in different, but equivalent manners depending of the
variables we use.
\newline
For example, the attractor eqs can stated by using as variables the dilaton $%
\sigma $, the dressed charges $R^{a}$ and $R^{I}$ and obviously the
Lagrange
multiplier $\lambda $. Then we have,%
\begin{equation}
\begin{tabular}{llll}
$\frac{\delta \widetilde{\mathcal{V}}_{BH}\ }{\delta \sigma }$ & $=$
& $0$ &
, \\
$\frac{\delta \widetilde{\mathcal{V}}_{BH}\ }{\delta R}$ & $=$ & $0$ & , \\
$\frac{\delta \widetilde{\mathcal{V}}_{BH}\ }{\delta \lambda }$ &
$=$ & $0$
& $.$%
\end{tabular}%
\end{equation}%
They can be also expressed by using as variables the dressed central
charges $Z^{a}=e^{-\sigma }R^{a}$ and $Z^{I}=e^{-\sigma }R^{I}$ and
the Lagrange
multiplier as follows,%
\begin{equation}
\begin{tabular}{llll}
$\left( 1+\mathrm{\lambda }\right) Z^{a}$ & $=$ & $0$ & , \\
$\left( 1-\mathrm{\lambda }\right) Z^{I}$ & $=$ & $0$ & , \\
$R_{a}R^{a}-R_{I}R^{I}$ & $=$ & $\mathrm{g}^{2}$ & ,%
\end{tabular}
\label{lam}
\end{equation}%
where now the dilaton has been absorbed in the Z's. \newline There
are three kinds of solutions of the eqs(\ref{lam}). These solutions
are given by:%
\begin{eqnarray}
\text{{\small solution} }\left( {\small 1}\right) &:&\quad
Z_{a}=0\quad
,\quad Z_{I}=0\quad ,\quad \lambda \neq \pm 1\text{ },  \label{sol1} \\
\text{{\small solution} }\left( {\small 2}\right) &:&\quad
Z_{a}=0\quad
,\quad Z_{I}\neq 0\quad ,\quad \lambda =+1\text{ },  \label{sol2} \\
\text{{\small solution} }\left( {\small 3}\right) &:&\quad Z_{a}\neq
0\quad ,\quad Z_{I}=0\quad ,\quad \lambda =-1\text{ }.  \label{sol3}
\end{eqnarray}%
The first one is a singular degenerate solution; while the two
others describe respectively a non BPS and a BPS black hole.
\newline
Moreover, since the dressed magnetic charge $Z$ is given by the product $%
e^{-\sigma }R$, the vanishing of the product%
\begin{equation}
e^{-\sigma }R=0\text{ },
\end{equation}%
can, in addition to the singular case $e^{-\sigma }=0=R$, be solved
as well by taking
\begin{equation}
e^{-\sigma }=0\quad ,\quad R\neq 0
\end{equation}%
or%
\begin{equation}
e^{-\sigma }\neq 0\quad ,\quad R=0.
\end{equation}%
Then, we have to distinguish the two following solutions:

\textbf{(a) }\emph{case 1}$:\sigma _{0}\rightarrow \infty $ \
\newline In this case, the value of the minimum of the potential at
the critical point is given by,\
\begin{equation}
\widetilde{\mathcal{V}}_{BH}^{\min }=e^{-2\sigma
_{0}}\mathrm{g}^{2}=0.
\end{equation}%
So the entropy $\mathcal{S}_{BH}$ of the black hole is zero,%
\begin{equation}
\mathcal{S}_{BH}=0.  \label{ebh}
\end{equation}
This configuration corresponds to the solution (\ref{sol1}).

\textbf{(b) }\emph{case 2}$:\sigma =\sigma _{0}<\infty $.\newline
Here the value of the dilaton at horizon $\sigma \left( r_{\text{horizon}%
}\right) =\sigma _{0}$ is given by a \emph{finite} number ($\sigma
_{0}<\infty $). The two solutions (\ref{sol2}-\ref{sol3}) read as follows%
\begin{equation}
\begin{tabular}{lllllll}
$\text{case }\left( 2a\right) :$ & $R_{a}=0$ & , & $R_{I}\neq 0$ & , & $%
\lambda =1$ & , \\
$\text{case }\left( 2b\right) :$ & $R_{a}\neq 0$ & , & $R_{I}=0$ & , & $%
\lambda =-1$ & .%
\end{tabular}%
\end{equation}%
The corresponding values of the 6D black potential $\widetilde{\mathcal{V}}%
_{BH}^{\min }$ at the minimum are%
\begin{equation}
\begin{tabular}{lllll}
$\text{case }\left( 2a\right) :$ &
$\widetilde{\mathcal{V}}_{BH}^{\min
}=-e^{-2\sigma _{0}}\mathrm{g}^{2}>0$ & , & $\mathrm{g}^{2}<0$ & , \\
$\text{case }\left( 2b\right) :$ &
$\widetilde{\mathcal{V}}_{BH}^{\min
}=+e^{-2\sigma _{0}}\mathrm{g}^{2}>0$ & , & $\mathrm{g}^{2}>0$ & .%
\end{tabular}%
\end{equation}%
In these relations, $\sigma _{0}$ is a \emph{free parameter}. To fix
it, we need an extra constraint. We will see later on that $\sigma
_{0}$ can be
indeed fixed in the case of the dyonic pair \emph{DP }$\equiv $ \emph{BH-BM}%
. \newline Before that let us complete this analysis by considering
also the effective
potential $\mathcal{V}_{BM}$ of the black membrane and its entropy $\mathcal{%
S}_{BM}$.

\subsection{Effective potential $\mathcal{V}_{BM}$}

\qquad The electrically charged black membrane \emph{BM} is the dual
of the magnetic black hole \emph{BH} considered above. Its effective
scalar
potential $\mathcal{V}_{BM}$ depends on the electric charges $\mathrm{q}%
_{\Lambda }$ and the field variables of the moduli space
(\ref{mds}). The 24
bare electric charges $\mathrm{q}_{\Lambda }$\ are given by,%
\begin{equation}
\begin{tabular}{llll}
$\mathrm{q}^{\Lambda }$ & $=$ &
$\int_{S^{4}}\mathcal{F}_{4}^{\Lambda }$ & ,
\\
$\mathcal{F}_{4}^{\Lambda }$ & $=$ & $\text{ }^{\star }\left( \mathcal{F}%
_{2}^{\Lambda }\right) $ & .%
\end{tabular}%
\end{equation}%
The $\mathrm{q}_{\Lambda }$'s with $\Lambda =1,...,24$, form a 24-
vector charge of the group $SO\left( 4,20\right) $. \newline The
explicit expression of potential $\mathcal{V}_{BM}$ of the black
membrane can be read like
\begin{equation}
\mathcal{V}_{BM}=\left( W_{a}W^{a}+W_{I}W^{I}\right) \text{ },
\label{b2b}
\end{equation}%
where now $W_{a}$ and $W_{I}$ are respectively the dressed electric
charges of the bare ones $\mathrm{q}_{\Lambda }$. These dressed
charges can be
expressed as linear combination as follows,%
\begin{equation}
\begin{tabular}{llll}
$W^{a}$ & $=$ & $\dsum\limits_{\Lambda =1}^{24}\mathrm{q}_{\Lambda
}P^{\Lambda a}$ & , \\
$W^{I}$ & $=$ & $\dsum\limits_{\Lambda =1}^{24}\mathrm{q}_{\Lambda
}P^{\Lambda I}$ & ,%
\end{tabular}
\label{p}
\end{equation}%
where, like for $U_{\Lambda \Sigma }$ of eq(\ref{u}), the field matrix $%
P^{\Lambda \Sigma }$ parameterizes the moduli space (\ref{mds}).
\newline The matrices $P^{\Lambda \Sigma }$ and $U_{\Lambda \Sigma
}$ are then related to each other. To obtain this relation, we use
the electric/magnetic duality exchanging the black hole and the
black membrane charges. \newline Formally, the electric/magnetic
duality can be stated at the level of the
effective scalar potentials $\mathcal{V}_{BH}$ and $\mathcal{V}_{BM}$ like%
\begin{equation}
\begin{tabular}{llll}
$\mathrm{g}^{\Lambda }$ & $\leftrightarrow $ & $\mathrm{q}_{\Lambda
}$ & ,
\\
$\mathcal{V}_{BH}$ & $\leftrightarrow $ & $\mathcal{V}_{BM}$ & ,%
\end{tabular}%
\end{equation}%
and then%
\begin{equation}
\begin{tabular}{llll}
$R_{a}$ & $\leftrightarrow $ & $W^{a}$ & , \\
$R_{I}$ & $\leftrightarrow $ & $W^{I}$ & .%
\end{tabular}%
\end{equation}%
Extending the electric/magnetic duality relation eqs(\ref{qa}),
which we
rewrite as follows,%
\begin{equation}
\begin{tabular}{llll}
$\sum_{\Lambda =1}^{24}\mathrm{q}_{\Lambda }\mathrm{g}^{\Lambda }$ & $=$ & $%
2\pi k$ & , \\
$\sum_{\Lambda =1}^{24}k_{\Lambda }$ & $=$ & $k\in \mathbb{Z}$ & ,%
\end{tabular}%
\end{equation}%
to the dressed charges,%
\begin{equation}
\sum_{\Lambda =1}^{24}W^{\Lambda }Z_{\Lambda }=k\text{ },
\label{wz}
\end{equation}%
we can determine the relation between $U_{\Lambda \Sigma }$ and
$P_{\Lambda \Sigma }$ matrices. Indeed putting
\begin{equation}
W^{\Lambda }=\sum\limits_{\Sigma }P^{\Lambda \Sigma
}\mathrm{q}_{\Sigma },
\end{equation}%
and
\begin{equation}
Z_{\Lambda }=\sum\limits_{\Upsilon }\mathrm{g}^{\Upsilon
}L_{\Upsilon \Lambda },
\end{equation}%
back into above relation, we find that the matrix $P^{\Upsilon
\Lambda }$ is just the inverse of the matrix $U_{\Sigma \Upsilon }$.
\newline Therefore, electric/magnetic duality mapping the black hole
\emph{BH} to the black membrane \emph{BM} is given by

\begin{equation}
\begin{tabular}{|llll|}
\hline
$\text{black hole BH}$ & $\qquad \underleftrightarrow{\text{{\small %
electric/magnetic}}}\qquad $ & $\text{black membrane BM}$ &  \\
&  &  &  \\
$\qquad \mathrm{g}^{\Lambda }$ & $\qquad \qquad \leftrightarrow \qquad $ & $%
\qquad \mathrm{q}_{\Lambda }$ &  \\
&  &  &  \\
$\qquad U_{\Sigma \Upsilon }$ & $\qquad \qquad \leftrightarrow \qquad $ & $%
\qquad P^{\Upsilon \Lambda }=\left( U_{\Sigma \Upsilon }\right) ^{-1}$ &  \\
&  &  &  \\
$\qquad Z_{\Lambda }$ & $\qquad \qquad \leftrightarrow \qquad $ &
$\qquad
W^{\Lambda }$ &  \\
&  &  &  \\
$\qquad \lambda $ & $\qquad \qquad \leftrightarrow \qquad $ &
$\qquad \xi $
&  \\
&  &  &  \\ \hline
\end{tabular}
\label{cr}
\end{equation}

\ \newline Notice that, like for the matrix $U_{\Lambda \Sigma }$,
we also have the following properties:\newline (\textbf{i}) the
factorization of the moduli space (\ref{mds}) as $SO\left(
1,1\right) \times \frac{SO\left( 4,20\right) }{SO\left( 4\right)
\times SO\left( 20\right) }$ allows to factorize $P^{\Lambda \Sigma
}$ like,
\begin{equation}
P^{\Lambda \Sigma }=e^{+\sigma }\left( L^{-1}\right) ^{\Lambda
\Sigma }.
\end{equation}%
Multiplying both sides of this equation by $\mathrm{q}_{\Sigma }$,
we obtain
the dressed central charges $W^{\Sigma }=\left( W^{a},W^{I}\right) $%
\begin{equation}
\begin{tabular}{llll}
$W^{a}$ & $=$ & $\mathrm{q}_{\Lambda }P^{\Lambda a}$ & , \\
$W^{I}$ & $=$ & $\mathrm{q}_{\Lambda }P^{\Lambda I}$ & .%
\end{tabular}%
\end{equation}%
As in the case of 6D black hole eq(\ref{fz}), these charges
factorize as well like
\begin{equation}
\begin{tabular}{llll}
$W^{a}$ & $=$ & $e^{+\sigma }T^{a}$ & , \\
$W^{I}$ & $=$ & $e^{+\sigma }T^{I}$ & ,%
\end{tabular}
\label{tq}
\end{equation}%
with%
\begin{equation}
\begin{tabular}{llll}
$T^{a}$ & $=$ & $\mathrm{q}_{\Lambda }\left( L^{-1}\right) ^{\Lambda
a}$ & ,
\\
$T^{I}$ & $=$ & $\mathrm{q}_{\Lambda }\left( L^{-1}\right) ^{\Lambda I}$ & .%
\end{tabular}
\label{tt}
\end{equation}%
The dressed charges $T^{a}$ and $T^{I}$\ are the dual of the $R_{a}$ and $%
R_{I}$.\newline (\textbf{ii}) the orthogonality property of the non
compact $SO\left(
4,20\right) $ group, which we can be written as%
\begin{equation}
\left( \sum_{a=1}^{4}\left( L^{-1}\right) ^{a\Lambda }\left(
L^{-1}\right) ^{a\Sigma }-\sum_{K=1}^{20}\left( L^{-1}\right)
^{K\Lambda }\left( L^{-1}\right) ^{K\Sigma }\right) =\eta ^{\Lambda
\Sigma },
\end{equation}%
allows to get more information on the dressed electric charges.
Multiplying
both sides of this algebraic constraint equation by $\mathrm{q}_{\Lambda }%
\mathrm{q}_{\Sigma }$, we obtain
\begin{equation}
W_{a}W^{a}-W_{I}W^{I}=e^{+2\sigma }\mathrm{q}^{2}\text{ },
\label{cstr}
\end{equation}%
with%
\begin{equation}
\mathrm{q}^{2}=\mathrm{q}_{\Lambda }\eta ^{\Lambda \Sigma }\mathrm{q}%
_{\Sigma }=\left( \sum_{a=1}^{4}\mathrm{q}_{a}^{2}-\sum_{I=1}^{20}\mathrm{q}%
_{I}^{2}\right) \text{ }.
\end{equation}%
This is the electric analogue of the constraint relation (\ref{or})
concerning the dressed magnetic charges of the black hole. This
condition
will be implemented in the effective potential $\mathcal{V}_{BM}$ eq(\ref%
{b2b}) by using a Lagrange multiplier $\xi $. The auxiliary field
$\xi $ should be thought as the analogue of the Lagrange multiplier
$\lambda $ used in the black hole case; see also table
(\ref{cr}).\newline Combining eq(\ref{b2b}) with eq(\ref{cstr}), we
end with the following
generalized effective scalar potential for the black membrane,%
\begin{equation}
\widetilde{\mathcal{V}}_{BM}=\left( W_{a}W^{a}+W_{I}W^{I}\right) +\mathrm{%
\xi }\left( W_{a}W^{a}-W_{I}W^{I}-e^{+2\sigma }\mathrm{q}^{2}\right)
. \label{ks}
\end{equation}%
Notice that lowering and rising indices of $SO\left( 4\right) $ and $%
SO\left( 20\right) $ are done with the usual Kronecker metric, that is $%
W_{a}=W^{a}$ and $W_{I}=W^{I}$. Those of $SO\left( 4,20\right) $ are
done with the metric $\eta _{\Lambda \Sigma }$.

\subsubsection{Black membrane attractor equations}

\qquad The effective scalar potential of the 6D black membrane can
be also
put in the form%
\begin{equation}
\widetilde{\mathcal{V}}_{BM}=\left( 1+\mathrm{\xi }\right)
W_{a}W^{a}+\left( 1-\mathrm{\xi }\right) W_{I}W^{I}-\mathrm{\xi
}e^{+2\sigma }\mathrm{q}^{2}. \label{vef}
\end{equation}%
The variation of $\widetilde{\mathcal{V}}_{BM}$ with respect to
$\mathrm{\xi }$ gives precisely the condition (\ref{cstr}); while
the variation with
respect to $W_{a}$ and $W_{I}$ give constraint eqs on the field moduli,%
\begin{equation}
\begin{tabular}{llll}
$\frac{\delta \widetilde{\mathcal{V}}_{BM}}{\delta W^{a}}$ & $=$ & $\left( 1+%
\mathrm{\xi }\right) W_{a}$ & , \\
$\frac{\delta \widetilde{\mathcal{V}}_{BM}}{\delta W^{I}}$ & $=$ & $\left( 1-%
\mathrm{\xi }\right) W_{I}$ & , \\
$\frac{\delta \widetilde{\mathcal{V}}_{BM}}{\delta \mathrm{\xi }}$ & $=$ & $%
W_{a}W^{a}-W_{I}W^{I}-e^{+2\sigma }\mathrm{q}^{2}$ & .%
\end{tabular}
\label{c23}
\end{equation}%
The attractor eqs for the black membrane corresponds to the extremum
(minimum) of this potential. These eqs read as follows
\begin{eqnarray}
\left( 1+\mathrm{\xi }\right) W_{a} &=&0\text{ \ },  \label{1} \\
\left( 1-\mathrm{\xi }\right) W_{I} &=&0\text{ \ },  \label{2} \\
\left( W_{a}W^{a}-W_{I}W^{I}\right) -e^{+2\sigma }\mathrm{q}^{2}
&=&0\text{ \ }.  \label{3}
\end{eqnarray}%
Like for the black hole, there are three solutions extremizing the
effective scalar potential $\widetilde{\mathcal{V}}_{BM}$. These
solutions, which are
classified by the sign of the semi-norm of the electric charge vector $%
\mathrm{q}_{\Lambda }$, are listed below:\newline
(1) \emph{first solution} ($\mathrm{q}^{2}=0$)$:$%
\begin{equation}
\begin{tabular}{llll}
$W_{a}$ & $=$ & $0$ & , \\
$W_{I}$ & $=$ & $0$ & , \\
$\mathrm{\xi }$ & $\neq $ & $\pm 1$ & .%
\end{tabular}%
\end{equation}%
This is a singular solution.\newline
(2) \emph{second solution} ($\mathrm{q}^{2}<0$)$:$%
\begin{equation}
\begin{tabular}{llll}
$W_{a}$ & $=$ & $0$ & , \\
$W_{I}$ & $\neq $ & $0$ & , \\
$\mathrm{\xi }$ & $=$ & $+1$ & .%
\end{tabular}
\label{ga}
\end{equation}%
This solution corresponds to \emph{non BPS} black membrane.\newline
(3) \emph{third solution} ($\mathrm{q}^{2}>0$)$:$%
\begin{equation}
\begin{tabular}{llll}
$W_{a}$ & $\neq $ & $0$ & , \\
$W_{I}$ & $=$ & $0$ & , \\
$\mathrm{\xi }$ & $=$ & $-1$ & .%
\end{tabular}
\label{gb}
\end{equation}%
This solution corresponds to \emph{BPS} black membrane.\newline
Putting these solutions back into (\ref{vef}), we can determine the value $%
\widetilde{\mathcal{V}}_{BM}^{\min }$ of the effective potential at
the extremum. For the three solutions, the extremal values can be
combined
altogether in a unique form given by:%
\begin{equation}
\widetilde{\mathcal{V}}_{BM}^{\min }=-\mathrm{\xi }e^{+2\sigma }\mathrm{q}%
^{2}\text{ .}
\end{equation}%
Notice that due to the constraint eq(\ref{3}) which requires $\mathrm{q}%
^{2}=0$ for $W_{a}=W_{I}=0$, the potential at the first extremum (
first
solution) should vanish:%
\begin{equation}
\left( 1\right) :\qquad \widetilde{\mathcal{V}}_{BM}^{\min }=0\text{
}. \label{tr}
\end{equation}%
For the two other cases (2) and (3) with $\mathrm{q}^{2}\neq 0$, the
values
of the effective potential at the corresponding extremum read as follows:%
\begin{equation}
\widetilde{\mathcal{V}}_{BM}^{\min }=e^{+2\sigma }\left\vert \mathrm{q}%
^{2}\right\vert >0\text{ },
\end{equation}%
where the dependence into the Lagrange parameter $\xi $ has been
also fixed as $\xi =\pm 1$. \newline Notice that $\xi =+1$
corresponds to the non BPS black membrane while $\xi =1 $ is a BPS
state. \newline Notice also that the value of the effective
potential at the extremums depends on the factor $e^{+2\sigma }$
which, like in the case of the black hole, is an unfixed
modulus.\newline Below, we give more details concerning the above
solutions; in particular those solutions with
\begin{equation}
\widetilde{\mathcal{V}}_{BM}^{\min }>0\text{ }.
\end{equation}%
Then, we turn to study the free factor $e^{\pm 2\sigma }$ and show
how it can be fixed in the case of the dyonic attractor pair
\emph{BH-BM}.

\subsubsection{Solving eq(\protect\ref{c23})}

\qquad Recall that $W^{a}$ and $W^{I}$ depend, in addition to
$\left( L^{-1}\right) ^{\Lambda \Sigma }$, on the dilaton $\sigma $
in the following manner eq(\ref{tq}),
\begin{equation}
\begin{tabular}{llll}
$W^{a}$ & $=$ & $e^{+\sigma }T^{a}$ & , \\
$W^{I}$ & $=$ & $e^{+\sigma }T^{I}$ & ,%
\end{tabular}%
\end{equation}%
with%
\begin{equation}
\begin{tabular}{llll}
$T^{a}$ & $=$ & $\mathrm{q}_{\Lambda }\left( L^{-1}\right) ^{\Lambda
a}$ & ,
\\
$T^{I}$ & $=$ & $\mathrm{q}_{\Lambda }\left( L^{-1}\right) ^{\Lambda I}$ & .%
\end{tabular}%
\end{equation}%
Using this factorization, we will show that there are various ways
to solve the attractor eqs of the black membrane. \newline
Among these solutions, we have the degenerate one associated with $%
W_{a}=0=W_{I}$ and leading to
\begin{equation}
\widetilde{\mathcal{V}}_{BM}^{\min }=0.
\end{equation}%
This solution will be ignored hereafter. \newline The two other
solutions are those associated with eqs(\ref{ga}-\ref{gb}). We have:

\emph{A. case:} $W_{a}=0$ $,$ $W_{I}\neq 0$ $,$ $\xi =1$.\newline
Since the $W$'s depends on the moduli and the bare charges; i.e,%
\begin{equation}
W=W\left( \sigma ,L,q\right) \text{ },
\end{equation}%
the conditions $W_{a}=0$ and $W_{I}\neq 0$ allows then to give the
relation
between the field moduli of (\ref{mds}) and electric charges \textrm{q}$%
_{\Lambda }$ of the black membrane. \newline
Substituting $W_{a}$ and $W_{I}$ in terms of $T_{a}$ and $T_{I}$, we have%
\begin{equation}
\begin{tabular}{llllll}
$W_{a}$ & $=$ & $e^{+\sigma }T^{a}$ & $=$ & $0$ & , \\
$W_{I}$ & $=$ & $e^{+\sigma }T^{I}$ & $\neq $ & $0$ & .%
\end{tabular}
\label{wa}
\end{equation}%
Obviously, the solutions of the above relations should satisfy the
constraint equation%
\begin{equation}
W_{a}W^{a}-W_{I}W^{I}=e^{+2\sigma }\mathrm{q}^{2}\text{ ,}
\end{equation}%
which, by substituting $W_{a}=0$, reduces to%
\begin{equation}
\dsum\limits_{I=1}^{20}W_{I}W^{I}=\dsum\limits_{I=1}^{20}W_{I}^{2}=-e^{+2%
\sigma }\mathrm{q}^{2}>0\text{ }.  \label{30}
\end{equation}%
As we see, definite positivity of the norm
$\sum_{I=1}^{20}W_{I}^{2}$ requires
\begin{equation}
\mathrm{q}^{2}=-\left\vert \mathrm{q}^{2}\right\vert <0\text{ }.
\end{equation}%
Eq(\ref{wa}) can be solved in two basic ways as follows:\newline
(\textbf{a}) either by taking $\sigma \rightarrow -\infty $ whatever
the values of $T^{a}$; in particular $T^{a}\neq 0$. But this
solution should be ruled out since we should have
\begin{equation}
W_{I}=e^{+\sigma }T_{I}\neq 0\text{ },
\end{equation}%
which violates eq(\ref{30}).\newline (\textbf{b}) or by taking
$\sigma =\sigma _{2}$, an arbitrary but \emph{a finite} number (say
$\sigma _{2}<\infty $), and $T^{a}=0$ but $T^{I}\neq 0$.
\newline
A solution for $T^{a}=0$ depends of the value of $\mathrm{q}^{2}=\mathrm{q}%
_{\Lambda }\eta ^{\Lambda \Sigma }\mathrm{q}_{\Sigma }$ and can, a
priori, be split into two situations (i) and (ii) corresponding
respectively to:

\textbf{(i)} \emph{a} \emph{light like charge vector} $\mathrm{q}^{2}=0$%
\newline
We already know that this case should be ruled out; but it is
interesting to
see the explicit relation between the field moduli $L_{\Lambda \Sigma }$ of (%
\ref{mds}) and the electric charges of the black membrane. We have
\begin{equation}
\begin{tabular}{lllll}
$\left( L^{-1}\right) ^{\Lambda a}$ & $=$ & $\#\text{ }q^{\Lambda
}q^{a}$ &
& , \\
$T^{a}$ & $=$ & $\#\text{ }\mathrm{q}^{2}q^{a}$ & $=0$ & .%
\end{tabular}%
\end{equation}%
However, because of eq(\ref{30}) which requires
\begin{equation}
\begin{tabular}{llllll}
$\dsum\limits_{I=1}^{20}W_{I}W^{I}$ & $=$ & $e^{+2\sigma
_{2}}\dsum\limits_{I=1}^{20}T_{I}T^{I}$ & $=$ & $-e^{+2\sigma _{2}}\mathrm{q}%
^{2}$ & ,%
\end{tabular}%
\end{equation}%
we get%
\begin{equation}
\dsum\limits_{I=1}^{20}T_{I}T^{I}=0\qquad \Rightarrow \qquad
T_{I}=0\text{ }.
\end{equation}%
This solution should be then ruled out since $T_{I}\neq 0$.

(\textbf{ii}) \emph{a non zero} \emph{semi-norm} $\mathrm{q}^{2}\neq 0$%
\newline
We have the following:%
\begin{equation}
\begin{tabular}{lllll}
$\left( L^{-1}\right) ^{\Lambda a}$ & $=$ & $\frac{\left( q^{\Lambda }q^{a}-%
\mathrm{q}^{2}\eta ^{\Lambda a}\right) }{\mathrm{q}^{2}}$ &  & , \\
$T_{a}$ & $=$ & $\frac{\left(
\mathrm{q}^{2}q^{a}-\mathrm{q}^{2}q^{a}\right)
}{\mathrm{q}^{2}}$ & $=0$ & .%
\end{tabular}%
\end{equation}%
This solution is acceptable provided $\mathrm{q}^{2}<0$ since
eq(\ref{30})
requires%
\begin{equation}
\dsum\limits_{I=1}^{20}T_{I}T^{I}=-\text{ }\mathrm{q}^{2}>0\text{ }.
\end{equation}%
From this relation we can determine $T^{I}$; i.e%
\begin{equation}
T^{I}=\frac{q^{I}\sqrt{-\mathrm{q}^{2}}}{\left(
\sum_{J=1}^{20}q_{J}^{2}\ \right) }\qquad ,\qquad I=1,...,20\text{
,}
\end{equation}%
which leads in turns to%
\begin{equation}
\left( L^{-1}\right) ^{\Lambda J}=\mathrm{q}^{\Lambda }\mathrm{q}^{J}\frac{%
\sqrt{-\mathrm{q}^{2}}}{\mathrm{q}^{2}\left(
\sum_{I=1}^{20}q_{I}^{2}\right) }\qquad ,\qquad I=1,...,20\text{ }.
\end{equation}%
In this case, the values of the effective scalar potential $\widetilde{%
\mathcal{V}}_{BM}$ at the minimum is given by
\begin{equation}
\widetilde{\mathcal{V}}_{BM}^{\min }=-e^{+2\sigma _{2}}\mathrm{q}%
^{2}=e^{+2\sigma _{2}}\left\vert \mathrm{q}^{2}\right\vert >0\text{
}.
\end{equation}%
It depends on the electric charges. but it has a free dependence in
the value $\sigma _{2}$\ of the dilaton.

\emph{B. case :} \ $W_{a}\neq 0$ $,$ $W_{I}=0$ $,$ $\xi =-1$\newline
We have to solve%
\begin{equation}
\begin{tabular}{llllll}
$W_{a}$ & $=$ & $e^{+\sigma }T^{a}$ & $\neq $ & $0$ & , \\
$W_{I}$ & $=$ & $e^{+\sigma }T^{I}$ & $=$ & $0$ & , \\
$\xi $ & $=$ & $-1$ &  &  & ,%
\end{tabular}%
\end{equation}%
with the constraint relation%
\begin{equation}
\begin{tabular}{llllll}
$\dsum\limits_{a=1}^{4}W_{a}W^{a}$ & $=$ &
$\dsum\limits_{a=1}^{4}W_{a}^{2}$
& $=$ & $e^{+2\sigma }\mathrm{q}^{2}>0$ & .%
\end{tabular}%
\end{equation}%
From this constraint relation we see that the electric charges of
the black membrane should be $\mathrm{q}^{2}>0$. \newline The method
is quite similar to the one used for the black hole case. After
some straightforward calculations, we end with the following%
\begin{equation}
\begin{tabular}{llllll}
$\text{case }\left( 3\right) :$ & $\widetilde{\mathcal{V}}_{BM}^{\min }$ & $%
= $ & $e^{+2\sigma _{2}}\mathrm{q}^{2}$ & $>0$ & ,%
\end{tabular}%
\end{equation}%
where we still have the unfixed factor $e^{+2\sigma _{2}}$.

\section{Entropy of the pair \emph{BH-BM}}

\qquad We start by recalling the various expressions of the
effective scalar potentials of the 6D black attractors that have
been obtained so far. These are collected in the following table

\begin{equation}
\begin{tabular}{lllll}
\underline{\emph{6D black attractors}} & : &  &
\underline{\emph{effective
scalar potential}} &  \\
&  &  &  &  \\
dyonic{\small \ }$\text{black string}$ & : &  &
$\mathcal{V}_{BS}\left(
\sigma \right) =\frac{\mathrm{q}_{0}^{2}}{2}e^{4\sigma }+\frac{\mathrm{g}%
_{0}^{2}}{2}e^{-4\sigma }$ &  \\
&  &  &  &  \\
$\text{black hole}$ & : &  & $\mathcal{V}_{BH}\left( \sigma
,R,\lambda
\right) =e^{-2\sigma }\mathcal{V}_{0}\left( R,\lambda \right) $ &  \\
&  &  &  &  \\
$\text{black 2-brane}$ & : &  & $\mathcal{V}_{BM}\left( \sigma
,T,\xi \right) =e^{2\sigma }\mathcal{V}_{2}\left( T,\xi \right) $ &
\end{tabular}
\label{hb}
\end{equation}

\ \ \ \ \newline The entropy $\mathcal{S}_{BS}$ of the dyonic
\emph{6D} black string \emph{BS}
reads, in terms of the electric $\mathrm{q}_{0}$ and magnetic $\mathrm{g}%
_{0} $ charges of the 3-form field strength $\mathcal{H}_{3}$, as
follows:
\begin{equation}
\begin{tabular}{llll}
$\mathcal{S}_{BS}$ & $=$ & $\frac{\left\vert \mathrm{g}_{0}\mathrm{q}%
_{0}\right\vert }{4}$ & ,%
\end{tabular}%
\end{equation}%
or again like%
\begin{equation}
\begin{tabular}{llllll}
$\mathcal{S}_{BS}$ & $=$ & $\frac{\pi }{2}\left\vert \mathrm{k}%
_{0}\right\vert $ & , & $\mathrm{k}_{0}\in \mathbb{Z}^{\ast }$ & .%
\end{tabular}%
\end{equation}%
For the entropies $\mathcal{S}_{BH}$ \ and $\mathcal{S}_{BM}$ of the \emph{6D%
} black hole \emph{BH} and black membrane \emph{BM}, the situation
is a little bit different. \newline
Viewed separately, the corresponding entropies $\mathcal{S}_{BH}$ and $%
\mathcal{S}_{BM}$ are respectively given by:%
\begin{equation}
\begin{tabular}{llll}
$\mathcal{S}_{BH}$ & $=$ & $\frac{1}{4}e^{-2\sigma _{0}}\left\vert \mathrm{g}%
^{2}\right\vert $ & ,%
\end{tabular}
\label{71}
\end{equation}%
and%
\begin{equation}
\begin{tabular}{llll}
$\mathcal{S}_{BM}$ & $=$ & $\frac{1}{4}e^{2\sigma _{2}}\left\vert \mathrm{q}%
^{2}\right\vert $ & ,%
\end{tabular}
\label{72}
\end{equation}%
with%
\begin{equation}
\begin{tabular}{llll}
$\mathrm{g}^{2}$ & $=$ & $\left(
\sum_{a=1}^{4}g_{a}^{2}-\sum_{I=1}^{20}g_{I}^{2}\right) $ & , \\
$\mathrm{q}^{2}$ & $=$ & $\left(
\sum_{a=1}^{4}q_{a}^{2}-\sum_{I=1}^{20}q_{I}^{2}\right) $ & .%
\end{tabular}%
\end{equation}%
In these relations $\sigma _{0}$ and $\sigma _{2}$ stand
respectively for the value of the dilaton field at the horizon of
the black hole \emph{BH} and the black membrane \emph{BM}:
\begin{equation}
\begin{tabular}{lllllll}
$\sigma _{0}$ & $=$ & $\sigma \left( r_{bh}\right) $ & $,$ &
$r_{bh}=$ &
{\small black hole horizon} & , \\
$\sigma _{2}$ & $=$ & $\sigma \left( r_{bm}\right) $ & , & $r_{bm}=$ & $%
\text{{\small black 2-brane horizon}}$ & .%
\end{tabular}%
\end{equation}%
As we have noted before, $\sigma _{0}$ and $\sigma _{2}$ might also
take finite values but unfortunately cannot be fixed if dealing with
\emph{BH} and \emph{BM} as independent objects.\newline
A way to see why the effective potentials%
\begin{equation}
\mathcal{V}_{BH}=\mathcal{V}_{BH}\left( \sigma ,R,\lambda \right) ,
\end{equation}%
and
\begin{equation}
\mathcal{V}_{BM}=\mathcal{V}_{BM}\left( \sigma ,T,\xi \right) ,
\end{equation}%
cannot fix the dilaton at their extremum is to note the
following:\newline
\textbf{(i)} First the scalar potentials $\mathcal{V}_{BH}$ and $\mathcal{V}%
_{BM}$ are eigenfunctions of the operator $\frac{d}{d\sigma }$:
\begin{equation}
\begin{tabular}{llll}
$\frac{d\mathcal{V}_{BH}}{d\sigma }$ & $=$ & $-2\mathcal{V}_{BH}$ & , \\
$\frac{d\mathcal{V}_{BM}}{d\sigma }$ & $=$ & $+2\mathcal{V}_{BM}$ & .%
\end{tabular}%
\end{equation}%
\textbf{(ii)} the zeros of the effective potentials $\mathcal{V}_{BH}$ and $%
\mathcal{V}_{BM}$ can be obtained in three ways. \newline
In the case of the black hole \emph{BH}, the zeros are given by,%
\begin{equation}
\begin{tabular}{llllll}
$\mathcal{V}_{BH}$ & $=$ & $e^{-2\sigma _{0}}\mathcal{V}_{0}\left(
R,\lambda \right) $ & $=0$ & $\Rightarrow $ & $\left\{
\begin{array}{c}
\left( 1\right) :e^{-2\sigma _{0}}=0,\qquad \mathcal{V}_{0}\left(
R,\lambda
\right) =0 \\
\left( 2\right) :e^{-2\sigma _{0}}=0,\qquad \mathcal{V}_{0}\left(
R,\lambda
\right) \neq 0 \\
\left( 3\right) :e^{-2\sigma _{0}}\neq 0,\qquad
\mathcal{V}_{0}\left(
R,\lambda \right) =0%
\end{array}%
\right. $%
\end{tabular}
\label{81}
\end{equation}%
For the configurations (1) and (2), the value $\sigma _{0}$ of the
dilaton at the critical point is
\begin{equation}
\sigma _{0}\rightarrow +\infty \text{ }.
\end{equation}%
They lead to the degenerate relations\textrm{\
}(\ref{se}-\ref{ebd}).
\newline
However, for the third configuration, the critical value of the
dilaton is unfixed and can be any arbitrary; but finite, value. This
is the case we are interested in here.\newline
In the case of the black membrane \emph{BM}, we have%
\begin{equation}
\begin{tabular}{lllllll}
$\mathcal{V}_{BM}$ & $=$ & $e^{+2\sigma _{2}}\mathcal{V}_{2}\left(
T,\xi \right) $ & $=$ & $0$ & $\Rightarrow $ & $\left\{
\begin{array}{c}
\left( 1\right) :e^{+2\sigma _{2}}=0,\qquad \mathcal{V}_{2}\left(
T,\xi
\right) =0 \\
\left( 2\right) :e^{+2\sigma _{2}}=0,\qquad \mathcal{V}_{2}\left(
T,\xi
\right) \neq 0 \\
\left( 3\right) :e^{+2\sigma _{2}}\neq 0,\qquad
\mathcal{V}_{2}\left( T,\xi
\right) =0%
\end{array}%
\right. $%
\end{tabular}
\label{82}
\end{equation}%
The configurations (1) and (2) imply
\begin{equation}
\sigma _{2}\rightarrow -\infty \text{ ,}
\end{equation}%
while for the third configuration leaves $\sigma _{2}$ an arbitrary
finite number. \newline Notice that eq(\ref{81}) and (\ref{82})
exhibit very remarkable properties; in particular the two
following:\newline (\textbf{a}) They are exchanged under
electric/magnetic duality. \newline At the black hole and the black
membrane horizons, we then have
\begin{equation}
\begin{tabular}{llll}
$\pm \sigma _{0}$ & $\text{\quad }\leftrightarrow \text{\quad }$ &
$\mp
\sigma _{2}$ & $,$ \\
$g_{\Lambda }$ & $\text{ \quad }\mathrm{\leftrightarrow }\text{ \quad }$ & $%
q_{\Lambda }$ & .%
\end{tabular}
\label{so}
\end{equation}%
(\textbf{b}) The above relation (\ref{so}) should be associated with eq(\ref%
{si}) of the dyonic black string. \newline This property shows that
\begin{equation}
\sigma _{2}=-\sigma _{0}\text{ ,}
\end{equation}%
ending then with one unknown quantity; say $\sigma _{0}$, which
remains unfixed.\newline
Moreover, eq(\ref{so}) teaches us that the black hole potential (\ref{bh}-%
\ref{bo})%
\begin{equation}
\mathcal{V}_{BH}=e^{-2\sigma }\mathcal{V}_{0}\text{ ,}
\end{equation}%
and the black membrane potential (\ref{b2b}-\ref{hb})
\begin{equation}
\mathcal{V}_{BM}=e^{+2\sigma }\mathcal{V}_{2}\text{ ,}
\end{equation}%
as two limits of the potential of the dyonic black pair \emph{DP
}$\equiv $
\emph{BM-BH}.%
\begin{equation}
\mathcal{V}_{DP}\simeq e^{-2\sigma }\mathcal{V}_{0}+e^{+2\sigma }\mathcal{V}%
_{2}\text{ }.
\end{equation}%
In the limit $\sigma \rightarrow -\infty $, the potential
$\mathcal{V}_{DP}$ of the dyonic pair reduces as
\begin{equation}
\mathcal{V}_{DP}\qquad \rightarrow \qquad \mathcal{V}_{BH}\text{ },
\end{equation}%
and in the limit $\sigma \rightarrow +\infty $,\ it behaves like,
\begin{equation}
\mathcal{V}_{DP}\qquad \rightarrow \qquad \mathcal{V}_{BM}\text{ }.
\end{equation}%
To get the explicit expression of $\sigma _{0}$, we have to study
the attractor mechanism of the dyonic attractor \emph{DP }$\equiv $
\emph{BM-BH}.

\subsection{Attractor eqs for the dyonic \emph{DP}}

\qquad To begin, notice that the general moduli dependence of the
effective scalar potential $\mathcal{V}_{DP}$ of the dyonic black
attractor pair is given by,
\begin{equation}
\mathcal{V}_{DP}=\mathcal{V}\left( \sigma ;R,T;\lambda ,\xi ,\zeta
;q,g\right) .
\end{equation}%
The set of parameters $\left\{ \sigma ,R,T,\lambda ,\xi ,\zeta
,q,g\right\} $ is the general set of the possible moduli in which
may depend the effective scalar potential\ and which are supposed to
describe the attractor eqs of the \emph{DP} dyonic pair. They are as
follows:\newline
\textbf{(a)} the $R$'s and the $T$'s are the dressed charges as in eqs(\ref%
{fz}-\ref{ry}) and (\ref{tq}-\ref{tt});\newline
\textbf{(b}) $\lambda $ and $\xi $ are the Lagrange multipliers given by eqs(%
\ref{la}-\ref{lm}) and (\ref{ks}); \newline
\textbf{(c)} $\mathrm{q=}\left( \mathrm{q}_{\Lambda }\right) $ and $\mathrm{%
g=}\left( \mathrm{g}_{\Lambda }\right) $ are the electric and
magnetic charges given by eqs(\ref{k0},\ref{gl},\ref{qa},\ref{ka})
\newline \textbf{(d)} the variable $\zeta $ is an extra Lagrange
multiplier that will be described below. \newline
Notice also that, like for the dyonic black string \emph{BS}, the potential $%
\mathcal{V}_{DP}$ of the dyonic pair should be also invariant under
the electric/magnetic duality (\ref{so}).

\emph{Expression of }$\mathcal{V}_{DP}$\newline The explicit
expression of $\mathcal{V}_{DP}$ is given by the sum of:
\newline
(\textbf{i}) the effective scalar potential of the black hole (\ref{bh}-\ref%
{bo}),
\begin{equation}
\mathcal{V}_{BH}=\mathcal{V}_{BH}\left( \sigma ,R,\lambda
,g_{\Lambda }\right) .
\end{equation}%
(\textbf{ii}) the effective scalar potential of the black membrane (\ref{b2b}%
) which is dual to $\mathcal{V}_{BH},$%
\begin{equation}
\mathcal{V}_{BM}=\mathcal{V}_{BM}\left( \sigma ,T,\xi ,q_{\Lambda
}\right) \text{ }.
\end{equation}%
(\textbf{iii}) an extra term depending on the dressed electric and
magnetic charges $Z$ and $W$. This term is given by\ the constraint
eq(\ref{wz})
\begin{equation}
\mathcal{C}=\mathcal{C}\left( Z,W\right) \text{ },
\end{equation}%
capturing the electric/magnetic duality between the dressed charges
of the black hole and the black membrane. It may be interpreted as
the interaction term.\newline
Then, we have%
\begin{equation}
\mathcal{V}_{DP}=\mathcal{V}_{BH}\ +\mathcal{V}_{BM}+\zeta
\mathcal{C}\text{ },
\end{equation}%
where $\zeta $ is a Lagrange multiplier used to implement the constraint (%
\ref{wz}) in the effective scalar potential of the \emph{DP} dyonic
pair.
\newline
In addition to the various electric and magnetic bare charges \textrm{q}$%
_{\Lambda }$ and \textrm{g}$_{\Lambda }$, the dyonic potential $\mathcal{V}%
_{DP}$ depends on the \emph{eighty one} field variables $\left(
\sigma ,L_{\Lambda \Sigma }\right) $ of the moduli space; and on the
three Lagrange multipliers $\lambda ,$ $\xi $ and $\zeta $. \newline
While the dilaton appears in $\mathcal{V}_{DP}$ as $e^{\pm 2\sigma
}$, the \emph{eighty} field moduli $L_{\Lambda \Sigma }$ are
involved in the game
through the dressed charges,%
\begin{equation}
\begin{tabular}{llll}
$R_{a}$ & $=$ & $R_{a}\left( L_{\Lambda \Sigma }\right) $ & , \\
$R_{I}$ & $=$ & $R_{I}\left( L_{\Lambda \Sigma }\right) $ & ,%
\end{tabular}%
\end{equation}%
and%
\begin{equation}
\begin{tabular}{llll}
$T_{a}$ & $=$ & $T_{a}\left( L_{\Lambda \Sigma }^{-1}\right) $ & , \\
$T_{I}$ & $=$ & $T_{I}\left( L_{\Lambda \Sigma }^{-1}\right) $ & .%
\end{tabular}%
\end{equation}%
By substituting $\mathcal{V}_{BH}$ and $\mathcal{V}_{BM}$ by their
explicit
expression, we can put the \emph{DP} effective scalar potential $\mathcal{V}%
_{DP}$ in the form%
\begin{equation}
\mathcal{V}_{DP}=e^{-2\sigma }\mathcal{V}_{0}+e^{+2\sigma }\mathcal{V}%
_{2}+\zeta \mathcal{C}\text{ },  \label{veff}
\end{equation}%
where we have set%
\begin{equation}
\begin{tabular}{llllll}
$\mathcal{V}_{0}$ & $=$ & $\left( 1+\lambda \right) \sum_{a}R_{a}R^{a}$ & $%
+\left( 1-\lambda \right) \sum_{I}R_{I}R^{I}$ & $-\lambda \mathrm{g}^{2}$ & ,%
\end{tabular}
\label{v02}
\end{equation}%
and%
\begin{equation}
\begin{tabular}{llllll}
$\mathcal{V}_{2}$ & $=$ & $\left( 1+\xi \right) \sum_{a}T_{a}T^{a}$ & $%
+\left( 1-\xi \right) \sum_{I}T_{I}T^{I}$ & $-\xi \mathrm{q}^{2}$ & ,%
\end{tabular}%
\end{equation}%
as well as%
\begin{equation}
\mathcal{C}=-\left( 1-\sum_{\Lambda ,\Sigma =1}^{24}\eta ^{\Lambda
\Sigma }Z_{\Lambda }W_{\Sigma }\right) =-\left( 1-\sum_{\Lambda
,\Sigma =1}^{24}\eta ^{\Lambda \Sigma }R_{\Lambda }T_{\Sigma
}\right) .
\end{equation}%
The equations defining the extremum (minimum) of the scalar potential $%
\mathcal{V}_{DP}$ are then given by the four following systems of eqs:%
\begin{equation}
\begin{tabular}{llll}
$\frac{\delta \mathcal{V}_{DP}}{\delta R^{a}}$ & $=$ & $0$ & , \\
$\frac{\delta \mathcal{V}_{DP}}{\delta R^{I}}$ & $=$ & $0$ & , \\
$\frac{\delta \mathcal{V}_{DP}}{\delta \lambda }$ & $=$ & $0$ & ,%
\end{tabular}
\label{vbh}
\end{equation}%
and%
\begin{equation}
\begin{tabular}{llll}
$\frac{\delta \mathcal{V}_{DP}}{\delta T^{a}}\ $ & $=$ & $0$ & , \\
$\frac{\delta \mathcal{V}_{DP}}{\delta T^{I}}\ $ & $=$ & $0$ & , \\
$\frac{\delta \mathcal{V}_{DP}}{\delta \xi }\ $ & $=$ & $0$ & ,%
\end{tabular}
\label{vbb}
\end{equation}%
as well as%
\begin{equation}
\frac{\delta \mathcal{V}_{DP}}{\delta \zeta }=0  \label{vd}
\end{equation}%
and finally%
\begin{equation}
\frac{\delta \mathcal{V}_{DP}}{\delta \sigma }=0.  \label{vsi}
\end{equation}%
Eqs(\ref{vbh}) give relative extremums (minimums) associated with
the black hole \emph{BH} contribution. \newline Eqs(\ref{vbb})
define relative extremums (minimums) associated with the black
membrane \emph{BM}. \newline Eq(\ref{vd}) captures the duality
relation between the black hole \emph{BH} and the black membrane
\emph{BM}. \newline Eq(\ref{vsi}) is in some sense special; it gives
the values of $\sigma _{0}$ and $\sigma _{2}$ we are after. \newline
Below, we give the details on the solutions of these eqs.

\subsection{Extremums of $\mathcal{V}_{DP}$}

\qquad Here we study the extremums (minimums) of the potential
(\ref{veff}). Since $\mathcal{V}_{DP}$ is multi-variables function,
we shall proceed by steps in order to get these minimums:\newline
(\textbf{1}) First we solve successively the eq(\ref{vbh}),
eq(\ref{vbb}) and (\ref{vd}). These solutions fix the critical
values of the field moduli
and the Lagrange multipliers in terms of the electric and magnetic charges $%
q_{0},$ $g_{0}$, $q_{a},$ $g_{a}$ and $q_{I},$ $g_{I}$;%
\begin{equation}
\begin{tabular}{llll}
$R^{\min }$ & $=$ & $R\left( q,g\right) $ & , \\
$T^{\min }$ & $=$ & $T\left( q,g\right) $ & , \\
$\lambda ^{\min }$ & $=$ & $\lambda \left( q,g\right) $ & , \\
$\xi ^{\min }$ & $=$ & $\xi \left( q,g\right) $ & , \\
$\zeta ^{\min }$ & $=$ & $\zeta \left( q,g\right) $ & .%
\end{tabular}%
\end{equation}%
(\textbf{2}) Then, we substitute the obtained solutions back into eq(\ref%
{veff}) to get the new effective potential $\widetilde{\mathcal{V}}_{\text{%
{\small DP}}}$ namely%
\begin{equation}
\widetilde{\mathcal{V}}_{\text{{\small DP}}}=e^{-2\sigma }\mathcal{V}%
_{0}^{\min }+e^{+2\sigma }\mathcal{V}_{2}^{\min }+\left( \zeta \mathcal{C}%
\right) ^{\min },  \label{bhbm}
\end{equation}%
where now%
\begin{equation}
\begin{tabular}{llll}
$\mathcal{V}_{0}^{\min }$ & $=$ & $\mathcal{V}_{0}\left( R^{\min
},T^{\min
},\lambda ^{\min },\zeta ^{\min }\right) $ & , \\
$\mathcal{V}_{2}^{\min }$ & $=$ & $\mathcal{V}_{2}\left( T^{\min
},R^{\min
},\xi ^{\min },\zeta ^{\min }\right) $ & .%
\end{tabular}%
\end{equation}%
(\textbf{3}) After that, we solve the attractor equation given by
the minimization of (\ref{bhbm}), i.e
\begin{equation}
\frac{\delta \widetilde{\mathcal{V}}_{DP}}{\delta \sigma }=0\text{
}, \label{bht}
\end{equation}%
in order to determine the critical values of $\sigma $ at the
extremums.

\subsubsection{Solving eqs(\protect\ref{vbh}-\protect\ref{vbb}-\protect\ref%
{vd})}

\textbf{A}) \emph{solution of eqs(\ref{vbh})}:\qquad \newline By
substituting eq(\ref{veff}) and (\ref{v02}), we can be put
eqs(\ref{vbh}) in the form
\begin{equation}
\begin{tabular}{llll}
$\left( 1+\lambda \right) R_{a}+\zeta T_{a}$ & $=$ & $0$ & $,$ \\
$\left( 1-\lambda \right) R_{I}-\zeta T_{I}$ & $=$ & $0$ & $,$ \\
$R_{a}R^{a}-R_{I}R^{I}$ & $=$ & $\mathrm{g}^{2}$ & $.$%
\end{tabular}%
\end{equation}%
These equations have three types of solutions which can be
classified
according to whether the sign of $\mathrm{g}^{2}$; that is $\mathrm{g}^{2}=0$%
, $\mathrm{g}^{2}>0$ or $\mathrm{g}^{2}<0$.

\underline{\emph{Case A1}}\textbf{\ (}$\mathrm{g}^{2}=0$\textbf{)}:
\newline
In this case, the solution reads as:%
\begin{equation}
\begin{tabular}{llll}
$R_{a}^{\min }$ & $=$ & $0$ & , \\
$R_{I}^{\min }$ & $=$ & $0$ & , \\
$\lambda ^{\min }$ & $=$ & $-1$ & , \\
$\zeta ^{\min }$ & $=$ & $0$ & ,%
\end{tabular}%
\end{equation}%
and all remaining other moduli are free.

\underline{\emph{Case A2}}\textbf{\ (}$\mathrm{g}^{2}>0$\textbf{)}:
\newline
Here the solution reads as:%
\begin{equation}
\begin{tabular}{llll}
$R_{a}^{\min }$ & $=$ & $\mathrm{g}_{a}\sqrt{\left\vert \mathrm{g}%
^{2}\right\vert }\left( \sum_{b=1}^{4}\mathrm{g}_{b}^{2}\right) ^{-\frac{1}{2%
}}$ & , \\
$\left( R_{a}R^{a}\right) ^{\min }$ & $=$ & $\left\vert \mathrm{g}%
^{2}\right\vert $ & , \\
$R_{I}^{\min }$ & $=$ & $0,$ & , \\
$\lambda ^{\min }$ & $=$ & $-1$ & , \\
$\zeta ^{\min }$ & $=$ & $0$ & ,%
\end{tabular}
\label{ra}
\end{equation}%
and all remaining other moduli are free.

\underline{\emph{Case A3}}\textbf{\ (}$\mathrm{g}^{2}<0$\textbf{)}:
\newline
In this case the solution is given by:%
\begin{equation}
\begin{tabular}{llll}
$R_{I}^{\min }$ & $=$ & $\mathrm{g}_{I}\sqrt{\left\vert \mathrm{g}%
^{2}\right\vert }\left( \sum_{J=1}^{20}g_{J}^{2}\right)
^{-\frac{1}{2}}$ & ,
\\
$\left( R_{I}R^{I}\right) ^{\min }$ & $=$ & $-\mathrm{g}^{2}$ & , \\
$R_{a}^{\min }$ & $=$ & $0$ & , \\
$\lambda ^{\min }$ & $=$ & $+1$ & , \\
$\zeta ^{\min }$ & $=$ & $0$ & ,%
\end{tabular}%
\end{equation}%
and all remaining other moduli are free. \newline
In all cases \textbf{A1}, \textbf{A2} and \textbf{A3} ($\mathrm{g}^{2}=0$, $%
\mathrm{g}^{2}>0$ and $\mathrm{g}^{2}<0$), we have
\begin{equation}
\mathcal{V}_{0}^{\min }=\left( R_{I}R^{I}\right) ^{\min }=\left\vert \mathrm{%
g}^{2}\right\vert .
\end{equation}

\textbf{B)} \emph{Solution of }eqs(\ref{vbb}): \newline Using
eq(\ref{veff}) and (\ref{v02}), we can be put eqs(\ref{vbb}) in the
form
\begin{equation}
\begin{tabular}{llll}
$\left( 1+\xi \right) T_{a}+\zeta R_{a}$ & $=$ & $0$ & , \\
$\left( 1-\xi \right) T_{I}-\zeta R_{I}$ & $=$ & $0$ & , \\
$T_{a}T^{a}-T_{I}T^{I}$ & $=$ & $\mathrm{q}^{2}$ & .%
\end{tabular}%
\end{equation}%
Here also we have three kinds of solutions depending on the signs of $%
\mathrm{q}^{2}$. The solutions are quite similar to the previous
cases; they are given by:

\underline{\emph{Case B1}}\textbf{\ (}$\mathrm{q}^{2}=0$\textbf{)}:
\newline
In this case the solution reads as:%
\begin{equation}
\begin{tabular}{llll}
$T_{a}^{\min }$ & $=$ & $0$ & , \\
$T_{I}^{\min }$ & $=$ & $0$ & , \\
$\xi ^{\min }$ & $=$ & $-1$ & , \\
$\zeta ^{\min }$ & $=$ & $0$ & ,%
\end{tabular}%
\end{equation}%
and all remaining moduli are free.

\underline{\emph{Case B2}}\textbf{\ (}$\mathrm{q}^{2}>0$\textbf{)}:
\newline
Here the solution is given by:%
\begin{equation}
\begin{tabular}{llll}
$T_{a}^{\min }$ & $=$ & $\mathrm{q}_{a}\sqrt{\left\vert \mathrm{q}%
^{2}\right\vert }\left( \sum_{b=1}^{4}q_{b}^{2}\right)
^{-\frac{1}{2}}$ & ,
\\
$\left( T_{a}T^{a}\right) ^{\min }$ & $=$ & $\mathrm{q}^{2}$ & , \\
$T_{I}^{\min }$ & $=$ & $0$ & , \\
$\xi ^{\min }$ & $=$ & $-1$ & , \\
$\zeta ^{\min }$ & $=$ & $0$ & ,%
\end{tabular}
\label{ta}
\end{equation}%
and all remaining other moduli are free.

\underline{\emph{Case B3}}\textbf{\ (}$\mathrm{q}^{2}<0$\textbf{)}:
\newline
In this case the solution reads as:%
\begin{equation}
\begin{tabular}{llll}
$T_{I}^{\min }$ & $=$ & $\mathrm{q}_{I}\sqrt{-\mathrm{q}^{2}}\left(
\sum_{J=1}^{20}q_{J}^{2}\right) ^{-\frac{1}{2}}$ & , \\
$\left( T_{I}T^{I}\right) ^{\min }$ & $=$ & $-\mathrm{q}^{2}$ & , \\
$T_{a}^{\min }$ & $=$ & $0$ & , \\
$\xi ^{\min }$ & $=$ & $+1$ & , \\
$\zeta ^{\min }$ & $=$ & $0$ & ,%
\end{tabular}%
\end{equation}%
and all remaining moduli are free. \newline
In all cases \textbf{B1}, \textbf{B2} and \textbf{B3} ($\mathrm{q}^{2}=0$, $%
\mathrm{q}^{2}>0$ and $\mathrm{q}^{2}<0$), we have
\begin{equation}
\mathcal{V}_{2}^{\min }=\left( T_{a}T^{a}\right) ^{\min }=\left\vert \mathrm{%
q}^{2}\right\vert .
\end{equation}

\textbf{C}. \emph{solution of }eqs(\ref{vd}): \ \newline
Eq(\ref{vd}) gives%
\begin{equation}
T^{a}R_{a}-T^{I}R_{I}=\sum_{a=1}^{4}\left( T_{a}R_{a}\right)
-\sum_{I=1}^{20}\left( T_{I}R_{I}\right) =k,
\end{equation}%
and is solved as:

\underline{\emph{Case\ C1}}\textbf{: }\newline
In this case, the solution corresponds to \textbf{\ }%
\begin{equation}
\begin{tabular}{llll}
$T^{a}R_{a}$ & $=$ & $k$ & , \\
$T^{I}R_{I}$ & $=$ & $0$ & ,%
\end{tabular}%
\end{equation}%
and requires that $T^{a}R_{a}\neq 0$ and $T^{a}\neq 0$. Consistency
with the
solutions of eqs(\ref{vbh}-\ref{vbb}) implies:%
\begin{equation}
\begin{tabular}{llll}
$T_{a}^{\min }$ & $=$ & $kR_{a}^{\min }\left( \sum_{b=1}^{4}\left(
R_{b}R_{b}\right) ^{\min }\right) ^{-1}$ & , \\
$T_{I}^{\min }$ & $=$ & $0$ & , \\
$R_{I}^{\min }$ & $=$ & $0$ & .%
\end{tabular}%
\end{equation}%
Moreover equating the expression of $T_{a}^{\min }$ given by case
\textbf{B2}
with the expression of $T_{a}^{\min }$ which we obtain by substituting $%
R_{a}^{\min }$ by its value given by case \textbf{A2}, we get the
following
identity%
\begin{equation}
\left( \sum_{b=1}^{4}\left( T_{b}^{\min }T_{b}^{\min }\right)
\right) \left( \sum_{b=1}^{4}R_{b}^{\min }R_{b}^{\min }\right)
=k^{2}.
\end{equation}%
Using eq(\ref{ta}) and eq(\ref{ra}), we can obtain the following
electric and magnetic duality relation
\begin{equation}
\mathrm{q}^{2}.\mathrm{g}^{2}=k^{2},\qquad \mathrm{q}^{2}>0,\qquad \mathrm{g}%
^{2}>0.
\end{equation}%
This electric/magnetic duality relation involves the squares of the
vector charges \textrm{q}$_{\Lambda }$ and \textrm{g}$_{\Lambda }$.

\underline{\emph{Case\ C2}}: \newline
In this case, the solution is given by%
\begin{equation}
\begin{tabular}{llll}
$T^{a}R_{a}$ & $=$ & $0$ & , \\
$T^{I}R_{I}$ & $=$ & $-k$ & ,%
\end{tabular}%
\end{equation}%
and corresponds to:%
\begin{equation}
\begin{tabular}{llll}
$T_{I}^{\min }$ & $=$ & $-kR_{I}^{\min }\left( \sum_{J=1}^{20}\left(
R_{J}R_{J}\right) ^{\min }\right) ^{-1}$ & , \\
$T_{a}^{\min }$ & $=$ & $0$ & , \\
$R_{a}^{\min }$ & $=$ & $0$ & .%
\end{tabular}%
\end{equation}%
Substituting the solution of $T_{I}^{\min }$ given by case \textbf{B3} and $%
R_{I}^{\min }$ \ given by case \textbf{A3}, we get the following identity%
\begin{equation}
\left( \sum_{I=1}^{20}\left( T_{I}^{\min }T_{I}\right) ^{\min
}\right) \left( \sum_{I=1}^{20}\left( R_{I}^{\min }R_{I}\right)
^{\min }\right) =k^{2},  \label{qg}
\end{equation}%
which corresponds to the electric/magnetic duality $\mathrm{q}^{2}.\mathrm{g}%
^{2}=k^{2}$; but now with $\mathrm{q}^{2}<0$ and $\mathrm{g}^{2}<0.$

\subsubsection{Solving eq(\protect\ref{bht})}

\qquad Substituting $\mathcal{V}_{0}^{\min }$,
$\mathcal{V}_{2}^{\min }$ and $\left( \zeta \mathcal{C}\right)
^{\min }$ by their expressions, we get the
following effective scalar potential for the dilaton field%
\begin{equation}
\widehat{\mathcal{V}}_{DP}\left( \sigma \right) =e^{-2\sigma
}\left\vert \mathrm{g}^{2}\right\vert +e^{+2\sigma }\left\vert
\mathrm{q}^{2}\right\vert \text{ }.  \label{ve}
\end{equation}%
This is as positive definite effective dyonic potential
\begin{equation}
\widehat{\mathcal{V}}_{DP}\left( \sigma \right) >0\text{ },
\end{equation}%
that depends, in addition to the dilaton $\sigma $, on the semi-norms $%
\mathrm{q}^{2}$ and $\mathrm{g}^{2}$ of the bare electric and
magnetic charges of the dyonic pair \emph{DP}.\newline Since the
second derivative
\begin{equation}
\frac{d^{2}\widehat{\mathcal{V}}_{DP}}{d\sigma ^{2}}\geq 0\text{ },
\end{equation}%
the minimum of eq(\ref{ve}) is obtained by solving
\begin{equation}
\left( e^{+2\sigma }\left\vert \mathrm{q}^{2}\right\vert
-e^{-2\sigma }\left\vert \mathrm{g}^{2}\right\vert \right) =0.
\label{de}
\end{equation}%
The critical value $\sigma _{c}$ of the dilaton solving this
constraint
relation is%
\begin{equation}
\begin{tabular}{llll}
$e^{+2\sigma _{c}}$ & $=$ & $\sqrt{\frac{\left\vert \mathrm{g}%
^{2}\right\vert }{\left\vert \mathrm{q}^{2}\right\vert }}$ & , \\
$\sigma _{c}$ & $=$ & $\frac{1}{4}\left( \ln \left\vert \mathrm{g}%
^{2}\right\vert -\ln \left\vert \mathrm{q}^{2}\right\vert \right) $ & .%
\end{tabular}%
\end{equation}%
Putting this solution back into eq(\ref{de}), we get%
\begin{equation}
\widehat{\mathcal{V}}_{{\small BH-BM}}^{\min }=2\sqrt{\left\vert \mathrm{g}%
^{2}\mathrm{q}^{2}\right\vert }\text{ .}
\end{equation}%
This relation should be compared with eqs(\ref{bfe}-\ref{bff}).

In the end, we would like to add that the analysis given in this
section extends directly to the the dyonic pairs
\begin{equation}
DP\equiv \emph{BH-B3B,}
\end{equation}%
\ and
\begin{equation}
DP\equiv \emph{BS-BM,}
\end{equation}
of (\ref{za}-\ref{az}) of the \emph{7D} $\mathcal{N}=2$ supergravity
embedded in \emph{11D} M-theory on K3.

\section{Conclusion and discussion}

In this paper, we have studied the extremal black brane attractors
in the
\emph{6D} (resp. \emph{7D}) $\mathcal{N}=2$ supergravity limit of the \emph{%
10D} type IIA superstring (resp. \emph{11D} M-theory) on K3. In
these limits, the classical entropy of electrically charge black
branes \emph{EBB} (resp. magnetically charged branes \emph{MBB})
have \emph{degenerate} values; see
eqs(\ref{se},\ref{sm},\ref{ebd},\ref{ebh}).\newline In trying to
understand this classical degeneracy, we have been lead to make a
proposal where the degenerate value of the 6D black hole \emph{BH}
entropy
\begin{equation}
\begin{tabular}{llll}
$\mathcal{S}_{BH}^{entropy}$ & $=$ & $0$ & ,%
\end{tabular}
\label{deh}
\end{equation}%
and the entropy of black membrane \emph{BM}
\begin{equation}
\begin{tabular}{llll}
$\mathcal{S}_{BM}^{entropy}$ & $=$ & $0$ & ,%
\end{tabular}
\label{deb}
\end{equation}%
appear as two singular limits of the classical entropy
\begin{equation}
\mathcal{S}_{BH-BM}^{entropy}\equiv \mathcal{S}_{DP}^{entropy}
\end{equation}%
of the bound state dual pair $\left( \emph{BH-BM}\right) _{6D}\equiv
DP$. This result applies as well for the \emph{6D} dyonic black
string $\left(
\emph{BS}\right) _{6D}$ and for the dual pair attractors $\left( \emph{BH-B3B%
}\right) _{7D}$ and $\left( \emph{BS-BM}\right) _{7D}$ of the
$\mathcal{N}=2$ \emph{7D} supergravity theory.\newline
In analyzing the degeneracy of $\mathcal{S}_{EBB}^{entropy}=\mathcal{S}%
_{MBB}^{entropy}=0$, we have also found that electric/magnetic
duality is a \emph{universal symmetry} playing a central in the
physics of \emph{6D/7D} black attractors. Among our results, we
mention the following:\newline First, by using the electric/magnetic
duality (\emph{e/m} symmetry for short), we have given a refined
classification of the black attractors in \emph{6D} and \emph{7D}.
These black branes are classified into two representations of the
\emph{e/m} symmetry: dyonic singlets and pairs as follows:

\textbf{(1) }\emph{Six dimensions}\newline In \emph{6D} non chiral
supergravity theory with sixteen supercharges, we have:\newline
(\textbf{a}) \emph{An attractor singlet}, corresponding to the
dyonic black string denoted as $\left( \emph{BS}\right) $. This
dyonic attractor carries an electric charge $q_{0}=\left(
\int_{S^{3}}^{\ast }\mathcal{F}_{3}\right) $ and a magnetic charge
$g_{0}=\left( \int_{S^{3}}\mathcal{F}_{3}\right) $
with $\mathcal{F}_{3}$ being the field strength of the {\small NS-NS} $%
\mathcal{B}_{\mu \nu }$- field.\newline (\textbf{b}) \emph{An
attractor pair }describing the dual pair
\begin{equation}
\begin{tabular}{llllll}
$DP$ & $\equiv $ & $\emph{BH-BM}$ & $\equiv $ & $\left(
\begin{array}{c}
\emph{BH} \\
\emph{BM}%
\end{array}%
\right) $ & ,%
\end{tabular}%
\end{equation}%
carrying \emph{24} electric and \emph{24} magnetic charges $\left\{
q_{\Lambda },g_{\Lambda }\right\} $.\newline The black hole \emph{BH
}carry \emph{24} magnetic charge $g^{\Lambda }=\left(
\int_{S^{4}}\mathcal{F}_{4}^{\Lambda }\right) $ and corresponds to
the singular limit
\begin{equation}
\begin{tabular}{llll}
$\sigma $ & $\rightarrow $ & $+\infty $ & , \\
$e^{-\sigma }$ & $\rightarrow $ & $0$ & ,%
\end{tabular}%
\end{equation}%
of the $SO\left( 1,1\right) $ factor of the moduli space\textrm{\
}$SO\left( 1,1\right) \times \frac{SO\left( 4,20\right) }{SO\left(
4\right) \times
SO\left( 20\right) }$. This singular limit may be formally stated as,%
\begin{equation}
\begin{tabular}{llll}
$\emph{DP}$ & $\underrightarrow{\sigma \rightarrow +\infty }$ &
$\left(
\begin{array}{c}
\emph{BH} \\
0%
\end{array}%
\right) $ & $.$%
\end{tabular}
\label{sbh}
\end{equation}%
The same feature is valid for the electrically charged black membrane \emph{%
BM }carrying \emph{24} electric charges $\left\{ q_{\Lambda
}\right\} $. The \emph{BM}, which is e/m dual to \emph{BH},
corresponds to the singular limit
\begin{equation}
\begin{tabular}{llll}
$\sigma $ & $\rightarrow $ & $-\infty $ & , \\
$e^{+\sigma }$ & $\rightarrow $ & $0$ & ,%
\end{tabular}%
\end{equation}%
in the moduli space. We also have%
\begin{equation}
\begin{tabular}{llll}
$\emph{DP}$ & $\underrightarrow{\sigma \rightarrow -\infty }$ &
$\left(
\begin{array}{c}
0 \\
\emph{BM}%
\end{array}%
\right) $ & $.$%
\end{tabular}
\label{mb}
\end{equation}%
\textbf{(2)} \emph{Seven dimensions}\newline In the 7D
$\mathcal{N}=2$ supergravity theory we have no attractor singlet;
but two pairs $\left( DP\right) _{1}$ and $\left( DP\right)
_{2}$:\newline The first pair is given by the bound state\emph{\
BH-B3B} carrying \emph{22} electric and \emph{22} magnetic charge
$\left\{ q_{\Lambda },g_{\Lambda }\right\} $.
\begin{equation}
\begin{tabular}{llllll}
$\left( DP\right) _{1}$ & $\equiv $ & \emph{BH-B3B} & $\equiv $ &
$\left(
\begin{array}{c}
\emph{BH} \\
\emph{B3B}%
\end{array}%
\right) $ & ,%
\end{tabular}
\label{dp1}
\end{equation}%
The second attractor pair is given by the dual pair%
\begin{equation}
\begin{tabular}{llllll}
$\left( DP\right) _{2}$ & $\equiv $ & \emph{BS-BM} & $\equiv $ &
$\left(
\begin{array}{c}
\emph{BS} \\
\emph{BM}%
\end{array}%
\right) $ & .%
\end{tabular}
\label{dp2}
\end{equation}%
This pair carries an electric charge $q_{0}$ and a magnetic one
$g_{0}$; it
should compared with the 6D black string $\left( \emph{BS}\right) _{6D}$.%
\newline
Notice that the black hole $\left( \emph{BH}\right) _{7D}$\emph{\ }(resp.%
\emph{\ B3B})\emph{\ }with the\emph{\ 22} magnetic charges
$g_{\Lambda }$ (resp. \emph{22} electric charges $q_{\Lambda }$)
follows as the singular limit $\sigma \rightarrow +\infty $ (resp.
$\sigma \rightarrow +\infty $) of $\left( DP\right) _{1}$.\newline
The same property holds for the \emph{7D} black string $\left( \emph{BS}%
\right) _{7D}$ and the black membrane $\left( \emph{BM}\right)
_{7D}$. They are singular limits of the $\left( DP\right) _{2}$
pair.

\ \newline
Then we have considered the question of computing the entropies $\mathcal{S}%
_{black-brane}$ of the above \emph{6D} and \emph{7D} black
attractors.
\newline
For the dyonic \emph{6D} black string $\left( \emph{BS}\right)
_{6D}$ with
an electric charge $q_{0}$ and a magnetic charge $g_{0}$, the entropy $%
\mathcal{S}_{BS}^{6D}$ is given by eqs(\ref{bfe}-\ref{bff}) namely,%
\begin{equation}
\mathcal{S}_{BS}^{6D}=\frac{1}{2}\mathrm{g}_{0}\mathrm{q}_{0}>0,
\end{equation}%
which, for later use, we prefer to rewrite as follows
\begin{equation}
\mathcal{S}_{BS}^{6D}=\frac{1}{2}\sqrt{\mathrm{g}_{0}^{2}\mathrm{q}_{0}^{2}}%
\text{ }>0.  \label{sbs}
\end{equation}%
Clearly $\mathcal{S}_{BS}^{6D}$ is invariant under e/m
symmetry.\newline For the case of \emph{6D}\ black hole $\left(
\emph{BH}\right) _{6D}$ and the \emph{6D} black membrane $\left(
\emph{BM}\right) _{6D}$, the corresponding entropies
$\mathcal{S}_{BH}^{6D}$ and $\mathcal{S}_{BM}^{6D}$ take degenerate
values as in eqs(\ref{deh}-\ref{deb}). \newline Recall that this
property of the classical entropy has been point out in
literature many years ago \textrm{\cite{B3};} see also\textrm{\ \cite%
{BDSS,S2}}. It is due to the specific structure of the scalar manifolds $%
\boldsymbol{M}_{6D}^{N=2}$ and $\boldsymbol{M}_{7D}^{N=2}$ of these
the \emph{6D} and \emph{7D} theories which contain an ambiguous
$SO\left(
1,1\right) $ factor as shown below,%
\begin{equation}
\begin{tabular}{llll}
$\boldsymbol{M}_{6D}^{N=2}$ & $=$ & $SO\left( 1,1\right) \times \frac{%
SO\left( 4,20\right) }{SO\left( 4\right) \times SO\left( 20\right) }$ & , \\
$\boldsymbol{M}_{7D}^{N=2}$ & $=$ & $SO\left( 1,1\right) \times \frac{%
SO\left( 3,19\right) }{SO\left( 3\right) \times SO\left( 19\right) }$ & .%
\end{tabular}
\label{msev}
\end{equation}%
The $SO\left( 1,1\right) $ factor, which is associated with the
dilaton, puts a very restrictive constraint on the critical value of
the effective scalar potential and on the entropy.\newline Moreover,
by freezing the dilaton to a some constant value; say $\sigma
=\sigma _{BH}$ for the BH and $\sigma =\sigma _{BM}$ for the black
membrane, the corresponding entropies are no longer zero; but they
depend on these free constant parameters.\newline To overcome this
difficulty, we have proposed that, classically, the black hole
\emph{BH} and the black membrane \emph{BM} of the \emph{6D} space
time
should be thought of as an attractor bound state with the singular limits (%
\ref{sbh},\ref{mb}). In this view, all the difficulties are overcome
and \emph{e/m} duality appears as a universal symmetry.

\emph{Entropy of dual pair DP }\newline With the attractor bound
state picture in mind, we have studied the
attractor mechanism of the \emph{6D} dual pair \emph{DP }$\equiv $ $\emph{%
BH-BM}$ and we have found, amongst others, the following:\newline
(\textbf{i}) the values $\sigma _{BH}$ and $\sigma _{BM}$ of the
dilaton at the horizons of $\left( \emph{BH}\right) _{6D}$ and
$\left( \emph{BM}\right) _{7D}$\emph{\ }are as follows
\begin{equation}
\begin{tabular}{llll}
$\sigma _{BH}$ & $=$ & $-\sigma _{BM}$ & ,%
\end{tabular}%
\end{equation}%
in agreement with e/m duality. Moreover we have been able to compute
$\sigma
_{BH}$ which is given by%
\begin{equation}
\begin{tabular}{llll}
$\sigma _{BH}$ & $=$ & $\frac{1}{4}\left( \ln \mathrm{g}^{2}\right) -\frac{1%
}{4}\left( \ln \mathrm{q}^{2}\right) $ & ,%
\end{tabular}%
\end{equation}%
with%
\begin{equation}
\begin{tabular}{llll}
$\mathrm{g}^{2}$ & $=$ & $\mathrm{g}^{\Lambda }\eta _{\Lambda \Sigma }%
\mathrm{g}^{\Sigma }$ & , \\
$\mathrm{q}^{2}$ & $=$ & $\mathrm{q}^{\Lambda }\eta _{\Lambda \Sigma }%
\mathrm{q}^{\Sigma }$ & ,%
\end{tabular}
\label{sbn}
\end{equation}%
where $\eta _{\Lambda \Sigma }$ is the metric of the tangent space $\mathbb{R%
}^{4,20}$. \newline (\textbf{ii}) the entropy of the six space-time
dimension \emph{DP} dual
pair is given by%
\begin{equation}
\mathcal{S}_{\emph{DP}}=\frac{1}{2}\sqrt{\left\vert \mathrm{g}^{2}\mathrm{q}%
^{2}\right\vert }\text{ ,}  \label{sbm}
\end{equation}%
Notice that the relation (\ref{sbm}) of the entropy
$\mathcal{S}_{\emph{DP}}$
is quite similar to the relation (\ref{sbs}) giving the entropy of the \emph{%
6D} black string.\newline At the end, we would like to add the two
following: \newline First, the explicit analysis we have made for
\emph{6D} applies as well for the black pairs (\ref{dp1}-\ref{dp2})
in \emph{7D} $\mathcal{N}=2$ supergravity embedded in 11D M-theory
on K3. The entropy of the attractor bounds $\left( DP\right) _{1}$
and $\left( DP\right) _{1}$ have similar expression as in
eqs(\ref{sbm}-\ref{sbn}) with $\eta _{\Lambda \Sigma }$ being the
metric of the flat space $\mathbb{R}^{3,19}$. \newline Second, the
Lagrange multiplier method we have developed in section 5 seems to
be the appropriate way to deal with the study of the critical points
of the black branes effective potentials.

 \acknowledgments
\qquad\ \ \newline EHS would like to thank M. Assorey, L. Boya, L.
R. Cortes for scientific discussions and Departemento di Fisica
Teorica, Universidad Zaragoza for kind hospitality. He also thanks
A. Belhaj, L.B\ Drissi, H. Jehjouh for collaboration in this
direction. AS has been supported by MCYT ( Spain) under grant FPA
2003-02948, CICYT (grant FPA-2006-02315) and DGIID-DGA (grant
2007-E24/2), Spain. This research work is supported by " Fisica de
altas energias: Particulas, Cuerdas y Cosmologia, A 9335/07" and the
program Protars D12/25/CNRST.

\section{Appendix: On effective potential in 6D and 7D}

We begin by recalling that, with the exception of $D=4$,
$\mathcal{N}=1,$ $2$ and $D=5,$ $\mathcal{N}=2$, all supergravity
theories contain scalar fields
whose kinetic Lagrangian is described by $\sigma $- models of the form $G/H$%
. The symmetry group $G$ is a non compact group acting as an
isometry group
on the scalar manifold and $H$ is the isotropy subgroup having the form $%
H=H_{\text{aut}}\otimes H_{\text{matter}}$. The subsymmetry
$H_{\text{aut}}$
is the automorphism group of the extended supersymmetric algebra and $H_{%
\text{matter}}$ is related to the matter supermultiplets. For the
list of the coset manifolds $G/H$ and the automorphism groups of the
various supergravity theories for any dimension D and number
$\mathcal{N}$, see
\textrm{\cite{salam,cast}}. For $D=6$, $\mathcal{N}=2$ and $D=7$, $\mathcal{N%
}=2$, these are given by eqs(\ref{0}) and (\ref{01}).\newline We aso
recall that generic $D$ supergravity theories with moduli space
$G/H$ have several specific properties shared by most of these
theories. Amongst these features, we quote the three
following:\newline (\textbf{1}) the group $G$ acts linearly on the
$\left( p+2\right) $- forms field strengths
$\mathcal{F}_{a_{1\text{\textperiodcentered \textperiodcentered
\textperiodcentered }}a_{p+2}}$ corresponding to the
various $\left( p+1\right) $- forms $\mathcal{A}_{a_{1\text{%
\textperiodcentered \textperiodcentered \textperiodcentered
}}a_{p+1}}$ appearing in the gravitational and matter multiplets.
\newline (\textbf{2}) the properties of a given supergravity theory
with fixed $D$ and $\mathcal{N}$ are completely specified by the
geometry of $G/H$, in particular in terms of the coset
representatives $L=L_{\Lambda \Sigma }$
satisfying the gauge symmetry relation%
\begin{equation*}
\begin{tabular}{llll}
$L\left( \xi ^{\prime }\right) =gL\left( \xi \right) h\left( g,\xi
\right) $ & , & $g\in G,$ $\ h\in H,$ $\ \xi ^{\prime }=\xi ^{\prime
}\left( \xi
\right) $ & ,%
\end{tabular}%
\end{equation*}%
with $\xi $ being the coordinates of the coset G/H. In particular,
the matrix $\mathcal{N}_{\Lambda \Sigma }$ capturing the field
coupling metric
of the $\left( p+2\right) $- forms $\mathcal{F}_{a_{1\text{%
\textperiodcentered \textperiodcentered \textperiodcentered }%
}a_{p+2}}^{\Lambda }$ in the supergravity Lagrangian density is
fixed in
terms of $L$. The \emph{physical} field strengths $\mathcal{T}_{a_{1\text{%
\textperiodcentered \textperiodcentered \textperiodcentered }%
}a_{p+2}}^{\Lambda }$ of the \emph{interacting} theories are also \emph{%
dressed}\ with scalar fields as explicitly developed in the
literature; especially in a series of papers by Ferrara and
collaborators; see for
instance sections 3 and 5 of the study \textrm{\cite{B3} and \cite{10}-\cite%
{12} }for a geometrical approach dealing with the so called $\widehat{%
\mathrm{F}}_{4}$ supergravity containing the \emph{6D}$\
\mathcal{N}=2$
superalgebra as a subsymmetry; see also the appendix B of \textrm{\cite{S1}}%
. \newline (\textbf{3}) Like in 4D $\mathcal{N}=2$ theory, higher
dimensional supergravity exhibits as well two kinds of central
charges: $Z_{geo}$ coming from gravity multiplet (geometry) and
$Z_{matter}$ arising from the matter
sector. The dressing property allows to write down the central charges $%
Z_{geo}\equiv Z_{a_{1}\text{\textperiodcentered \textperiodcentered
\textperiodcentered }a_{p}}$ associated to the $\left( p+1\right) $- forms $%
\mathcal{A}_{a_{1\text{\textperiodcentered \textperiodcentered
\textperiodcentered }}a_{p+1}}^{gravity}$ in \emph{the gravitational
multiplet} in terms of the geometrical structure of the moduli
space. The
matter $\left( p+1\right) $- forms $\mathcal{A}_{a_{1\text{%
\textperiodcentered \textperiodcentered \textperiodcentered }%
}a_{p+1}}^{matter}$ of the matter multiplets give rise to charges
that are closely related to the central charges.\newline Notice in
passing that when $p>1$, the central charges do not appear in the
usual supersymmetry algebra, but in its extended version containing
the central generators $Z_{a_{1}\text{\textperiodcentered
\textperiodcentered \textperiodcentered }a_{p}}$ associated to p-
dimensional extended objects. Notice also that besides the fact that
they satisfy differential relations of Maurer- Cartan type, the
central charges $Z$ satisfy as well sum rules quite analogous to
those for the $\mathcal{N}=2$ special geometry case
\textrm{\cite{cer}}. These sum rules, which define in particular the
effective potential,
\begin{equation}
\mathcal{V}_{eff}\sim \left\vert Z_{geo}\right\vert ^{2}+\left\vert
Z_{matter}\right\vert ^{2}.  \label{wei}
\end{equation}%
have been analyzed in details in \textrm{\cite{B3}}. Our main goal
below is
to write down the explicit form of the dressed charges $Z_{geo}$, $%
Z_{matter} $ in the 6D/7D supergravity cases and then $\mathcal{V}%
_{eff}^{6D/7D}$. We also give some useful relations between $Z_{geo}$ and $%
Z_{matter}$ which are analogous the familiar $D=4$, $\mathcal{N}=2$
supergravity using special geometry relations
\textrm{\cite{cer}}.\newline For concreetness, we shall first focus
on $\mathcal{N}=2$ supergravity in 6D and then move to 7D. These
theories have respectively \emph{81} (\emph{58}) scalars distributed
as follows: \textbf{(i)} the dilaton $\sigma $ belonging to the 6D
(7D) $\mathcal{N}=2$ gravity multiplet. \textbf{(ii)} the eighty
(fifty seven) other moduli $\phi _{aI}$ ($\rho _{aI}$) belonging to
the 6D (7D) $\mathcal{N}=2$ Maxwell multiplets.

\subsection{6D $\mathcal{N}=2$ supergravity}

The effective scalar potential $\mathcal{V}_{eff}$ of the 6D black
objects is given by the so called Weinhold potential (\ref{wei})
expressed as
quadratics of the \emph{dressed} charges \textrm{\cite{F0,F1,F2,F3}},%
\begin{equation}
\left( Z_{+},Z_{-},Z_{a},Z_{I}\right) ,\qquad a=1,...,4,\qquad
I=1,...,20. \label{cec}
\end{equation}%
These charges appear in the supersymmetric transformations of the
(fermionic) fields of the 6D supergravity theory.\newline
At the event horizon of the 6D black objects, the potential $\mathcal{V}%
_{eff}$ attains the minimum. The real $\left( \sigma ,\phi
_{aI}\right) $ moduli parameterizing $\frac{SO\left( 1,1\right)
\times SO\left( 4,20\right) }{SO\left( 4\right) \times SO\left(
20\right) }$ are generally fixed by the charges
\begin{equation}
g^{+},\text{ \ }g^{-},\text{ \ }g^{a},\text{ \ }h^{I},\text{ \
}q_{a},\text{ \ }p_{I},  \label{cah}
\end{equation}%
of the $\mathcal{N}=2$ 6D supergravity gauge field strengths%
\begin{equation*}
H_{3}^{+}\text{ \ },\text{ \ }H_{3}^{-}\text{ \ },\text{ \
}F_{2}^{a}\text{
\ },\text{ \ }F_{2}^{I}\text{ \ },\text{ \ }F_{4a}\text{ \ },\text{ \ }%
F_{4I}.
\end{equation*}%
The attractor equations of the 6D $\mathcal{N}=2$ black objects are
obtained
from the minimization of the $\left( \mathcal{V}_{eff}\right) _{\text{black}%
} $. Notice that from the field spectrum of the 6D $\mathcal{N}=2$
non chiral supergravity, one learns that two basic situations should
be distinguished:\newline
(\textbf{1}) 6D black string (\emph{BS}) with near horizon geometry $%
AdS_{3}\times S^{3}$. This is a 6D dyonic black F- string solution.
The electric/magnetic charges involved here are those of the gauge
invariant 3-
form field strengths%
\begin{equation*}
H_{3}^{+}=\frac{1}{2}\left( H_{3}+\text{ }^{\star }H_{3}\right)
\quad ,\quad H_{3}^{-}=\frac{1}{2}\left( H_{3}-\text{ }^{\star
}H_{3}\right) ,
\end{equation*}%
associated with the usual 2- form antisymmetric $B_{\mu \nu }^{\pm
}$ fields in 6D. The $\star $ conjugation stands for the usual
Poincar\'{e} duality interchanging $n$- forms with $\left(
6-n\right) $ ones. \newline (\textbf{2}) 6D black hole (\emph{BH})
and its black 2- brane (\emph{BM}) dual. The field strengths
involved in these objects are related by the Poincar\'{e} duality in
6D space time which interchanges the 2- and 4- form field
strengths.\newline Below, we study briefly and separately these two
configurations.

\emph{Black string in 6D}\newline The BPS black object of the 6D
$\mathcal{N}=2$ non chiral theory is a dyonic
string charged under both the self dual $H_{3}^{+}$ and anti-self dual $%
H_{3}^{-}$ field strengths of the NS-NS B$^{\pm }$-fields. Using the
following bare magnetic/electric charges,%
\begin{equation*}
g^{\pm }=\int_{S^{3}}H_{3}^{\pm },\qquad g^{\pm }=\frac{1}{2}\left(
g\pm e\right) ,
\end{equation*}%
where $g=\int_{S^{3}}H_{3}$ and $e=\int_{S^{3}}$ $^{\star }H_{3}$,
one can write down the physical charges in terms of the dressed
charges.

(a) \textit{Dressed charges}\newline The dressed charges play an
important role in the study of supergravity theories
\textrm{\cite{B3}}. They appear in the supersymmetric
transformations of the Fermi fields (here gravitinos), and generally
read
like%
\begin{equation}
Z^{\pm }=X_{+}^{\pm }g^{+}+X_{-}^{\pm }g^{-},  \label{cz}
\end{equation}%
where the real $2\times 2$ matrix
\begin{equation*}
X=\left(
\begin{array}{cc}
X_{+}^{+} & X_{-}^{+} \\
X_{-}^{+} & X_{+}^{+}%
\end{array}%
\right) ,
\end{equation*}%
parameterizes the $SO\left( 1,1\right) $ factor of the moduli space $%
\widehat{G}$. Taking the $\eta _{rs}$ flat metric as $\eta
=diag\left( 1,-1\right) $, we can express all the four real
parameters $X_{-}^{\pm }$ and $X_{+}^{\pm }$ in terms of the dilaton
$\sigma =\sigma \left( x\right) $ by solving the constraint eqs
$X^{t}\eta X=\eta $ which split into four
constraint relations like%
\begin{equation}
\begin{tabular}{llll}
$X_{+}^{+}X_{+}^{+}-X_{+}^{-}X_{+}^{-}=1$ & $,$ & $%
X_{-}^{-}X_{-}^{-}-X_{-}^{+}X_{-}^{+}=1$ & $,$ \\
$X_{+}^{+}X_{-}^{+}-X_{+}^{-}X_{-}^{-}=0$ & , & $%
X_{-}^{+}X_{+}^{+}-X_{-}^{-}X_{+}^{-}=0$ & .%
\end{tabular}
\label{xeq}
\end{equation}%
These eqs can be solved by,
\begin{equation}
X_{+}^{+}=X_{-}^{-}=\cosh \left( 2\sigma \right) ,\qquad
X_{+}^{-}=X_{-}^{+}=\sinh \left( 2\sigma \right) .  \label{xx}
\end{equation}%
Putting these solutions back into the expressions of the central charges $%
Z^{+}$ and $Z^{-}$ (\ref{cz}), we get the following dilaton
dependent
quantities%
\begin{equation}
Z^{\pm }=\frac{1}{2}\left[ g\exp \left( -2\sigma \right) \pm e\exp
\left( 2\sigma \right) \right] .  \label{zz}
\end{equation}%
Notice that these dressed charges have no dependence on the $\omega
_{aI}$ field moduli of the coset $SO\left( 4,20\right) /SO\left(
4\right) \times SO\left( 20\right) $. This is because the NS-NS B-
fields is not charged under the isotropy group of the above coset
manifold.

(b) \textit{Black string potential}\newline With the dressed charges
$Z^{+}$ and $Z^{-}$, we can write down the gauge invariant effective
scalar potential $\mathcal{V}_{BFS}$. It is given by the
so called Weinhold potential,%
\begin{equation}
\mathcal{V}_{BS}=\left( Z^{+}\right) ^{2}+\left( Z^{-}\right) ^{2}.
\label{ana}
\end{equation}%
Notice that, as far symmetries are concerned, one also have the other \emph{%
"orthogonal"} combination namely $\left( Z^{+}\right) ^{2}-\left(
Z^{-}\right) ^{2}$. This combination corresponds just to the
electric/magnetic charge quantization condition. By substituting eq(\ref{cz}%
) into the relation (\ref{ana}), we get the following form of the potential,%
\begin{equation*}
\mathcal{V}_{BS}=\left( g^{+},g^{-}\right) \mathcal{M}\left(
\begin{array}{c}
g^{+} \\
g^{-}%
\end{array}%
\right) ,
\end{equation*}%
with%
\begin{equation*}
\mathcal{M}=\left(
\begin{array}{cc}
\left( X_{+}^{+}\right) ^{2}+\left( X_{+}^{-}\right) ^{2} &
2X_{+}^{+}X_{-}^{+} \\
2X_{+}^{-}X_{-}^{-} & \left( X_{-}^{-}\right) ^{2}+\left(
X_{-}^{+}\right)
^{2}%
\end{array}%
\right) .
\end{equation*}%
From this matrix and using the transformations given in
\textrm{\cite{F1}},
we can read the gauge field coupling metric $\mathcal{N}_{+-}$ and $\mathcal{%
N}_{-+}$ that appear in the 6D $\mathcal{N}=2$ supergravity
component field
Lagrangian density%
\begin{equation*}
\frac{\mathcal{L}_{6D}^{N=2\text{ sugra}}}{\sqrt{-g}}=\mathcal{R}%
_{6}1+\left( \frac{1}{2}\mathcal{N}_{+-}H^{+}\wedge H^{-}+\frac{1}{2}%
\mathcal{N}_{-+}H^{-}\wedge H^{+}\right) +\cdots
\end{equation*}%
In this eq, $\mathcal{R}_{6}$ is the usual 6D scalar curvature and
$g=\det \left( g_{\mu \nu }\right) $. By further using (\ref{zz}),
we can put the
potential $\mathcal{V}_{BFS}$ into the following form%
\begin{equation}
\mathcal{V}_{BS}\left( \sigma \right) =\frac{g^{2}}{2}\exp \left(
-4\sigma \right) +\frac{e^{2}}{2}\exp \left( 4\sigma \right) .
\label{bfs}
\end{equation}%
Notice that the self and anti- self duality properties of the field
strengths H$_{3}^{+}$ and H$_{3}^{-}$ imply that the corresponding
magnetic/electric charges are related as $g^{+}=e^{+},$
$g^{-}=-e^{-}$.
Using the quantization condition for the dyonic 6D black F- string namely $%
\left( e^{+}g^{+}+g^{-}e^{-}\right) =2\pi k$, $k$ integer, one gets,%
\begin{equation}
\left( g^{+}g^{+}-g^{-}g^{-}\right) =eg=2\pi k.  \label{qn}
\end{equation}%
Then the the quantity $\left( Z^{+}\right) ^{2}-\left( Z^{-}\right)
^{2}$ becomes $\left( Z^{+}\right) ^{2}-\left( Z^{-}\right)
^{2}=2eg$, being just the quantization condition of the
electric/magentic charges of the F- string in 6D space time.

\textbf{(2)} \emph{6D Black Hole}\newline Contrary to the dyonic BS,
the 6D black hole is magnetically charged under the $U^{4}\left(
1\right) \times U^{20}\left( 1\right) $ gauge group symmetry
generated by the gauge transformations of the $\left( 4+20\right) $
gauge fields of the 6D $\mathcal{N}=2$ gravity fields spectrum.
Recall that
in 6D, the electric charges are given, in terms of the field strength $%
F_{4a} $ and $F_{4I}$, by,%
\begin{equation*}
\begin{tabular}{llll}
$q_{a}=\int_{S^{4}}F_{4a}$ & $,$ & $p_{I}=\int_{S^{4}}F_{4I}$ & $.$%
\end{tabular}%
\end{equation*}%
with $a=1,...,4$ and .$I=1,...,20$ The corresponding magnetic duals,
which
concern the black 2- brane, involve the 2- form field strengths $%
F_{2}^{\Lambda }$ integrated over 2- sphere,
\begin{equation*}
\begin{tabular}{llll}
$g^{a}=\int_{S^{2}}F_{2}^{a}$ & , & $h^{I}=\int_{S^{2}}F_{2}^{I}$ & .%
\end{tabular}%
\end{equation*}%
Like for black string, the charges $Q_{\Lambda }=\left(
q_{a},p_{I}\right) $ are not the physical ones. The physical
charges; to be denoted like $Z_{a},$ $Z_{I},$ appear dressed by the
6D scalar fields parameterizing the moduli space of the 10D type IIA
superstring on K3. Recall that the charges $Z_{a}$ and $Z_{I}$
appear respectively in the supersymmetric transformations of the
\emph{four} gravi-photinos/dilatinos and the \emph{twenty} photinos of the U$%
^{20}\left( 1\right) $ Maxwell multiplet of the gauge-matter sector.

(a) \emph{Dressed charges}\newline The dressing of the \emph{twenty
four} electric charges $\left(
q^{a},p^{I}\right) $ of the gauge fields $\left( \mathcal{A}_{\mu }^{a},%
\mathcal{A}_{\mu }^{I}\right) $ read as follows:%
\begin{eqnarray}
Z_{a} &=&e^{-\sigma }\left( Y_{ab}q^{b}+\phi _{aJ}p^{J}\right) ,  \notag \\
Z_{I} &=&e^{-\sigma }\left( V_{Ib}q^{b}+Y_{IJ}p^{J}\right) .
\label{czz}
\end{eqnarray}%
Using the real $24\times 24$ matrix $M_{\Lambda \Sigma }=e^{-\sigma
}\times L_{\Lambda \Sigma }$,
\begin{equation*}
L_{\Lambda \Sigma }=\left(
\begin{array}{cc}
Y_{ab} & \phi _{aJ} \\
V_{Ia} & Y_{IJ}%
\end{array}%
\right) ,
\end{equation*}%
that defines the moduli space $\widehat{G}$, the dressed charges
$Z_{\Lambda
}=\left( Z_{a},Z_{I}\right) $ can be put in the condensed form%
\begin{eqnarray}
Z_{a} &=&M_{a\Sigma }Q^{\Sigma }=e^{-\sigma }L_{a\Sigma }Q^{\Sigma
},  \notag
\\
Z_{I} &=&M_{I\Sigma }Q^{\Sigma }=e^{-\sigma }L_{I\Sigma }Q^{\Sigma
}. \label{dzz}
\end{eqnarray}%
Obviously not all the parameters carried by $L_{\Lambda \Sigma }$\
are independent; the extra dependent degrees of freedom are fixed by
imposing the $SO\left( 4,20\right) $ orthogonality constraint eqs
and requiring gauge
invariance under $SO\left( 4\right) \times SO\left( 20\right) $. The factor $%
e^{-\sigma }$ of eq(\ref{czz}) is then associated with the non
compact abelian factor $SO\left( 1,1\right) $ considered
previously.\newline
Taking the $\eta _{\Lambda \Sigma }$ flat metric of the non compact group $%
SO\left( 4,20\right) $ as $\eta _{\Lambda \Sigma }=diag\left(
4\left( +\right) ,20\left( -\right) \right) $, we can express all
the $24\times 24=576$ real parameters $L_{\Lambda \Sigma }$ in terms
of eighty of them
only; that is in terms of $\phi _{aJ}$. Notice moreover that setting,%
\begin{eqnarray}
Z_{a} &=&e^{-\sigma }R_{a},\qquad R_{a}=\left( L_{a\Sigma }Q^{\Sigma
}\right) ,  \notag \\
Z_{I} &=&e^{-\sigma }R_{I},\qquad R_{I}=\left( L_{I\Sigma }Q^{\Sigma
}\right) ,  \label{lz}
\end{eqnarray}%
as well as $L_{\Sigma }^{\Upsilon }\cdot E_{\digamma }^{\Sigma
}=\left( L_{a}^{\Upsilon }E_{\digamma }^{a}-L_{I}^{\Upsilon
}E_{\digamma }^{I}\right) =\delta _{\digamma }^{\Upsilon }$, one can
compute a set of useful
relations. In particular we have%
\begin{eqnarray}
dL_{\digamma \Lambda } &=&L_{\Upsilon \Lambda }\cdot \left(
dL_{\Sigma
}^{\Upsilon }\right) \cdot P_{\digamma }^{\Sigma },  \notag \\
\nabla Z_{a} &=&\left( D^{H_{1}}Z_{a}+Z_{a}d\sigma \right) ,  \label{dz} \\
\nabla Z_{I} &=&\left( D^{H_{2}}Z_{I}+Z_{I}d\sigma \right) ,  \notag
\end{eqnarray}%
where%
\begin{eqnarray}
D^{H_{1}}Z_{a} &=&\left( dZ_{a}-\Omega _{a}^{b}Z_{b}\right) ,\qquad H_{1}=%
\mathcal{O}\left( 4\right) ,  \notag \\
D^{H_{2}}Z_{I} &=&\left( dZ_{I}-\Omega _{I}^{J}Z_{J}\right) ,\qquad H_{2}=%
\mathcal{O}\left( 4\right) ,  \label{bz}
\end{eqnarray}%
and where $\Omega _{a}^{b}$ and $P_{a}^{I}$\ are given by%
\begin{equation*}
\Omega _{a}^{b}=E_{a}^{\Sigma }\cdot \left( dL_{\Sigma }^{b}\right)
,\qquad P_{a}^{I}=E_{a}^{\Sigma }\cdot \left( dL_{\Sigma
}^{I}\right) ,
\end{equation*}%
together with similar relation for $\Omega _{I}^{J}$ and $P_{I}^{a}$. Using (%
\ref{dz}), we can write down the Maurer-Cartan eqs for the dressed
charge.
They read as follows,%
\begin{equation}
\nabla Z_{a}=P_{a}^{I}Z_{I},\qquad \nabla Z_{I}=P_{I}^{a}Z_{a}.
\label{cm}
\end{equation}%
Notice in passing that $Z_{I}=0$ is a solution of $\nabla Z_{a}=0$.
The same property is valid for $Z_{a}=0$ which solves $\nabla
Z_{I}=0$.

(b) \emph{Effective black hole potential}\newline Using the dressed
charges (\ref{czz}-\ref{dzz}), we can write down the gauge
invariant effective scalar potential $\mathcal{V}_{BH}$. Following \textrm{%
\cite{B3,F2}}, $\mathcal{V}_{BH}$ reads as,%
\begin{equation}
\mathcal{V}_{BH}\left( \sigma ,L\right) =\left( Z_{a}Z^{a}\right)
+\left( Z_{I}Z^{I}\right) ,  \label{pot}
\end{equation}%
which can be also put in the form%
\begin{equation*}
\mathcal{V}_{BH}\left( \sigma ,L\right) =e^{-2\sigma }\left[ \left(
R_{a}R^{a}\right) +\left( R_{I}R^{I}\right) \right] .
\end{equation*}%
Clearly $\mathcal{V}_{BH}$, which is positive, is manifestly gauge
invariant under both: \newline (\textbf{a}) the $U^{4}\left(
1\right) \times U^{20}\left( 1\right) $ gauge transformations since
the vectors $Z_{a}$ and $Z_{I}$ depend on the electric
charges of the field strengths only which, as we know, are gauge invariant.%
\newline
(\textbf{b}) the gauge transformations of the $SO\left( 4\right)
\times SO\left( 20\right) $ isotropy group of the moduli space.
$\mathcal{V}_{BH}$ is given by scalar products of the vectors
$Z_{a}$ and $Z^{a}$ (resp $Z^{I}$ and $Z_{I}$).\newline
Using eqs(\ref{czz}), we can express the black hole potential as follows:%
\begin{equation*}
\mathcal{V}_{BH}=e^{-2\sigma }\left( q^{a}\mathcal{N}_{ab}q^{b}+q^{a}%
\mathcal{N}_{aJ}p^{J}+p^{I}\mathcal{N}_{Ib}q^{b}+p^{I}\mathcal{N}%
_{IJ}p^{J}\right) ,
\end{equation*}%
or in a condensed manner like $\mathcal{V}_{BH}=e^{-2\sigma }Q^{\Lambda }%
\mathcal{N}_{\Lambda \Sigma }Q^{\Sigma }$ with%
\begin{equation*}
\mathcal{N}_{\Lambda \Sigma }=\left(
\begin{array}{cc}
\mathcal{N}_{ab} & \mathcal{N}_{aJ} \\
\mathcal{N}_{aJ} & \mathcal{N}_{IJ}%
\end{array}%
\right)
\end{equation*}%
Notice that, like for BS, $\mathcal{N}_{\Lambda \Sigma }$ has a 6D
filed theoretical interpretation in terms of the {\LARGE \ }gauge
coupling of the
gauge field strengths $\mathcal{F}_{\mu \nu }^{\Lambda }$; i.e a term like $%
\frac{1}{4}\sqrt{-g}\mathcal{N}_{\Lambda \Sigma }\mathcal{F}_{\mu
\nu }^{\Lambda }\mathcal{F}^{\mu \nu \Sigma }$ appears in the
component fields of the 6D $\mathcal{N}=2$ supergravity Lagrangian
density.

\subsection{7D N=2 supergravity}

Here we discuss briefly the effective scalar potential of the black
objects in 7D.\ This study is quite similar to the previous 6D
analysis. Recall that the moduli space of this theory is given by
$\frac{SO\left( 3,19\right) \times SO\left( 1,1\right) }{SO\left(
3\right) \times SO\left( 19\right) }$. In 7D space time, the bosonic
fields content of the $\mathcal{N}=2$
supergravity multiplet is given by%
\begin{equation*}
\left( g_{\mu \nu },\text{ \ }B_{\left[ \mu \nu \right] },\text{ \ }\mathcal{%
A}_{\mu }^{a},\text{ \ }\sigma \right) ,\qquad a=1,2,3,\qquad \mu
,\nu ,\rho =0,...,6,
\end{equation*}%
where $B_{\left[ \mu \nu \right] }$ is dual to a 3- form gauge field\ $C_{%
\left[ \mu \nu \sigma \right] }$. There is also nineteen U$\left(
1\right) $ Maxwell with the following 6D bosons:
\begin{equation*}
\left( \mathcal{A}_{\mu }^{I}\text{ \ },\text{ \ }\rho ^{aI}\right)
,\qquad a=1,2,3,\qquad I=1,...,19,\text{\ }
\end{equation*}%
where $\rho ^{aI}$ capture $3\times 19$ degrees of freedom. The
gauge invariant $\left( p+2\right) $- forms of the 7D
$\mathcal{N}=2$ supergravity
are given by%
\begin{equation*}
H_{3}\sim dB_{2},\qquad \mathcal{F}_{2}^{a}\sim
d\mathcal{A}^{a},\qquad \mathcal{F}_{2}^{I}\sim d\mathcal{A}^{I}.
\end{equation*}%
Extending the above 6D study to the 7D case, one distinguishes:
\newline (\textbf{i}) 7D black 2- brane (black membrane BM). The
effective scalar potential of the BM is
\begin{equation*}
\mathcal{V}_{BM}^{7D}\left( \sigma \right) \sim Z^{2}=e^{-4\sigma
}g^{2},
\end{equation*}%
with $g=\int_{S^{3}}H_{3}$. The extremum of this potential is given by $%
\sigma =\infty $. The value of the potential at the minimum is
$\left[ \mathcal{V}_{BM}^{7D}\left( \infty \right) \right] _{\min
}=0$ and so the entropy vanishes identically. \newline (\textbf{ii})
7D black hole: The effective potential of this black hole is
given by%
\begin{equation}
\mathcal{V}_{BH}^{7D}\left( \sigma ,L\right)
=\sum_{a=1}^{3}Z_{a}Z^{a}+\sum_{I=1}^{19}Z_{I}Z^{I},  \label{ral}
\end{equation}%
where%
\begin{equation}
Z_{a}=e^{-\sigma }L_{a\Lambda }g^{\Lambda },\qquad Z_{I}=e^{-\sigma
}L_{a\Lambda }g^{\Lambda },  \label{rel}
\end{equation}%
satisfying the constraint relation,%
\begin{equation*}
\sum_{a=1}^{3}Z_{a}Z^{a}-\sum_{I=1}^{19}Z_{I}Z^{I}=Q^{2},\qquad
\left( \sum_{a=1}^{3}q_{a}q^{a}-\sum_{I=1}^{19}p_{I}p^{I}\right)
=Q^{2}
\end{equation*}%
and $Q^{\Lambda }=\left( q^{a},p^{I}\right) $ with $q^{a}=\int_{S^{2}}%
\mathcal{F}_{2}^{a},$ $p^{I}=\int_{S^{2}}\mathcal{F}_{2}^{I},$ $a=1,2,3,$ $%
I=1,...,19$. The real $22\times 22$ matrix
\begin{equation}
L_{a\Lambda }=\left(
\begin{array}{cc}
L_{ab} & \rho _{aI} \\
V_{Ia} & L_{IJ}%
\end{array}%
\right) ,
\end{equation}%
is associated with the group manifold $SO\left( 3,19\right)
/SO\left(
3\right) \times SO\left( 19\right) $. It is an orthogonal matrix satisfying $%
L^{t}\eta L=\eta $ with $\eta =diag\left[ 3\left( +\right) ,19\left(
-\right) \right] .$ The $SO\left( 3\right) \times SO\left( 19\right)
$
symmetry can be used to choose $L_{ab}$ and $L_{IJ}$\ matrices as $%
L_{ab}-L_{ba}=0,$ $L_{IJ}-L_{JI}=0$. Putting the relations
(\ref{rel}) back into (\ref{ral}), we get
$\mathcal{V}_{BH}^{7D}\left( \sigma ,L\right)
=e^{-2\sigma }Q^{\Lambda }\mathcal{N}_{\Lambda \Sigma }Q^{\Sigma }$ where $%
\mathcal{N}_{\Lambda \Sigma }=\left( L_{a\Lambda }L_{\Sigma
}^{a}+L_{I\Lambda }L_{\Sigma }^{I}\right) $.

\end{document}